\documentclass[11pt]{article}

\usepackage[a4paper,margin=1in]{geometry}

\usepackage{amsmath,amssymb,amsthm}
\usepackage{graphicx}
\usepackage{subcaption}
\usepackage{multirow}
\usepackage{nicematrix}
\usepackage{tikz}
\usetikzlibrary{arrows.meta}
\usepackage{booktabs}
\usepackage{threeparttable}
\usepackage{algorithm2e}
\usepackage{pdflscape}
\usepackage{natbib}
\usepackage{xurl}
\usepackage{url}
\usepackage{xcolor}

\definecolor{linkblue}{RGB}{0,80,160}

\usepackage[
    colorlinks=true,
    linkcolor=linkblue,
    citecolor=linkblue,
    urlcolor=linkblue
]{hyperref}
\usepackage{authblk}

\definecolor{LightBlue}{rgb}{0.7,0.85,1}
\definecolor{Coral}{rgb}{1,0.6,0.6}
\definecolor{MixedColor}{rgb}{0.85,0.725,0.8}
\definecolor{MixedColor1}{rgb}{0.925,0.8625,0.9}
\definecolor{MixedColor2}{rgb}{0.8875,0.84375,0.9}
\definecolor{MixedColor3}{rgb}{0.8515,0.82575,0.9}
\definecolor{MixedColor4}{rgb}{0.925,0.7625,0.8}
\definecolor{MixedColor5}{rgb}{0.8875,0.74375,0.8}
\definecolor{MixedColor6}{rgb}{0.85,0.725,0.8}
\definecolor{correct}{RGB}{31,119,180}
\definecolor{over}{RGB}{255,127,14}
\definecolor{under}{RGB}{44,160,44}

\makeatletter
\newtheoremstyle{myremark}
  {\topsep}
  {\topsep}
  {\normalfont}
  {}
  {\itshape}
  {.}
  {.5em}
  {\thmname{#1}\thmnumber{ #2}\thmnote{ {\itshape(#3)}}}
\makeatother

\theoremstyle{myremark}
\newtheorem{definition}{Definition}
\newtheorem{thm}{Proposition}
\newtheorem{rmk}{Remark}

\title{Decomposing Co-movements in Matrix-Valued Time Series:
A Pseudo-Structural Reduced Rank Approach}

\author[1]{Alain Hecq}
\author[1]{Ivan Ricardo\thanks{Corresponding author. Email:
\href{mailto:iu.ricardo@maastrichtuniversity.nl}{iu.ricardo@maastrichtuniversity.nl}}}
\author[1]{Ines Wilms}

\affil[1]{Department of Quantitative Economics,
Maastricht University,
Maastricht, The Netherlands}

\date{July 7, 2026}

\begin{document}

\maketitle

\begin{abstract}
A pseudo-structural framework is proposed for analyzing contemporaneous co-movements in stationary reduced-rank matrix autoregressive (RRMAR) models.
Unlike conventional vector autoregressive (VAR) models that discard the matrix structure, the formulation preserves it, enabling a decomposition of co-movements into three interpretable components: row-specific, column-specific, and joint (row--column) interactions across the matrix-valued time series.
The estimator admits standard asymptotic inference and a BIC-type criterion is proposed for the joint selection of the reduced ranks and the autoregressive lag order.
The method's finite-sample performance in terms of estimation accuracy, coverage, and rank selection is validated through simulation experiments, including cases of rank misspecification.
Practical usefulness is illustrated through an application to labor market data from nine Midwestern U.S. states, revealing distinct row-, column-, and joint co-movement patterns.
\end{abstract}


\section{Introduction}

Macroeconomic and financial time series have expanded both in number and complexity.
This complexity is inherently multidimensional -- researchers routinely observe 
entities (e.g., countries, firms) across multiple indicators (e.g., GDP growth, inflation, unemployment) over time. The time series thus oftentimes displays a matrix structure: a series of matrix data are observed over time.
Traditional methods, such as vector autoregressions (VARs; e.g., \citealp{lutkepohl2005new}) would typically vectorize matrix-valued data into a single high-dimensional vector which loses the row- and column-specific information of the data.
Panel VARs \citep{holtz1988estimating, koop_forecasting_2019} can handle multiple cross-sections but do not naturally model the two-way dependencies that are intrinsic to matrix-valued data.
The methods for matrix-valued time series (MVTS; \citealp{chen_autoregressive_2021, tsay_matrix-variate_2024, zhang2024additive, samadi_matrix-valued_2025}) address this gap by preserving and exploiting the matrix structure of the data.
However, MVTS models typically still encounter difficulties with the large number of parameters that need to be estimated.
Recently, however, reduced-rank matrix autoregressive models (RRMAR, \citealp{xiao_reduced_forth}) 
have shown their promise to this end, by exploiting low-rank decompositions of autoregressive coefficient matrices.

The RRMAR is used to decompose contemporaneous co-movements in matrix-valued time series into three interpretable components, in line with economic intuition: row-specific, column-specific, and joint (row and column) co-movements, thereby isolating co-movement relations within and across the matrix dimensions.
This decomposition is obtained by casting the RRMAR in a \textit{pseudo-structural} form,
imposing identification restrictions (analogous to those in structural VARs) that rotate factor matrices so that co-movements partition into the three components.
These co-movements are interpreted through the lens of serial correlation common features \citep{engle_testing_1993}, where a common serial correlation occurs if a linear combination of a series is free of serial correlation even though each individual
series displays it.
The pseudo-structural specification then permits standard asymptotic inference on row-, column-, and joint co-movement parameters.

Furthermore, a carefully designed algorithm is proposed to identify the co-movements, and practical guidance on selecting the reduced ranks is offered through a BIC-type criterion.
The finite-sample properties of estimation, inference, and rank selection procedures are evaluated through Monte Carlo simulations, including experiments with rank under- and over-specification.
The simulations demonstrate good accuracy in parameter recovery in cases of rank under- and over-specification, while valid coverage of confidence intervals is only present in cases of rank over-specification.
Moreover, rank selection procedures using BIC often identify the true rank, even in cases of rank over-specification for one of the matrix-valued time series dimensions.
Finally, the pseudo-structural framework is illustrated through an empirical application to coincident indexes across nine Midwestern U.S. states, uncovering distinct row-, column-, and joint co-movement patterns reflecting both within-state labor market dynamics and cross-state linkages.

The decomposition builds on existing work on matrix-valued time series (MVTS) and reduced-rank regression but differs in both emphasis and methodology.
\citet{xiao_reduced_forth} introduce the reduced-rank matrix autoregressive model (RRMAR), provide estimation procedures, establish parameter consistency, propose an extended BIC for rank selection, and evaluate model performance in forecasting exercises.
In contrast, it is shown that the RRMAR can be used to characterize contemporaneous co-movements in matrix-valued time series and to decompose those co-movements into three interpretable components (row, column, and joint interactions).
The simulation study examines the kernel densities of the decomposed parameters under three rank-specification scenarios — overestimation, underestimation, and correct specification — and complements \citet{xiao_reduced_forth} by reporting results for the traditional BIC and for models with longer lag lengths.

Beyond the RRMAR literature, MVTS factor models \citep{wang_factor_2019, chen_factor_2022, chen2023inference, gao2023two} have been used to extract common row/column factors for dimension reduction, and reduced-rank regression methods \citep{cubadda_reduced_2022, cubadda_vector_2017, cubadda_representation_2019, cubadda_dimension_2022} have been applied to forecasting and co-movement detection \citep{escribano1994cointegration, cubadda_studying_2009}.
Moreover, this dimension reduction can be extended to the nonstationary case, where cointegration may occur over the row and column dimensions \citep{li_cointegrated_2024, chen2025inference, hecq2025cointegrated, lopetuso_cointegrated_2025}.
By contrast, the pseudo-structural reduced-rank formulation explicitly separates contemporaneous interactions of stationary matrix-valued time series into row, column, and joint components — a decomposition not provided by existing high-dimensional methods that separate lagged from contemporaneous predictors \citep{wang_high-dimensional_2022, wang_high-dimensional_2023}.
The inferential procedure targets these distinct contemporaneous components directly, yielding interpretable measures of how rows and columns interact contemporaneously in matrix-valued time series.

The remainder of the paper is structured as follows. Section \ref{sec:representation} starts by building the foundation for the pseudo-structural representation of the RRMAR.
Section \ref{sec:estimationselection} details the estimation of the pseudo-structural model and how the ranks of the coefficient matrices are selected.
Section \ref{sec:simulation} contains a simulation study evaluating the estimation and inference of pseudo-structural parameters and the proposed rank selection procedure.
Section \ref{sec:empirical} illustrates the pseudo-structural framework through an application to coincident indexes across nine Midwestern U.S. states.
Section \ref{sec:conclusion} concludes and discusses future research avenues.

A word on notation.
Throughout the paper, scalars are denoted by lowercase letters $x$, vectors by boldface lowercase letters $\mathbf{x}$, and matrices by boldface capital letters $\mathbf{X}$.
For a generic matrix $\mathbf{X}$, $\mathbf{X}^\top$, $\|\mathbf{X}\|_F$, and $\text{vec}(\mathbf{X})$ denote, respectively, the transpose, the Frobenius norm, and the column-wise vectorization.
Finally, the nullspace and column space of a matrix are denoted by $\mathcal{N}(\cdot)$ and $\mathcal{C}(\cdot)$, respectively.

\section{Theoretical Framework and Pseudo-Structural Representation}
\label{sec:representation}
Section \ref{sec:mvtslit} reviews the reduced-rank matrix-valued time series models under study and then defines contemporaneous co-movements within the framework of serial correlation common features.
Section \ref{sec:pseudostructural} presents the equivalent pseudo-structural form for these reduced rank matrix-valued time series models.
Section \ref{sec:sccf} intuitively discusses the link between the serial correlation common features and the reduced rank matrix autoregressive model through an example of countries and economic indicators.

\subsection{Co-movements in Reduced Rank Matrix-Valued Models}
\label{sec:mvtslit}

The reduced-rank matrix autoregressive (RRMAR) model, which forms the basis of the analysis of contemporaneous co-movements, is reviewed in this section.
The model, with one autoregressive lag, is 
given by
\begin{align}
    \label{eq:rrmar}
    \mathbf{Y}_{t} = \mathbf{U}_{1} \mathbf{U}_{3}^{\top} \mathbf{Y}_{t-1} \mathbf{U}_{4} \mathbf{U}_{2}^{\top} + \mathbf{E}_{t},
\end{align}
where $\mathbf{Y}_{t} \in \mathbb{R}^{N_{1} \times N_{2}}$ is the response matrix, $\mathbf{U}_{1}, \mathbf{U}_{3} \in \mathbb{R}^{N_{1} \times r_{1}}$, and $\mathbf{U}_{2}, \mathbf{U}_{4} \in \mathbb{R}^{N_{2} \times r_{2}}$ are the coefficient matrices with (reduced) row rank $1 \leq r_{1} \leq N_{1}$ and column rank $1 \leq r_{2} \leq N_{2}$.
The errors are assumed to follow a matrix-valued normal distribution \citep{dawid1981matrix}, namely
\begin{align*}
    \mathbf{E}_t \sim MVN(\mathbf{0}, \boldsymbol{\Sigma}_{1}, \boldsymbol{\Sigma}_{2}) \Leftrightarrow \mathbf{e}_{t} = \operatorname{vec}(\mathbf{E}_{t})  \sim N (\operatorname{vec}(\mathbf{0}), \boldsymbol{\Sigma}_{2} \otimes \boldsymbol{\Sigma}_{1}),
\end{align*}
where $\boldsymbol{\Sigma}_{1} \in \mathbb{R}^{N_1 \times N_1}$ and $\boldsymbol{\Sigma}_{2} \in \mathbb{R}^{N_2 \times N_2}$ are positive definite row- and column-covariance matrices, and
the notation $MVN(\cdot,\cdot,\cdot)$ denotes the matrix-valued normal distribution and $N(\cdot,\cdot)$ denotes the multivariate normal distribution.
To simplify the analysis, each series is assumed to have been demeaned over time to remove the constant term.

\begin{rmk}
    The Kronecker structure $\boldsymbol{\Sigma}_2 \otimes \boldsymbol{\Sigma}_1$ on the error covariance matrix imposes separability between the row and column covariance, reducing the number of free parameters.
    This is a strong assumption which may be restrictive in practice.
    In empirical work, separability can be assessed prior to estimation using dedicated tests (e.g., \citealp{nelson2005likelihood, mitchell2006likelihood}).
    If separability is misspecified, the full information maximum likelihood estimators may retain consistency, as the model does not itself rely on the Kronecker structure. 
    However, this may invalidate the inference procedure proposed here.
\end{rmk}

Defining $\mathbf{y}_{t} = \operatorname{vec}(\mathbf{Y}_{t})$, the equivalent vectorized form is 
\begin{align}
    \label{eq:vec_rrmar}
    \mathbf{y}_{t} = \underbrace{(\mathbf{U}_{2} \otimes \mathbf{U}_{1}) (\mathbf{U}_{4} \otimes \mathbf{U}_{3})^\top}_{\mathbf{A}} \mathbf{y}_{t-1} + \mathbf{e}_{t}.
\end{align}
Stationarity of the model requires the spectral radius of $\mathbf{A}  \in \mathbb{R}^{N_1N_2 \times N_1 N_2}$ to be strictly less than one.
The Kronecker structure of $\mathbf{A}$ suggests the presence of shared dynamic behavior across the different series, which is formalized using the concept of common co-movements introduced by \citet{engle_testing_1993}.
\begin{definition}[Serial Correlation Common Feature (SCCF), \citealp{engle_testing_1993}]
    A feature will be said to be \textit{common} if a linear combination of the series fails to have the feature even though each of the series individually has the feature.
\end{definition}

\begin{rmk}
Definition 1 excludes \emph{trivial} serial correlation common features (see e.g., \citealp{ englegranger1987cointegration} or more recently \citealp{hualde2026trivial}). 
A trivial SCCF arises when a linear combination removes serial correlation only because one of the series in the combination is already serially uncorrelated (white noise), so the null vector merely selects that series rather than revealing a genuine co-movement.
\end{rmk}

Two left null space matrices, $\boldsymbol{\delta} \in \mathbb{R}^{N_{1} \times (N_{1} - r_{1})}$ and $\boldsymbol{\gamma} \in \mathbb{R}^{N_{2} \times (N_{2} - r_{2})}$, are introduced which annihilate the row and column dynamics, respectively -- that is, they satisfy $\boldsymbol{\delta}^{\top} \mathbf{U}_{1} = \mathbf{0}$ and $\boldsymbol{\gamma}^{\top} \mathbf{U}_{2} = \boldsymbol{0}$, where $\boldsymbol{0}$ denotes a conformable matrix of zeros.
In economic terms, the null space matrices $\boldsymbol{\delta}$ and $\boldsymbol{\gamma}$ identify linear combinations of rows and columns of $\mathbf{Y}_t$ that remove the serial correlation generated by $\mathbf{U}_1$ and $\mathbf{U}_2$, and these linear combinations reveal the presence of common contemporaneous co-movements.
However, due to the Kronecker structure of the coefficient matrix, three matrices in total annihilate the serially correlated component $(\mathbf{U}_{2} \otimes \mathbf{U}_{1}) (\mathbf{U}_{4} \otimes \mathbf{U}_{3})^{\top} \mathbf{y}_{t-1}$:
\begin{align*}
    \left(\mathbf{I}_{N_{2}} \otimes \boldsymbol{\delta}\right)^{\top} \mathbf{y}_{t} &= \left(\mathbf{I}_{N_{2}} \otimes \boldsymbol{\delta}\right)^{\top} \mathbf{e}_{t}, \\
    \left(\boldsymbol{\gamma} \otimes \mathbf{I}_{N_{1}}\right)^{\top} \mathbf{y}_{t} &= \left(\boldsymbol{\gamma} \otimes \mathbf{I}_{N_{1}}\right)^{\top} \mathbf{e}_{t}, \\
    \left(\boldsymbol{\gamma} \otimes \boldsymbol{\delta}\right)^{\top} \mathbf{y}_{t} &= \left(\boldsymbol{\gamma} \otimes \boldsymbol{\delta}\right)^{\top} \mathbf{e}_{t}.
\end{align*}

This raises the question: How do the individual null spaces $\boldsymbol{\delta}$ and $\boldsymbol{\gamma}$ interact to annihilate the Kronecker product dynamics $\mathbf{U}_{2} \otimes \mathbf{U}_{1}$?
Proposition \ref{prop:one} resolves this by providing an explicit form for the null space of $(\mathbf{U}_{2} \otimes \mathbf{U}_{1})^{\top}$ which decomposes into three orthogonal components; its proof is given in Appendix \ref{sec:proofprop1}.

\begin{thm}
    \label{prop:one}
    For a reduced rank matrix autoregressive model with coefficient matrix $(\mathbf{U}_{2} \otimes \mathbf{U}_{1}) (\mathbf{U}_{4} \otimes \mathbf{U}_{3})^{\top}$ in its vectorized form, where $\mathbf{U}_{1} \in \mathbb{R}^{N_{1} \times r_{1}}$ and $\mathbf{U}_{2} \in \mathbb{R}^{N_{2} \times r_{2}}$, let $\boldsymbol{\delta} \in \mathbb{R}^{N_{1} \times (N_{1} - r_{1})}$ and $\boldsymbol{\gamma} \in \mathbb{R}^{N_{2} \times (N_{2} - r_{2})}$ satisfy $\boldsymbol{\delta}^{\top} \mathbf{U}_{1} = \mathbf{0}$ and $\boldsymbol{\gamma}^{\top} \mathbf{U}_{2} = \mathbf{0}$. 
    Then, 
    \begin{align*}
    \mathcal{N}\!\left( (\mathbf{U}_2 \otimes \mathbf{U}_1)^\top \right) 
    &= \left( \mathcal{N}(\mathbf{U}_2^\top) \otimes \mathcal{C}(\mathbf{U}_1) \right) 
       \oplus \left( \mathcal{C}(\mathbf{U}_2) \otimes \mathcal{N}(\mathbf{U}_1^\top) \right) \\
    &\quad \oplus \left( \mathcal{N}(\mathbf{U}_2^\top) \otimes \mathcal{N}(\mathbf{U}_1^\top) \right),
    \end{align*}
    where $\oplus$ is the direct sum of subspaces and $\otimes$ is the Kronecker product.
\end{thm}

The three orthogonal components in the decomposition of Proposition \ref{prop:one} correspond to (i) column-specific co-movements $\left(\mathcal{N}(\mathbf{U}_{2}^{\top}) \otimes \mathcal{C}(\mathbf{U}_{1})\right)$, (ii) row-specific co-movements $\left(\mathcal{C}(\mathbf{U}_{2}) \otimes \mathcal{N}(\mathbf{U}_{1}^{\top})\right)$, and (iii) joint co-movements $\left(\mathcal{N}(\mathbf{U}_{2}^{\top}) \otimes \mathcal{N}(\mathbf{U}_{1}^{\top})\right)$.
However, the interpretation of the joint co-movement component is not immediately obvious.
In what follows, it is first shown how each of the three components can be formalized in a pseudo-structural model; which is an equivalent representation of the RRMAR model in \eqref{eq:rrmar}.
The section ends with a toy example to illustrate how each component can be intuitively interpreted.

\begin{rmk}
The three pseudo-structural components form a direct sum of orthogonal 
subspaces, as established in Proposition \ref{prop:one}, and are therefore 
algebraically non-redundant. However, they share a common generating 
structure: both the row-specific component 
$\mathcal{C}(\mathbf{U}_2) \otimes \mathcal{N}(\mathbf{U}_1^\top)$ and the 
joint component $\mathcal{N}(\mathbf{U}_2^\top) \otimes \mathcal{N}(\mathbf{U}_1^\top)$ 
involve $\mathcal{N}(\mathbf{U}_1^\top)$, and similarly both the column-specific 
and joint components involve $\mathcal{N}(\mathbf{U}_2^\top)$. This shared 
dependence on the same null spaces means that the three components are not 
statistically independent, even though they are orthogonal as subspaces. 
The joint component is distinguished by requiring \emph{simultaneous} 
annihilation of both dimensions, a condition that neither the row- nor 
column-specific component imposes on its own. Each component therefore 
captures co-movements that are not recoverable from the others without 
imposing additional restrictions.
\end{rmk}

\subsection{Pseudo-Structural Form}
\label{sec:pseudostructural}

\citet{vahid_common_1993} construct a \emph{pseudo-structural} form that is algebraically equivalent to a reduced-rank VAR and decomposes dynamics into contemporaneous and lagged components.
To keep the exposition self-contained, their approach is first reviewed and then extended to matrix-valued time series by deriving an analogous pseudo-structural representation for the RRMAR.
To keep notation (relatively) compact, the pseudo-structural model corresponding to the RRMAR in \eqref{eq:rrmar} is presented for a single lag.
The correspondence also holds for multi-lag RRMAR models, albeit with additional notation (see Remark \ref{remark:RRMAR-lags}).

Let the reduced-rank VAR($1$) for an \(N\)-dimensional stationary vector process with rank \(r\) be given by
\[
    \mathbf{y}_{t} = \mathbf{A}\mathbf{B}^\top \mathbf{y}_{t-1} + \mathbf{e}_{t},
\]
where \(\mathbf{y}_t\in\mathbb{R}^N\), \(\mathbf{A},\mathbf{B}\in\mathbb{R}^{N\times r}\), and \(\operatorname{rank}(\mathbf{A}\mathbf{B}^\top)= 1 \leq r<N\).
Hence, there exists a left null space basis \(\boldsymbol{\psi}\in\mathbb{R}^{N\times(N-r)}\) with \(\boldsymbol{\psi}^\top\mathbf{A}=\mathbf{0}\).
Rotating \(\boldsymbol{\psi}\) so its top block is the identity, it can be written as
\[
    \boldsymbol{\psi}=\begin{bmatrix} \mathbf{I}_{N-r} \\ \boldsymbol{\psi}^* \end{bmatrix},
\]
with $\boldsymbol{\psi}^*\in\mathbb{R}^{r\times (N-r)}$, possibly after reordering the variables.
Because $\boldsymbol{\psi}$ has full column rank $N-r$, the rows can always be permuted so that the top $N-r$ rows form an invertible block, which is then right-multiplied by its inverse to obtain the displayed form.
Then \(\boldsymbol{\psi}^\top\mathbf{y}_t\) yields \((N-r)\) pseudo-structural equations, while the remaining \(r\) equations form reduced-form regressions that use unrestricted lags as instruments.
Stacking these relations gives
\begin{align}
    \label{eq:vec-pseudo}
    \underbrace{
    \begin{bmatrix}
	\mathbf{I}_{N-r} & \boldsymbol{\psi}^{*\top} \\
	\mathbf{0} & \mathbf{I}_{r}
    \end{bmatrix}}_{\boldsymbol{\Omega}} 
    \begin{bmatrix}
        \mathbf{y}_{1,t} \\
        \mathbf{y}_{2,t}
    \end{bmatrix} = 
    \underbrace{\begin{bmatrix}
	    \mathbf{0} & \mathbf{0}\\
	    \boldsymbol{\Pi}_{1}^{*} & \boldsymbol{\Pi}_{2}^{*}
    \end{bmatrix}}_{\boldsymbol{\Pi}} 
    \begin{bmatrix}
        \mathbf{y}_{1,t-1} \\
        \mathbf{y}_{2,t-1}
    \end{bmatrix} 
    +
    \begin{bmatrix}
	\mathbf{I}_{N-r} & \boldsymbol{\psi}^{*\top} \\
	\mathbf{0} & \mathbf{I}_{r}
    \end{bmatrix}
    \begin{bmatrix}
        \mathbf{e}_{1,t} \\
        \mathbf{e}_{2,t}
    \end{bmatrix},
\end{align}
where \(\boldsymbol{\Omega} \in \mathbb{R}^{N \times N}\) encodes contemporaneous 
relations and \(\boldsymbol{\Pi} \in \mathbb{R}^{N \times N}\) collects the lagged 
dynamics, with $\boldsymbol{\Pi}_{1}^{*} \in \mathbb{R}^{r \times (N-r)}$ and 
$\boldsymbol{\Pi}_{2}^{*} \in \mathbb{R}^{r \times r}$ denoting the unrestricted 
lag coefficient matrices on $\mathbf{y}_{1,t-1}$ and $\mathbf{y}_{2,t-1}$, 
respectively.

The same logic is now applied to the RRMAR in \eqref{eq:rrmar}.
Partition \(\mathbf{Y}_t\) into four (unbalanced) blocks:
\begin{align}
    \label{eq:partition}
    \mathbf{Y}_{t} = 
    \begin{bmatrix}
	\mathbf{Y}_{11,t} & \mathbf{Y}_{12,t} \\
	\mathbf{Y}_{21, t} & \mathbf{Y}_{22,t}
    \end{bmatrix},
\end{align}
where $\mathbf{Y}_{11,t} \in \mathbb{R}^{(N_{1} - r_{1}) \times (N_{2} - r_{2})}$, $\mathbf{Y}_{12,t} \in \mathbb{R}^{(N_{1} - r_{1}) \times r_{2}}$, $\mathbf{Y}_{21,t}\in \mathbb{R}^{r_{1} \times (N_{2} - r_{2})}$, and $\mathbf{Y}_{22,t} \in \mathbb{R}^{r_{1} \times r_{2}}$.
The parameters $\boldsymbol{\delta} \in \mathbb{R}^{N_1 \times (N_1 - r_1)}$ and $\boldsymbol{\gamma} \in \mathbb{R}^{N_2 \times (N_2 - r_2)}$ are rotated to have the $(N_{1} - r_{1})$- and $(N_{2} - r_{2})$-dimensional identity sub-matrix respectively
\begin{align*}
    \boldsymbol{\delta} = 
    \begin{bmatrix}
        \mathbf{I}_{N_{1} - r_{1}} \\
	\boldsymbol{\delta}^{*}
    \end{bmatrix} \quad \text{and} \quad 
    \boldsymbol{\gamma} = 
    \begin{bmatrix}
        \mathbf{I}_{N_{2} - r_{2}} \\
	\boldsymbol{\gamma}^{*}
    \end{bmatrix},
\end{align*}
where $\boldsymbol{\delta}^* \in \mathbb{R}^{r_1 \times (N_1-r_1)}$ and $\boldsymbol{\gamma}^* \in \mathbb{R}^{r_2 \times (N_2-r_2)}$.
There always exists a reordering of rows and columns for which the upper blocks of $\boldsymbol{\delta}$ and $\boldsymbol{\gamma}$ can be normalized as an identity matrix; this is assumed to hold for the adopted order.

This rotation comes from defining two rotation matrices $\mathbf{Q}_1$ and $\mathbf{Q}_2$ such that 
\begin{align*}
    \mathbf{Y}_{t} = \underbrace{\mathbf{U}_{1} \mathbf{Q}_{1}}_{\mathbf{U}_1^*} \underbrace{\mathbf{Q}_{1}^{-1} \mathbf{U}_{3}^{\top}}_{\mathbf{U}_3^{*\top}} \mathbf{Y}_{t-1} \underbrace{\mathbf{U}_{4} \mathbf{Q}_{2}^{-\top}}_{\mathbf{U}_4^{*}} \underbrace{\mathbf{Q}_2^{\top} \mathbf{U}_{2}^\top}_{\mathbf{U}_2^{*\top}} + \mathbf{E}_{t},
\end{align*}
yields the same result as equation \eqref{eq:rrmar}.
Setting these rotation matrices to be the inverse of the bottom $r_1 \times r_1$ or $r_2 \times r_2$ submatrix of $\mathbf{U}_1$ and $\mathbf{U}_2$ yields
\begin{align*}
    \mathbf{U}_{1}^* = 
\begin{bmatrix}
    - \boldsymbol{\delta}^{*\top} \\
    \mathbf{I}
\end{bmatrix} \qquad \text{and} \qquad 
\mathbf{U}_{2}^* = 
\begin{bmatrix}
    - \boldsymbol{\gamma}^{*\top} \\
    \mathbf{I}
\end{bmatrix}.
\end{align*}
Taking the left nullspace of $\mathbf{U}_{1}^*$ and $\mathbf{U}_{2}^*$ then yields 
\begin{align*}
    \mathcal{N}(\mathbf{U}_{1}^{*\top}) = 
\begin{bmatrix}
    \mathbf{I} &
    \boldsymbol{\delta}^{*\top}
\end{bmatrix} \qquad \text{and} \qquad 
    \mathcal{N}(\mathbf{U}_{2}^{*\top}) = 
\begin{bmatrix}
    \mathbf{I} & 
    \boldsymbol{\gamma}^{*\top}
\end{bmatrix}.
\end{align*}
Because left multiplying by the left nullspace yields the zero vector, this is a valid nullspace for each of $\mathbf{U}_{1}^*$ and $\mathbf{U}_{2}^*$.

Stacking the two nullspace matrices as in \citet{vahid_common_1993} then gives
\begin{align*}
    \boldsymbol{\Omega}_1 = 
    \begin{bmatrix}
        \mathbf{I}_{N_1 - r_1} & \boldsymbol{\delta}^{*\top} \\
        \mathbf{0} & \mathbf{I}_{r_1}
    \end{bmatrix} \quad \text{and} \quad
    \boldsymbol{\Omega}_2 = 
    \begin{bmatrix}
        \mathbf{I}_{N_2 - r_2} & \boldsymbol{\gamma}^{*\top} \\
        \mathbf{0} & \mathbf{I}_{r_2}
    \end{bmatrix}.
\end{align*}
This form then allows a pseudo-structural system of equations to be constructed, similar to the form of equation \eqref{eq:vec-pseudo}.
Pre- and post-multiplying the rotated equation \eqref{eq:rrmar} by these matrices thus yields
\begin{align*}
    \boldsymbol{\Omega}_1 \mathbf{Y}_t \boldsymbol{\Omega}_2^\top &= \boldsymbol{\Omega}_1 \mathbf{U}_{1}^* \mathbf{U}_{3}^{*\top} \mathbf{Y}_{t-1} \mathbf{U}_{4}^* \mathbf{U}_{2}^{*\top} \boldsymbol{\Omega}_2^\top + \boldsymbol{\Omega}_1 \mathbf{E}_{t} \boldsymbol{\Omega}_2^\top, \\
    &= 
    \begin{bmatrix}
        \mathbf{0} & \mathbf{0} \\
        \mathbf{0} & \mathbf{U}_3^{*\top} \mathbf{Y}_{t-1} \mathbf{U}_4^*
    \end{bmatrix} + \boldsymbol{\Omega}_1 \mathbf{E}_{t} \boldsymbol{\Omega}_2^\top,
\end{align*}
where the contemporaneous relations on the left hand side of the equation are given by
\begin{align*}
    \boldsymbol{\Omega}_1 \mathbf{Y}_t \boldsymbol{\Omega}_2^\top = 
    \begin{bmatrix}
        \mathbf{Y}_{11} + \boldsymbol{\delta}^{*\top} \mathbf{Y}_{21} + \mathbf{Y}_{12} \boldsymbol{\gamma}^* + \boldsymbol{\delta}^{*\top} \mathbf{Y}_{22} \boldsymbol{\gamma}^* & \mathbf{Y}_{12} + \boldsymbol{\delta}^{*\top} \mathbf{Y}_{22} \\
       \mathbf{Y}_{21} + \mathbf{Y}_{22} \boldsymbol{\gamma}^* & \mathbf{Y}_{22}
    \end{bmatrix}.
\end{align*}
Thus, there are four equations associated with four blocks, three of which are given by the three orthogonal components of Proposition \ref{prop:one}, while the fourth block is a block of lags as instruments, as in equation \eqref{eq:vec-pseudo}.
An example of different normalization choices is included in Appendix \ref{sec:norm_invariance} to illustrate how different normalizations affect the estimated pseudo-structural parameters, but not the reduced-form dynamics.

Interpretation can be difficult if done directly on the matrix forms.
Hence, the system is vectorized while keeping the structure intact to obtain a balanced form similar to that from equation \eqref{eq:vec-pseudo}.
Up to a permutation (detailed in Appendix \ref{sec:companionform}), $\boldsymbol{\Omega}_2 \otimes \boldsymbol{\Omega}_1= \boldsymbol{\Omega} \in \mathbb{R}^{N_1 N_2 \times N_1 N_2}$ and $\boldsymbol{\Pi} \in \mathbb{R}^{N_1 N_2 \times N_1 N_2}$ can be defined as
\begin{align*}
    \boldsymbol{\Omega} = 
    \begin{bmatrix}
	\mathbf{I}_{(N_{1}-r_{1})(N_{2}-r_{2})} & \mathbf{I}_{N_{2} - r_{2}} \otimes \boldsymbol{\delta}^{*\top} & \boldsymbol{\gamma}^{*\top} \otimes \mathbf{I}_{N_{1} - r_{1}} & \boldsymbol{\gamma}^{*\top} \otimes \boldsymbol{\delta}^{*\top} \\
	\mathbf{0} & \mathbf{I}_{r_{1} (N_{2} - r_{2})} & \mathbf{0} & \boldsymbol{\gamma}^{*\top} \otimes \mathbf{I}_{r_{1}} \\
	\mathbf{0} & \mathbf{0} & \mathbf{I}_{r_{2} (N_{1} - r_{1})} & \mathbf{I}_{r_{2}} \otimes \boldsymbol{\delta}^{*\top} \\
	\mathbf{0} & \mathbf{0}  & \mathbf{0} & \mathbf{I}_{r_{1} r_{2}}
    \end{bmatrix}
    \end{align*}
and
    \begin{align}
    \label{eq:omega-and-pi}
    \boldsymbol{\Pi} = 
    \begin{bmatrix}
	\mathbf{0} \\
	\mathbf{0} \\
	\mathbf{0} \\
	(\mathbf{U}_{4} \otimes \mathbf{U}_{3})^{\top}
    \end{bmatrix},
\end{align}
where 
$\boldsymbol{\Omega}$ contains the contemporaneous row- and column-specific relations in  $\boldsymbol{\delta}^{*}$ and $\boldsymbol{\gamma}^{*}$, while
$\boldsymbol{\Pi}$ holds the lags with a matrix autoregressive restriction (see e.g., \citealp{chen_autoregressive_2021}) as instruments.
This results in the pseudo-structural form
\begin{align}
    \label{eq:pseudo-structural}
    \boldsymbol{\Omega} \mathbf{y}_{t}^* = \boldsymbol{\Pi} \mathbf{y}_{t-1} + \boldsymbol{\Omega} \mathbf{e}_{t}^{*},
\end{align}
where $\mathbf{y}_t^* = \begin{bmatrix}
        \mathbf{y}_{11, t}^\top &
        \mathbf{y}_{21, t}^\top &
        \mathbf{y}_{12, t}^\top &
        \mathbf{y}_{22, t}^\top
\end{bmatrix}^\top$ is the stacked block-vectorized form of the block matrix in equation \eqref{eq:partition}.
Note that while $\mathbf{y}_t^*$ uses the block-permuted ordering to align with the block structure of $\boldsymbol{\Omega}$, the predictor $\mathbf{y}_{t-1} = \operatorname{vec}(\mathbf{Y}_{t-1})$ follows the standard column-wise ordering; since $\boldsymbol{\Pi}$ places all non-zero content in its bottom-right block, the two orderings are equivalent on the right hand side.
This pseudo-structural form has exactly the same number of parameters as the RRMAR, as given by the formula $2r_{1}N_{1} - r_{1}^2 + 2r_{2}N_{2} - r_{2}^2$.

To make the balancing of contemporaneous and lagged terms explicit, the pseudo-structural system can be written in a manner similar to that of \citet{vahid_common_1993}:
\begin{align*}
    \mathbf{y}_{11,t} &= -(\mathbf{I}_{N_{2} - r_{2}} \otimes \boldsymbol{\delta}^{*\top}) \,\mathbf{y}_{21,t}
    - (\boldsymbol{\gamma}^{*\top} \otimes \mathbf{I}_{N_{1} - r_{1}})\,\mathbf{y}_{12,t}
    - (\boldsymbol{\gamma}^{*\top} \otimes \boldsymbol{\delta}^{*\top})\,\mathbf{y}_{22,t}
    + (\boldsymbol{\Omega}\mathbf{e}_{t}^{*})_{1,:},\\
    \mathbf{y}_{21,t} &= -(\boldsymbol{\gamma}^{*\top} \otimes \mathbf{I}_{r_{1}})\,\mathbf{y}_{22,t}
    + (\boldsymbol{\Omega}\mathbf{e}_{t}^{*})_{2,:},\\
    \mathbf{y}_{12,t} &= -(\mathbf{I}_{r_{2}} \otimes \boldsymbol{\delta}^{*\top})\,\mathbf{y}_{22,t}
    + (\boldsymbol{\Omega}\mathbf{e}_{t}^{*})_{3,:},\\
    \mathbf{y}_{22,t} &= (\mathbf{U}_{4} \otimes \mathbf{U}_{3})^{\top}\mathbf{y}_{t-1}
    + (\boldsymbol{\Omega}\mathbf{e}_{t}^{*})_{4,:},
\end{align*}
where $(\boldsymbol{\Omega}\mathbf{e}_{t}^{*})_{i,:}$ denotes the $i$th row of $\boldsymbol{\Omega}\mathbf{e}_{t}^{*}$.
This structure is visualized in the diagram of Figure \ref{fig:dag-balancing}, where the lagged effects enter only through $\mathbf{y}_{22,t}$ and then propagate to $\mathbf{y}_{12,t}$, $\mathbf{y}_{21,t}$, and $\mathbf{y}_{11,t}$:
\begin{itemize}
    \item $\mathbf{y}_{11,t}$ is determined contemporaneously by $\mathbf{y}_{21,t}$, $\mathbf{y}_{12,t}$, and $\mathbf{y}_{22,t}$;
    \item $\mathbf{y}_{12,t}$ and $\mathbf{y}_{21,t}$ are determined contemporaneously by $\mathbf{y}_{22,t}$;
    \item $\mathbf{y}_{22,t}$ is determined dynamically by the lagged effect $(\mathbf{U}_{4} \otimes \mathbf{U}_{3})^{\top} \mathbf{y}_{t-1}$.
\end{itemize}

\begin{figure}[!t]
  \centering
    \begin{tikzpicture}[
      ->,
      >={Stealth[length=6pt,width=6pt]},
      line width=0.8pt,
      label style/.style={font=\small, fill=white, inner sep=1pt}
    ]
      \node[circle, draw] (A) at (0,3) {$\mathbf{y}_{11,t}$};
      \node[circle, draw] (B) at (0,0) {$\mathbf{y}_{21,t}$};
      \node[circle, draw] (C) at (4,3) {$\mathbf{y}_{12,t}$};
      \node[circle, draw] (D) at (4,0) {$\mathbf{y}_{22,t}$};
      \node[circle, draw] (E) at (8,0) {$\mathbf{y}_{t-1}$};
      \draw (B) -- (A) 
        node[midway, left, label style] {$-(\mathbf{I} \otimes \boldsymbol{\delta}^{*\top})$};
      \draw (D) -- (C) 
        node[midway, right, label style] {$-(\mathbf{I} \otimes \boldsymbol{\delta}^{*\top})$};
      \draw (C) -- (A) 
        node[midway, above, label style] {$-(\boldsymbol{\gamma}^{*\top} \otimes \mathbf{I})$};
      \draw (D) -- (A) 
        node[midway, above, label style] {$-(\boldsymbol{\gamma}^{*\top} \otimes \boldsymbol{\delta}^{*\top})$};
      \draw (D) -- (B) 
        node[midway, below, label style] {$-(\boldsymbol{\gamma}^{*\top} \otimes \mathbf{I})$};
      \draw (E) -- (D) 
        node[midway, below, label style] {$(\mathbf{U}_4 \otimes \mathbf{U}_3)^\top$};
    \end{tikzpicture}
    \caption{Diagram of the balanced pseudo-structural 
    form. Arrows indicate contemporaneous dependencies with their associated coefficient matrices; the sole lagged effect (from \(\mathbf{y}_{t-1}\)) 
    enters \(\mathbf{y}_{22,t}\).}
  \label{fig:dag-balancing}
\end{figure}

The balanced system decomposes $\mathbf{y}_{t}$ into row-specific, column-specific, and joint components.
The row-specific component, captured by $\mathbf{I} \otimes \boldsymbol{\delta}^{*\top}$, describes co-movements along the row dimension of the matrix-valued time series that are common across all columns.
The column-specific component, captured by $\boldsymbol{\gamma}^{*\top} \otimes \mathbf{I}$, describes co-movements along the column dimension that are common across all rows.
Finally, the joint component, captured by $\boldsymbol{\gamma}^{*\top} \otimes \boldsymbol{\delta}^{*\top}$, identifies co-movements that operate simultaneously along both dimensions, linking $\mathbf{y}_{22,t}$ directly to $\mathbf{y}_{11,t}$.
Crucially, this joint component is not redundant: it captures co-movements that require simultaneous annihilation of both row and column dynamics, and cannot be recovered from the row- and column-specific components alone.

\begin{rmk}
\label{rmk:permutation}
The partition in \eqref{eq:partition} requires selecting which series populate the four blocks, which directly affects interpretation and the estimated block matrices.
In practice, this reduces to a single choice: which $r_1$ rows and $r_2$ columns populate $\mathbf{y}_{22,t}$.
Once these are chosen, the remaining blocks are simply whatever rows and columns were not assigned to $\mathbf{y}_{22,t}$, arranged into the three remaining corners of the partition.
Any such choice that respects the row-and-column structure of the data, corresponding to a permutation of the form $(\mathbf{P}_2 \otimes \mathbf{P}_1)\mathbf{y}_t$, leaves the maximized likelihood unchanged and yields an observationally equivalent model.
A generic permutation that mixes rows and columns, however, breaks the matrix structure and defines a different model with a different likelihood.
In practice, the choice of $\mathbf{y}_{22,t}$ should be guided by economic considerations, as discussed in Section  \ref{sec:empirical}. 
\end{rmk}

\begin{rmk}
The term pseudo-structural is used deliberately to emphasize that it provides an algebraically equivalent decomposition of the RRMAR, isolating interpretable row-, column-, and joint co-movement components. 
The interpretation of the co-movement components should not be considered causal as this would require additional assumptions about the error structure, which are beyond the present scope.
\end{rmk}

\begin{rmk}
    The pseudo-structural form is not limited to the RRMAR of \citet{xiao_reduced_forth} considered here.
    A pseudo-structural form with different restrictions on \(\boldsymbol{\Pi}\) can, amongst others, be connected to reduced-rank tensor models \citep{hecq2024reduced,wang_high-dimensional_2024}, namely by replacing, for instance, \((\mathbf{U}_4\otimes\mathbf{U}_3)^\top\) by \(\mathbf{G}(\mathbf{U}_4\otimes\mathbf{U}_3)^\top\), thereby allowing for the introduction of a core tensor \(\mathbf{G}\).
\end{rmk}

\begin{rmk} \label{remark:RRMAR-lags}
The RRMAR has been presented with a single lag for notational clarity.
The correspondence between the RRMAR and the pseudo-structural form also holds for multiple \(p\) lags. For a setting with multiple lags, the matrix \(\boldsymbol{\Omega}\) would remain the same while \(\boldsymbol{\Pi}\) would need to be generalized to \(\{\boldsymbol{\Pi}_i\}_{i=1}^p\), each with a block structure analogous to the one above, where the lower block of \(\boldsymbol{\Pi}_i\) equals \((\mathbf{U}_{4,i}\otimes\mathbf{U}_{3,i})^\top\).
A companion form can then be constructed, and details for this companion form are given in Appendix \ref{sec:companionform}.
\end{rmk}

\subsection{Serial Correlation Common Feature}
\label{sec:sccf}

To illustrate the reduced-rank restriction and provide some intuition for the row- and column-specific co-movements as well as the joint co-movements, a toy example with $N_1=3$ economic indicators corresponding to $N_2 = 4$ countries is considered. 
The countries are denoted by the United States (USA), Canada (CAN), Germany (DEU), and France (FRA), and the economic indicators as the interest rate (IR), gross domestic product (GDP), and manufacturing production (PROD).
As throughout the paper, $\mathbf{Y}_t$ collects the stationary transformations of these indicators (the demeaned first differences / growth rates), so that the SCCF relations below, in which a linear combination reduces to white noise, are well defined.
Vectorizing the matrix-valued time series yields a $12 \times 12$ coefficient matrix $\mathbf{A}$, which is illustrated in Figure \ref{fig:sccfmat}.
Three cases of reduced rank matrix autoregressive models are discussed, (i) partially reduced only among the rows, (ii) partially reduced only among the columns, and (iii) reduced in both rows and columns.

\begin{figure}
\[
\hspace{-25pt}
\setlength\arraycolsep{3.5pt}
\renewcommand{\arraystretch}{0.73}
\begin{pNiceMatrix}%
[   margin,
    last-col,
    first-col,
  ]
\text{\small GDP} & \Block[fill=red!15,rounded-corners]{2-11}{} a_{1,1} & \Hdotsfor{9} & a_{1,12} & \\
\text{\small PROD} & a_{2,1} & & & & & & & & & & a_{2,12} & \\
\text{\small IR} &
& & & & & & & & & &
& \\
\text{\small GDP} & \Block[fill=cyan!15,rounded-corners]{2-11}{} a_{4,1} & & & & & & & & & & a_{4,12} & \\
\text{\small PROD} & a_{5,1} & & & & & & & & & & a_{5,12} & \\
\text{\small IR} &
& & & & & & & & & & & \\
\text{\small GDP} & \Block[fill=green!15,rounded-corners]{2-11}{} a_{7,1} & & & & & & & & & & a_{7,12} & \\
\text{\small PROD} & a_{8,1} & & & & & & & & & & a_{8,12} & \\
\text{\small IR} &
& & & & & & & & & & & \\
\text{\small GDP} & \Block[fill=violet!15,rounded-corners]{2-11}{} a_{10,1} & & & & & & & & & & a_{10,12} & \\
\text{\small PROD} & a_{11,1} & & & & & & & & & & a_{11,12} & \\
\text{\small IR} &
& & & & & & & & & & & \\
\end{pNiceMatrix}
\hspace{40pt}
\NiceMatrixOptions{xdots={horizontal-labels,line-style = <->}}
\begin{pNiceMatrix}%
[   margin,
    last-col,
  ]
\Block[fill=red!15,rounded-corners]{1-11}{} a_{1,1} & \Hdotsfor{9}[line-style = standard] & a_{1,12} & \Vdotsfor{3}^{\text{USA}} \\
\Block[fill=cyan!15,rounded-corners]{1-11}{} a_{2,1} & & & & & & & & & & a_{2,12} & \\
\Block[fill=green!15,rounded-corners]{1-11}{} a_{3,1} & & & & & & & & & & a_{3,12} & \\
\Vdotsfor{6}[line-style = standard] & & & & & & & & & & & \Vdotsfor{3}^{\text{CAN}} \\
& & & & & & & & & & & \\
& & & & & & & & & & & \\
& & & & & & & & & & & \Vdotsfor{3}^{\text{DEU}} \\
& & & & & & & & & & & \\
& & & & & & & & & & & \\
\Block[fill=red!15,rounded-corners]{1-11}{} a_{10,1} & & & & & & & & & & a_{10,12} & \Vdotsfor{3}^{\text{FRA}} \\
\Block[fill=cyan!15,rounded-corners]{1-11}{} a_{11,1} & & & & & & & & & & a_{11,12} & \\
\Block[fill=green!15,rounded-corners]{1-11}{} a_{12,1} & & & & & & & & & & a_{12,12} & \\
\end{pNiceMatrix}\]
    \caption{SCCF restriction for $\mathbf{A}$ either in the indicator 
    dimension or the state dimension. The left matrix shows the restrictions 
    along the indicator dimension (rank $r_1=2$), while the right matrix 
    shows the restrictions along the state dimension (rank $r_2=3$).}
    \label{fig:sccfmat}
\end{figure}

Firstly, the focus is on an example in which the rank of the economic indicator dimension (rows of the matrix-valued time series) of the coefficient matrix is reduced to two, hence $r_1 = 2$, while the country dimension is left at full rank with $r_2=4$.
Thus, only $\boldsymbol{\delta} = (1, \delta_1^*, \delta_2^*)^\top$ annihilates the dynamics of the system, and the rank of two implies one SCCF co-movement relation for all three economic indicators.
In this example, visualized in the left matrix of Figure \ref{fig:sccfmat}, the GDP and PROD of each country are, for simplicity, allowed to co-move (as highlighted through the same coloring for each of the four countries in Figure \ref{fig:sccfmat}, left panel), but the IR is unrestricted (i.e.\ unrestricted elements are neither displayed nor colored in Figure \ref{fig:sccfmat}, left panel).
The restriction $\delta_{2}^* = 0$ is thus applied since IR is left unrestricted (for simplicity).
These relationships can then be quantified as $a_{1, i} = -\delta_{1}^* a_{2, i}$, $a_{4, i} = -\delta_{1}^* a_{5, i}$, $a_{7, i} = -\delta_{1}^* a_{8, i}$, and $a_{10, i} = - \delta_{1}^* a_{11, i}$ where $\boldsymbol\delta^* = (\delta_{1}^*, \delta_{2}^*)^\top$ with $\delta_{1}^*$ an arbitrary constant representing the scale with respect to the GDP (and the negative sign will become clear once the link with the pseudo-structural form is discussed below).
The result is a restricted null space matrix for the rows of the matrix-valued time series given by $\boldsymbol{\delta} = (1, \delta_1^*, 0)^\top$.
This example illustrates that within-country co-movements can be captured, where (certain) indicators co-move for each country, but the countries themselves do not.
Equivalently, the serially correlated component can be removed through a linear combination of any country's GDP and the same country's PROD, indicative of a serial correlation common feature in the model.
These within-country co-movements, captured by the $\boldsymbol{\delta} = (1, \delta_1^*, \delta_2^*)^\top$, can similarly be represented by the pseudo-structural system of equations.
This form yields four relationships corresponding to each of the four countries in the example.
This yields
\begin{align} \label{eq:toy-delta-eqs}
    \boldsymbol{\delta}^{\top} \mathbf{Y}_t =
    \begin{bmatrix}
        y_{11t} + \delta_1^* y_{21t} + \delta_2^* y_{31t} \\
        y_{12t} + \delta_1^* y_{22t} + \delta_2^* y_{32t} \\
        y_{13t} + \delta_1^* y_{23t} + \delta_2^* y_{33t} \\
        y_{14t} + \delta_1^* y_{24t} + \delta_2^* y_{34t}
    \end{bmatrix}^\top.
\end{align}
This normalization provides the contemporaneous co-movement relations with respect to the first economic indicator (GDP), and if $\delta_{2}^{*}$ is zero, the toy example given above is recovered.
For instance, the co-movement relation for the economic indicators of the USA given in equation \eqref{eq:toy-delta-eqs} would be $y_{11t} = -\delta_1^* y_{21t}$ plus a white noise process represented by the first element of $\boldsymbol{\delta}^\top \mathbf{E}_t$. 
The connection with the toy example can now be directly made since the scaling term for any country's PROD to recover the same country's GDP series is given by $-\delta_1^*$.
GDP is thus the same linear function of PROD and IR across all countries, with the same scalar coefficients $\delta_1^*$ and $\delta_2^*$ applying to every country (with $\delta_2^*=0$ in the toy example).

Secondly, a low rank can arise in the country dimension (columns of the matrix-valued time series), as shown on the right matrix of Figure \ref{fig:sccfmat}.
Suppose the country dimension has reduced rank $r_2=3$, while the indicator dimension remains full rank with $r_1=3$. 
In this case, there exists a vector $\boldsymbol{\gamma} = (1, \gamma_1^*, \gamma_2^*, \gamma_3^*)^\top$ that annihilates the system’s dynamics. 
In the toy example, only the USA and FRA move together, as highlighted through the same coloring for each of the four indicators in Figure \ref{fig:sccfmat}, right panel;  CAN and DEU remain unrestricted (and are therefore neither colored nor displayed).
The restrictions $\gamma_1^* =  \gamma_2^* = 0$ are thus applied.
This relationship can be quantified as $a_{1, i} = - \gamma_{3}^* a_{10, i}$, $a_{2, i} = -  \gamma_{3}^* a_{11, i}$, and $a_{3, i} = -  \gamma_{3}^* a_{12, i}$ for $i = 1, \dots, 12$ and where $\boldsymbol\gamma^*=(\gamma_{1}^*, \gamma_{2}^*, \gamma_{3}^*)$, with $\gamma_{3}^*$ an arbitrary constant representing the scale of the last country (FRA) with respect to the first country (USA), and the other two elements are zero since the countries CAN and DEU are left unrestricted (for simplicity).
This again leads to a restricted null space matrix, but for the columns of the matrix-valued time series, given by $\boldsymbol{\gamma} = (1, 0, 0, \gamma_3^*)^\top$.
Given this normalization, USA and FRA co-move with one another by the scale of $\gamma_{3}^*$,
while 
CAN and DEU, on the other hand, are not bound to move in tandem with any other country in the example.
The same proportionality factor $\gamma_3^*$  governs the relationship between the USA and France for all three indicators simultaneously.
These dynamics thus represent instances of across-country co-movements and can be represented by the pseudo-structural system of equations in an analogous manner as the within-country co-movements discussed above.

Finally, consider the case in which the rank is reduced along both dimensions.
In line with the previous two examples, there are $N_1 = 3$ economic indicators and $N_2 = 4$ countries, but with a reduced rank in both dimensions, namely, $r_1 = 2$ and $r_2 = 3$.
As in the previous examples, the matrices $\boldsymbol{\delta}$ and $\boldsymbol{\gamma}$ are defined as
\setlength{\arraycolsep}{4pt}
\begin{align*}
    \boldsymbol{\delta} = 
    \begin{bNiceMatrix}
	\Block[fill=LightBlue!75, rounded-corners]{1-3}{} 1 &
	\delta_1^* &
	0
    \end{bNiceMatrix}^\top \quad \text{and} \quad \boldsymbol{\gamma} = 
    \begin{bNiceMatrix}
	\Block[fill=Coral!75, rounded-corners]{1-4}{} 1 &
	0 &
	0 &
	\gamma_3^*
    \end{bNiceMatrix}^\top,
\end{align*}
with the restriction $\delta_2^* = \gamma_1^* = \gamma_2^* = 0$ to align with the previous examples where GDP co-moves with manufacturing production and the USA co-moves with France.
The difference here being that the system has a rank reduction in both dimensions, thus an interaction between the two co-movement relations will occur.
Figure \ref{subfig:matrixa} displays the full left null space structure for such a configuration.
The following three groupings of the null space matrix can be observed:
\begin{itemize}
    \item \textit{Row-specific co-movements}: $\left(\mathcal{C}(\mathbf{U}_{2}) \otimes \mathcal{N}(\mathbf{U}_{1}^{\top})\right)$, represented by the blue blocks in $\boldsymbol{\delta}$.
    \item \textit{Column-specific co-movements}: $\left(\mathcal{N}(\mathbf{U}_{2}^{\top}) \otimes \mathcal{C}(\mathbf{U}_{1})\right)$, represented by the red blocks in $\boldsymbol{\gamma}$.
    \item \textit{Joint co-movements}: $\left(\mathcal{N}(\mathbf{U}_{2}^{\top}) \otimes \mathcal{N}(\mathbf{U}_{1}^{\top})\right)$, given by the purple color from the first column.

\end{itemize}

Here, the row-specific and the column-specific co-movements display a particular structure in the full left null space, as visualized by the blue and red coloring: GDP co-moves with the manufacturing production (row-specific in blue) and the USA co-moves with France (column-specific in red).
The joint co-movements form the overlap between the row-specific and the column-specific
structures, as can be seen from the first
column of the null space matrix that highlights (in purple) the interaction of the two null spaces.

This implies that the USA GDP co-moves not only with the USA manufacturing production, but also with the French GDP and the French manufacturing production.

As a final illustration, consider the special case where both $r_{1} = r_{2} = 1$.
Then the left null space becomes large (dimension $11$ in the $12$-series toy example).
The matrices $\boldsymbol{\delta}$ and $\boldsymbol{\gamma}$ are defined as
\begin{align*}
    \boldsymbol{\delta} = 
    \begin{bNiceMatrix}
	\Block[fill=LightBlue!50, rounded-corners]{3-1}{} 1 & \Block[fill=LightBlue, rounded-corners]{3-1}{} 0 \\
	0 & 1 \\
	\delta_{1}^* & \delta_{2}^*
    \end{bNiceMatrix} \quad \text{and} \quad 
    \boldsymbol{\gamma} = 
    \begin{bNiceMatrix}
	\Block[fill=Coral!50, rounded-corners]{4-1}{} 1 & \Block[fill=Coral!75, rounded-corners]{4-1}{} 0 & \Block[fill=Coral, rounded-corners]{4-1}{} 0 \\
	0 & 1 & 0 \\
	0 & 0 & 1 \\
	\gamma_{1}^* & \gamma_{2}^* & \gamma_{3}^*
    \end{bNiceMatrix}.
\end{align*}
One convenient construction of $\boldsymbol{\delta}$ and $\boldsymbol{\gamma}$ that yields an explicit null space matrix is given in Figure \ref{subfig:matrixb}.
As can be seen, the terms associated with $\boldsymbol{\delta}$ form a block structure in the null space (as indicated in blue).
The structure associated with $\boldsymbol{\gamma}$ is indicated in red.
The terms associated with $\boldsymbol{\gamma} \otimes \boldsymbol{\delta}$ form the overlap between the two structures as visible from the purple shading in the figure.
The figures thus show how the joint, row-specific, and column-specific null space components combine, which provides intuition for modeling the orthogonal subspaces explicitly, as in Proposition \ref{prop:one}.

\begin{figure}
    \hspace*{-1.5cm}
    \centering
    \begin{subfigure}[b]{0.45\textwidth}
	\footnotesize
        \centering
        \[
        \setlength\arraycolsep{4.5pt}
        \renewcommand{\arraystretch}{0.25}
        \begin{pNiceMatrix}%
        [   margin,
            first-col,
            last-col,
          ]
          \text{\small GDP} & \Block[fill=LightBlue!75, rounded-corners]{3-1}{} \Block[fill=MixedColor!75, rounded-corners]{1-1}{} 1 & 0 & 0 & 0 & 0 & 0 & \\
        \text{\small PROD} & \delta_1^* & 0 & 0 & 0 & \Block[fill=Coral!75, rounded-corners]{1-1}{} 1 & 0 & \\
        \text{\small IR} & 0 & 0 & 0 & 0 & 0 & \Block[fill=Coral!75, rounded-corners]{1-1}{} 1 & \\
        \text{\small GDP} & \Block[fill=Coral!75, rounded-corners]{1-1}{} 0 & \Block[fill=LightBlue!75, rounded-corners]{3-1}{} 1 & 0 & 0 & 0 & 0 & \\
        \text{\small PROD} & 0 & \delta_1^* & 0 & 0 & \Block[fill=Coral!75, rounded-corners]{1-1}{} 0 & 0 & \\
        \text{\small IR} & 0 & 0 & 0 & 0 & 0 & \Block[fill=Coral!75, rounded-corners]{1-1}{} 0 & \\
        \text{\small GDP} & \Block[fill=Coral!75, rounded-corners]{1-1}{} 0 & 0 & \Block[fill=LightBlue!75, rounded-corners]{3-1}{} 1 & 0 & 0 & 0 & \\
        \text{\small PROD} & 0 & 0 & \delta_1^* & 0 & \Block[fill=Coral!75, rounded-corners]{1-1}{} 0 & 0 & \\
        \text{\small IR} & 0 & 0 & 0 & 0 & 0 & \Block[fill=Coral!75, rounded-corners]{1-1}{} 0 & \\
        \text{\small GDP} & \Block[fill=Coral!75, rounded-corners]{1-1}{} \gamma_3^* & 0 & 0 & \Block[fill=LightBlue!75, rounded-corners]{3-1}{} 1 & 0 & 0 & \\
        \text{\small PROD} & \Block[fill=MixedColor!75, rounded-corners]{1-1}{} \gamma_3^* \delta_1^* & 0 & 0 & \delta_1^* & \Block[fill=Coral!75, rounded-corners]{1-1}{} \gamma_3^* & 0 & \\
        \text{\small IR} & 0 & 0 & 0 & 0 & 0 & \Block[fill=Coral!75, rounded-corners]{1-1}{} \gamma_3^* & \\
        \end{pNiceMatrix}
        \]
        \caption{Rank restriction $r_{1} = 2$ and $r_{2} = 3$.}
        \label{subfig:matrixa}
    \end{subfigure}
    \begin{subfigure}[b]{0.45\textwidth}
	\footnotesize
        \centering
        \[
        \NiceMatrixOptions{xdots={horizontal-labels,line-style = <->}}
        \setlength\arraycolsep{0.4pt}
        \renewcommand{\arraystretch}{0.64}
        \begin{pNiceMatrix}%
        [   margin,
            last-col,
          ]
	  \Block[fill=LightBlue!50, rounded-corners]{3-1}{} \Block[fill=MixedColor1, rounded-corners]{1-1}{} 1 & \Block[fill=LightBlue, rounded-corners]{3-1}{} 0 & \Block[fill=Coral!75, rounded-corners]{1-1}{} 0 & 0 & \Block[fill=Coral, rounded-corners]{1-1}{} 0 & 0 & 0 & 0 & 0 & 0 & 0 & \Vdotsfor{3}^{\text{USA}} \\
          0 & \Block[fill=MixedColor2, rounded-corners]{1-1}{} 1 & 0 & \Block[fill=Coral!75, rounded-corners]{1-1}{} 0 & 0 & \Block[fill=Coral, rounded-corners]{1-1}{} 0 & 0 & 0 & 0 & 0 & 0 & \\
          \delta_1^* & \delta_2^* & 0 & 0 & 0 & 0 & 0 & 0 & \Block[fill=Coral!50, rounded-corners]{1-1}{} 1 & \Block[fill=Coral!75, rounded-corners]{1-1}{} 0 & \Block[fill=Coral, rounded-corners]{1-1}{} 0 & \\
          \Block[fill=Coral!50, rounded-corners]{1-1}{} 0 & 0 & \Block[fill=LightBlue!50, rounded-corners]{3-1}{} \Block[fill=MixedColor3, rounded-corners]{1-1}{} 1 & \Block[fill=LightBlue, rounded-corners]{3-1}{} 0 & \Block[fill=Coral, rounded-corners]{1-1}{} 0 & 0 & 0 & 0 & 0 & 0 & 0 & \Vdotsfor{3}^{\text{CAN}}\\
          0 & \Block[fill=Coral!50, rounded-corners]{1-1}{} 0 & 0 & \Block[fill=MixedColor4, rounded-corners]{1-1}{} 1 & 0 & \Block[fill=Coral, rounded-corners]{1-1}{} 0 & 0 & 0 & 0 & 0 & 0 & \\
          0 & 0 & \delta_1^* & \delta_2^* & 0 & 0 & 0 & 0 & \Block[fill=Coral!50, rounded-corners]{1-1}{} 0 & \Block[fill=Coral!75, rounded-corners]{1-1}{} 1 & \Block[fill=Coral, rounded-corners]{1-1}{} 0 & \\
          \Block[fill=Coral!50, rounded-corners]{1-1}{} 0 & 0 & \Block[fill=Coral!75, rounded-corners]{1-1}{} 0 & 0 & \Block[fill=LightBlue!50, rounded-corners]{3-1}{} \Block[fill=MixedColor5, rounded-corners]{1-1}{} 1 & \Block[fill=LightBlue, rounded-corners]{3-1}{} 0 & 0 & 0 & 0 & 0 & 0 & \Vdotsfor{3}^{\text{DEU}}\\
          0 & \Block[fill=Coral!50, rounded-corners]{1-1}{} 0 & 0 & \Block[fill=Coral!75, rounded-corners]{1-1}{} 0 & 0 & \Block[fill=MixedColor6, rounded-corners]{1-1}{} 1 & 0 & 0 & 0 & 0 & 0 & \\
          0 & 0 & 0 & 0 & \delta_1^* & \delta_2^* & 0 & 0 & \Block[fill=Coral!50, rounded-corners]{1-1}{} 0 & \Block[fill=Coral!75, rounded-corners]{1-1}{} 0 & \Block[fill=Coral, rounded-corners]{1-1}{} 1 & \\
          \Block[fill=Coral!50, rounded-corners]{1-1}{} \gamma_1^* & 0 & \Block[fill=Coral!75, rounded-corners]{1-1}{} \gamma_2^* & 0 & \Block[fill=Coral, rounded-corners]{1-1}{} \gamma_3^* & 0 & \Block[fill=LightBlue!50, rounded-corners]{3-1}{} 1 & \Block[fill=LightBlue, rounded-corners]{3-1}{} 0 & 0 & 0 & 0 & \Vdotsfor{3}^{\text{FRA}}\\
          0 & \Block[fill=Coral!50, rounded-corners]{1-1}{} \gamma_1^* & 0 & \Block[fill=Coral!75, rounded-corners]{1-1}{} \gamma_2^* & 0 & \Block[fill=Coral, rounded-corners]{1-1}{} \gamma_3^* & 0 & 1 & 0 & 0 & 0 & \\
          \Block[fill=MixedColor1, rounded-corners]{1-1}{} \gamma_1^* \delta_1^* & \Block[fill=MixedColor2, rounded-corners]{1-1}{} \gamma_1^* \delta_2^* & \Block[fill=MixedColor3, rounded-corners]{1-1}{} \gamma_2^* \delta_1^* & \Block[fill=MixedColor4, rounded-corners]{1-1}{} \gamma_2^* \delta_2^* & \Block[fill=MixedColor5, rounded-corners]{1-1}{} \gamma_3^* \delta_1^* & \Block[fill=MixedColor6, rounded-corners]{1-1}{} \gamma_3^* \delta_2^* & \delta_{1}^* & \delta_{2}^* & \Block[fill=Coral!50, rounded-corners]{1-1}{} \gamma_{1}^* & \Block[fill=Coral!75, rounded-corners]{1-1}{} \gamma_{2}^* & \Block[fill=Coral, rounded-corners]{1-1}{} \gamma_{3}^* &
        \end{pNiceMatrix}
        \]
        \caption{Rank restriction $r_{1} = 1$ and $r_{2} = 1$.}
        \label{subfig:matrixb}
    \end{subfigure}
    \caption{SCCF restrictions under a reduced rank in both dimensions $N_{1}$ and 
$N_{2}$. Columns are color-coded by component of the direct-sum decomposition in Proposition~\ref{prop:one}. 
Pure blue columns correspond to row-specific co-movements $\left(\mathcal{C}(\mathbf{U}_{2}) \otimes \mathcal{N}(\mathbf{U}_{1}^{\top})\right)$, pure red columns to column-specific co-movements $\left(\mathcal{N}(\mathbf{U}_{2}^{\top}) \otimes \mathcal{C}(\mathbf{U}_{1})\right)$, and mixed blue, red, and purple columns correspond to joint co-movements $\left(\mathcal{N}(\mathbf{U}_{2}^{\top}) \otimes \mathcal{N}(\mathbf{U}_{1}^{\top})\right)$.}

    \label{fig:fullnull1}
\end{figure}

\section{Estimation and Selection of Ranks}
\label{sec:estimationselection}

This section describes the full information maximum likelihood (FIML) estimator for the pseudo-structural model (Section \ref{sec:estimation}) and then discusses practical rank selection (Section \ref{sec:rankselection}).

\subsection{Estimation of Pseudo-Structural Model for Fixed Ranks}
\label{sec:estimation}
From the pseudo-structural model with one lag in equation \eqref{eq:pseudo-structural}, a $p$ lag pseudo-structural model is defined as
\begin{align}
    \label{eq:lagp_pseudostructural}
    \boldsymbol{\Omega} \mathbf{y}_t^* = \sum_{j=1}^p \boldsymbol{\Pi}_j \mathbf{y}_{t-j} + \boldsymbol{\Omega} \mathbf{e}_t,
\end{align}
where $\boldsymbol{\Pi}_j$ has the same structure as $\boldsymbol{\Pi}$ in equation \eqref{eq:omega-and-pi}, but now in terms of $\mathbf{U}_{3,j}$ and $\mathbf{U}_{4,j}$.
For given ranks $r_{1}$ and $r_{2}$, $\boldsymbol{\Omega}$, the lag coefficients $\{\boldsymbol{\Pi}_{j}\}_{j=1}^{p}$, and the covariance matrices $\boldsymbol{\Sigma}_{1}$, and $\boldsymbol{\Sigma}_{2}$ are estimated by maximizing the full-information likelihood $\mathcal{L}(\boldsymbol{\theta})$ defined by
\begin{align}
    \label{eq:fiml}
    \mathcal{L}(\boldsymbol{\theta}) = -\frac{(T-p) N_{2}}{2}\log |\boldsymbol{\Sigma}_{1}| -\frac{(T-p) N_{1}}{2}\log |\boldsymbol{\Sigma}_{2}| - \nonumber \\
    \frac{1}{2} \sum_{t=p+1}^T \operatorname{tr}\left((\boldsymbol{\Omega} \mathbf{y}_{t}^* - \sum_{j=1}^p \boldsymbol{\Pi}_{j} \mathbf{y}_{t-j})^{\top} (\boldsymbol{\Omega} (\boldsymbol{\Sigma}_{2} \otimes \boldsymbol{\Sigma}_{1}) \boldsymbol{\Omega}^{\top})^{-1}(\boldsymbol{\Omega} \mathbf{y}_{t}^* - \sum_{j=1}^p \boldsymbol{\Pi}_{j} \mathbf{y}_{t-j})\right),
\end{align}
where $\boldsymbol{\theta}$ collects the parameters $\boldsymbol{\delta}^*$, $\boldsymbol{\gamma}^*$, $\boldsymbol{\Sigma}_{1}$, $\boldsymbol{\Sigma}_{2}$, $\mathbf{U}_{3,j}$, and $\mathbf{U}_{4,j}$.
Note that $\boldsymbol{\Omega}$ is omitted from the $\log | \cdot |$ terms because the determinant of $\boldsymbol{\Omega}$ is always one by construction (see Section \ref{sec:pseudostructural}).

Because the optimization problem is non-convex, the estimator is initialized both from the RRMAR solution (given from \citealp{xiao_reduced_forth}) and from several randomized starts to reduce the risk of convergence to local optima.
The RRMAR initialization exploits the multidimensional structure of the coefficient matrices and yields a consistent starting point close to a global optimum.
Optimization is performed with gradient-based methods (BFGS, \citealp{nocedal2006numerical}) and convergence is monitored across starts to avoid local maxima and saddle points.
Further algorithmic details appear in Appendix \ref{sec:algorithm}.

\begin{rmk}
  \label{rmk:asymptotic}
The asymptotic properties of the estimator follow from \citet[Theorem 2]{xiao_reduced_forth} through a rotation argument.
Applying the delta method and continuous mapping theorem under the identifying normalization 
    \[
	\mathbf U_1^*=\mathbf U_1\mathbf Q_1 = \begin{bmatrix}-\boldsymbol{\delta}^*\\ \mathbf{I}_{r_{1}} \end{bmatrix} \quad \text{and} \quad \mathbf U_2^*=\mathbf U_2\mathbf Q_2 = \begin{bmatrix}-\boldsymbol{\gamma}^*\\ \mathbf{I}_{r_{2}} \end{bmatrix},
    \]
    the pseudo-structural parameters satisfy
    \[
    \sqrt{T} \bigl( \widehat{\boldsymbol{\delta}}^* - \boldsymbol{\delta}^* \bigr) \overset{d}{\longrightarrow} N \bigl( \mathbf{0},  \boldsymbol{\Sigma}_{\delta^*} \bigr) \quad \text{and} \quad \sqrt{T} \bigl( \widehat{\boldsymbol{\gamma}}^* - \boldsymbol{\gamma}^* \bigr) \overset{d}{\longrightarrow} N \bigl( \mathbf{0},  \boldsymbol{\Sigma}_{\gamma^*} \bigr).
    \]
    Hence, $\widehat{\boldsymbol{\delta}}^*$ and $\widehat{\boldsymbol{\gamma}}^*$ are consistent and $\sqrt{T}$-asymptotically normal, with asymptotic covariance matrices $\boldsymbol{\Sigma}_{\delta^*}, \boldsymbol{\Sigma}_{\gamma^*}$ derived from the rotated asymptotic distribution of $\widehat{\mathbf{U}}_i^*$.
    This rotational argument preserves the over-specification invariance of the rank: if one rank (e.g., $r_1$) is overestimated ($\widehat{r}_1 \geq r_1^*$) while the other is correctly specified ($\widehat{r}_2 = r_2^*$), the estimators for $\boldsymbol{\gamma}^*$ retain consistency and $\sqrt{T}$-asymptotic normality. This justifies setting $\widehat{r}_1 = N_1$ ($\widehat{r}_{2} = N_{2}$) and selecting $\widehat{r}_2$ ($\widehat{r}_{1}$) via information criteria without invalidating inference on $\boldsymbol{\gamma}^*$ \citep{xiao_reduced_forth}.
\end{rmk}
\begin{rmk}
\label{rmk:triviality}
Definition~1 characterizes a \emph{non-trivial} common feature and therefore excludes degenerate cofeature vectors such as $\boldsymbol{\delta}=(1,0)^\top$, whereas the parameter space over which $\mathcal{L}(\boldsymbol{\theta})$ in \eqref{eq:fiml} is maximized is not restricted to exclude them.
Under the normalization given in Section \ref{sec:pseudostructural} these degenerate vectors correspond to $\boldsymbol{\delta}^*=\mathbf{0}$ (respectively $\boldsymbol{\gamma}^*=\mathbf{0}$).
The columns of $\mathbf{U}_i^*$ then reduce to unit vectors that pick out individual series, the trivial case ruled out by Definition~1.

This exclusion is not imposed on the estimator, because the degeneracy is resolved by inference.
The trivial configuration is therefore the testable null $H_0:\boldsymbol{\delta}^*=\mathbf{0}$ (respectively $\boldsymbol{\gamma}^*=\mathbf{0}$), and rejecting it at a chosen significance level establishes that the estimated common feature is non-trivial in the sense of Definition~1; the $p$-values reported in Section~\ref{sec:appestimates} assess exactly these restrictions.
This is in line with the common-feature literature, in which the cofeature structure is tested rather than imposed a priori \citep{engle_testing_1993, vahid_common_1993}.
\end{rmk}

\subsection{Selection of the Ranks}
\label{sec:rankselection}

In practice, the true ranks $r_1$ and $r_2$ are unknown and must be estimated.
Information criteria-- the Akaike Information Criterion (AIC, \citealp{akaike_new_1974}), the Bayesian Information Criterion (BIC, \citealp{schwarz_estimating_1978}), and the Extended Bayesian Information Criterion (EBIC, \citealp{chen2008ebic}) -- are used to jointly obtain an estimate for the ranks for the RRMAR.
These are therefore referred to as the rank selection criteria in the remainder of the paper.
Define
\begin{align}
    \text{IC}(r_1, r_2) = -2 \mathcal{L} (\widehat{\boldsymbol{\theta}}) + c_{T} \phi(r_1, r_2), \label{eq:IC}
\end{align}
where $\mathcal{L} (\widehat{\boldsymbol{\theta}})$ is the log likelihood value obtained from estimating the pseudo-structural form and $c_{T}$ is a penalty term such that $c_{T} = 2$ for the AIC, $c_{T} = \ln(T)$ for the BIC, and $c_T = \ln(T) + 2 \gamma \ln(N_1 N_2)$ for the EBIC, and $\phi(r_1, r_2)$ is the number of parameters in a model with $p$ lags and reduced ranks $r_1$ and $r_2$, given by
\begin{align*}
\phi(r_{1}, r_{2}) = r_{1}N_{1} (1+p) - r_{1}^2 + r_{2}N_{2}(1+p) - r_{2}^2.
\end{align*}
Following \citet{chen2012ebic}, the EBIC tuning parameter $\gamma$ is set to $0.5$, as they argue that the standard BIC can fail to select the correct variables when the number of parameters exceeds the sample size.
Then the selected ranks (for fixed $p$) can be obtained based on the minimum value derived from either AIC, BIC or EBIC.

\begin{thm}
    Given the FIML estimator is $\sqrt{T}$-consistent and asymptotically normal (up to orthonormal rotation) and that the error covariance $\boldsymbol{\Sigma} = \boldsymbol{\Sigma}_{2} \otimes \boldsymbol{\Sigma}_{1}$ is finite and positive definite, the BIC and EBIC provide weakly consistent estimators of the ranks $r_{1}$ and $r_{2}$.
\end{thm}

\begin{rmk}
    Rank selection consistency of the BIC and EBIC is proven for an RRMAR with fixed lag order $p$. 
    In practice, the IC criterion in equation \eqref{eq:IC} can be easily extended to jointly select the ranks and the lag order.
    Simulation results for the joint rank and lag selection using the three different information criteria are presented in Section \ref{sec:rankselectionsim}.
\end{rmk}

\section{Simulation Study}
\label{sec:simulation}

Two Monte Carlo experiments are conducted to assess (i) the inferential performance of the pseudo-structural parameter estimates and (ii) the effectiveness of the rank-selection procedure described in Section \ref{sec:rankselection}.
The first experiment examines the sampling densities and coverage probabilities of $\widehat{\boldsymbol{\delta}}$ and $\widehat{\boldsymbol{\gamma}}$.
The second evaluates how well the procedure recovers the true ranks and the autoregressive lag order.

For each experiment, the matrix dimensions are set to $N_1\times N_2 = 3\times 4$.
Data are generated from the pseudo-structural model \eqref{eq:lagp_pseudostructural} using specified matrices $\boldsymbol{\Omega}$ and $\boldsymbol{\Pi}$.
The true parameter vectors $\boldsymbol{\delta}^*$ and $\boldsymbol{\gamma}^*$ are drawn independently from standard normal distributions and $\boldsymbol{\Omega}$ is constructed as in \eqref{eq:omega-and-pi}.
The matrix $\boldsymbol{\Pi}$ is formed by sampling each column of $\mathbf{U}_{3,i}$ and $\mathbf{U}_{4,i}$ for $i = 1, \dots, p$ from independent standard normal distributions.
The error matrices $\mathbf{E}_t$ are i.i.d.\ from a standard matrix-normal with mean $\mathbf{0}$ and row/column covariances $\boldsymbol{\Sigma}_1$ and $\boldsymbol{\Sigma}_2$.
For each configuration $T+50$ observations are simulated, the first 50 are discarded as burn-in, and results are reported for $T=100$ and $T=250$.
Throughout, the signal-to-noise ratio (SNR), defined as the ratio of the largest eigenvalue of the coefficient matrix to the largest eigenvalue of the error covariance matrix, is fixed at $0.7$. This value is chosen as a conservative stress test relative to the empirical SNR of the dataset, which is approximately $1.8$.

\subsection{Estimation and Inference}
\label{sec:inference_and_coverage}

\begin{figure}[t]
  \centering
  \begin{subfigure}[b]{0.48\textwidth}
    \centering
    \includegraphics[width=\textwidth]{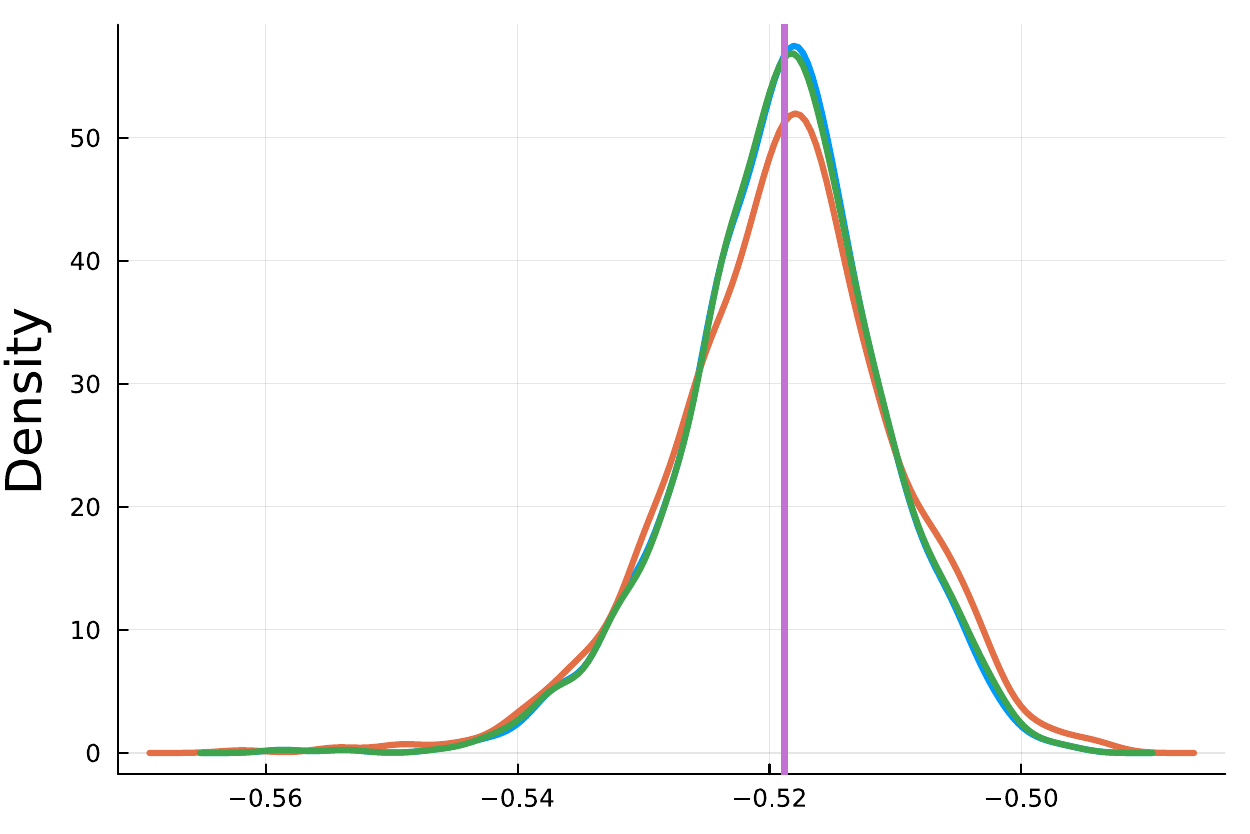}
    \caption{$\widehat{\delta}_1^*$, $T=100$}
    \label{fig:delta1_100}
  \end{subfigure}
  \hfill
  \begin{subfigure}[b]{0.48\textwidth}
    \centering
    \includegraphics[width=\textwidth]{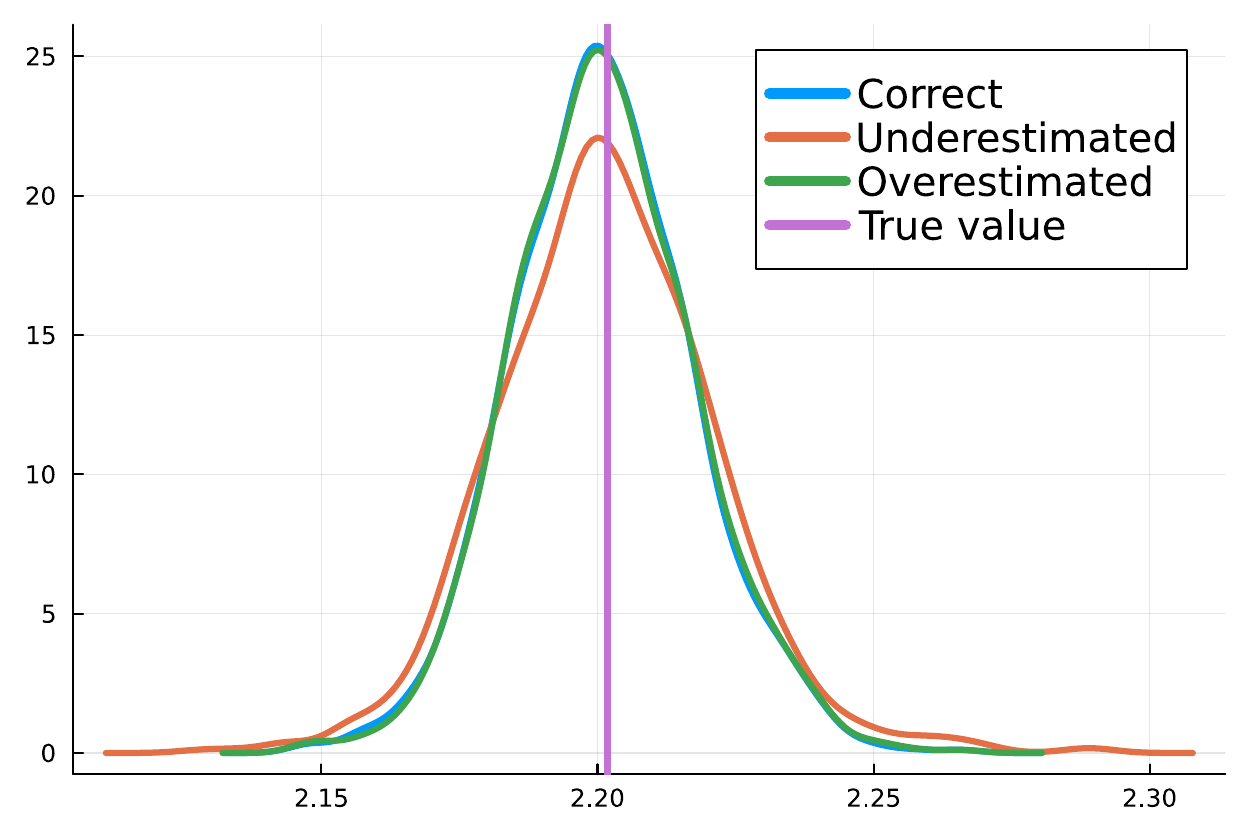}
    \caption{$\widehat{\delta}_2^*$, $T=100$}
    \label{fig:delta2_100}
  \end{subfigure}

  \vspace{0.5em}

  \begin{subfigure}[b]{0.48\textwidth}
    \centering
    \includegraphics[width=\textwidth]{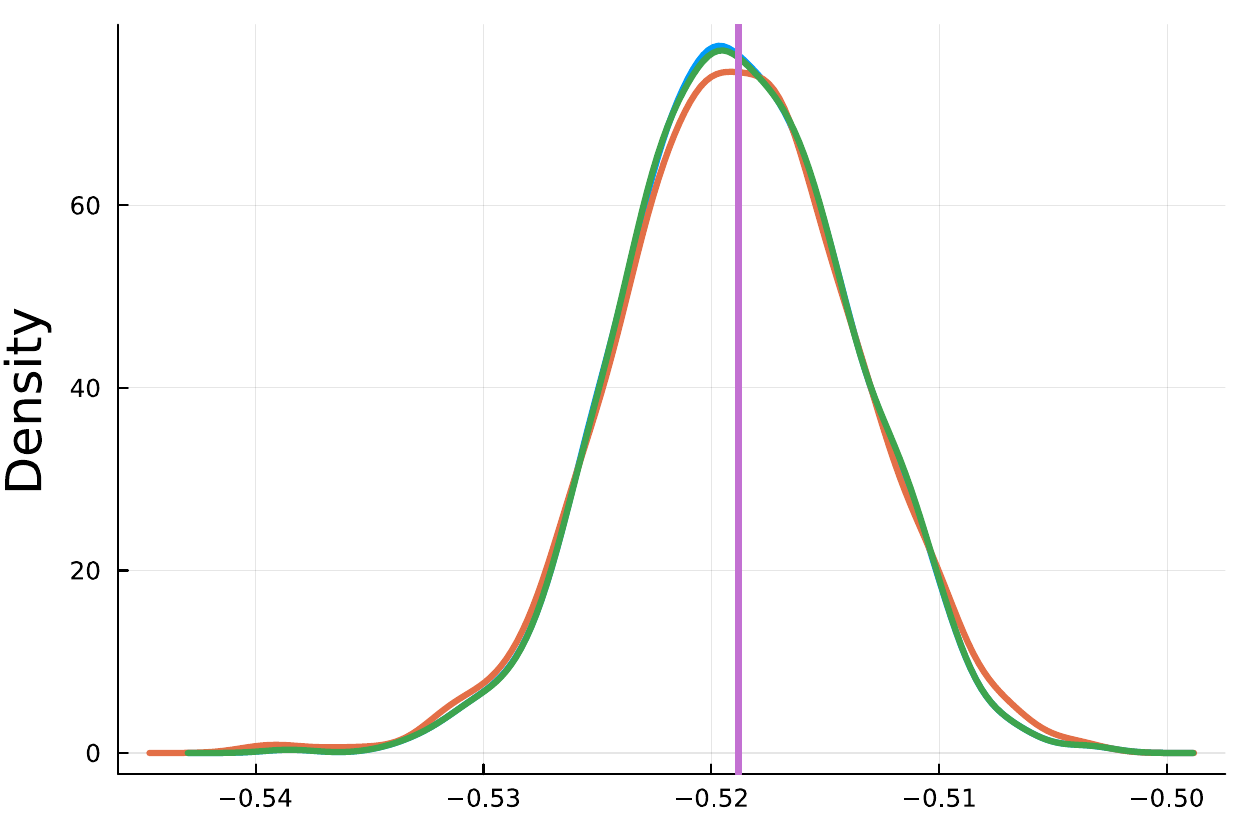}
    \caption{$\widehat{\delta}_1^*$, $T=250$}
    \label{fig:delta1_250}
  \end{subfigure}
  \hfill
  \begin{subfigure}[b]{0.48\textwidth}
    \centering
    \includegraphics[width=\textwidth]{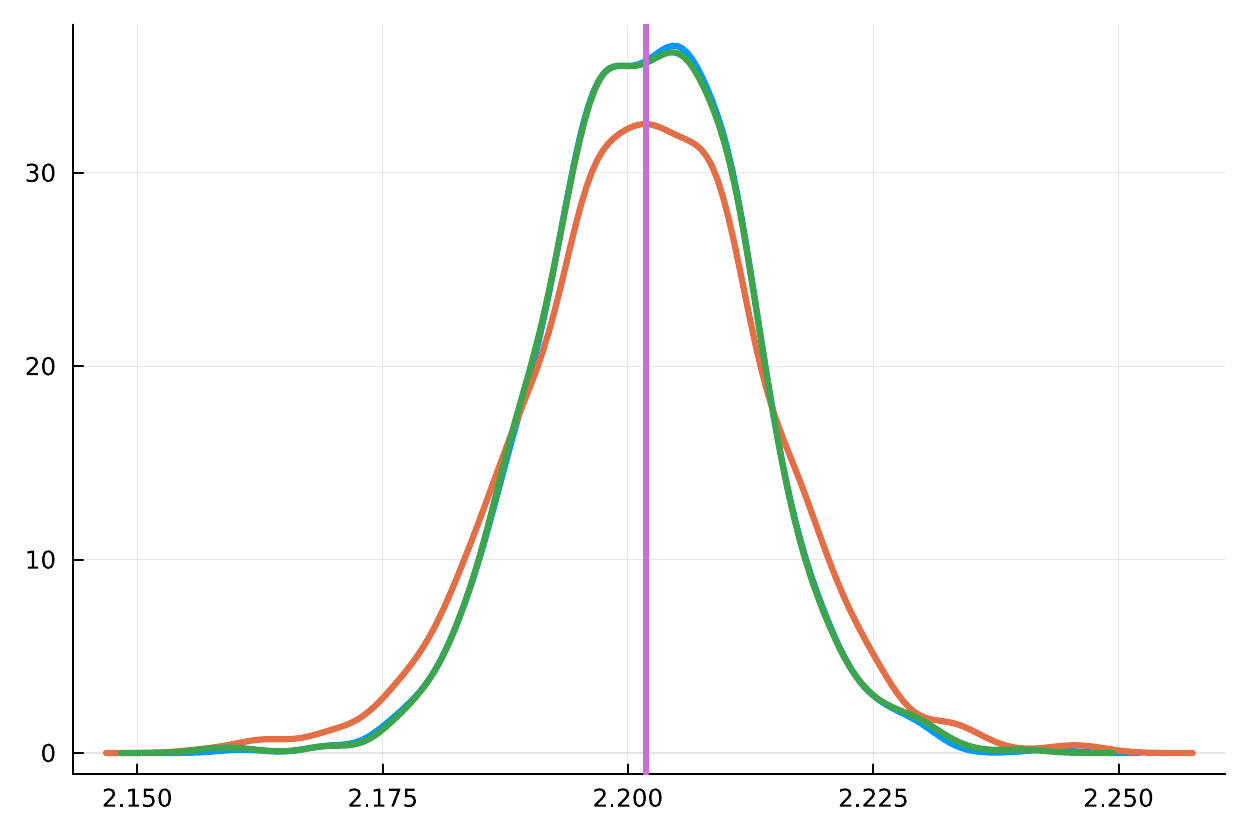}
    \caption{$\widehat{\delta}_2^*$, $T=250$}
    \label{fig:delta2_250}
  \end{subfigure}
  \caption{Sampling distributions of $\widehat{\delta}_1^*$ (left) and $\widehat{\delta}_2^*$ (right) under correct, under- and over-rank specifications in the second rank. Each panel shows the kernel density with the true value marked by a vertical line. Top row: $T=100$; bottom row: $T=250$.}
  \label{fig:delta_densities}
\end{figure}

To evaluate the estimation and inferential performance of the pseudo-structural estimates, $1000$ simulation runs are conducted for a matrix-valued time series with dimension $N_{1} \times N_{2} = 3 \times 4$ and ranks $r_1\times r_2 = 2\times 2$.
Results for the $N_{1} \times N_{2} = 3\times 6$ case with ranks $r_1\times r_2 = 2\times 2$ are similar and also available in Appendix \ref{sec:addsimresults}.

First, $\boldsymbol{\delta}^*$ is estimated under the rank settings: (i) correctly estimated ranks $\widehat{r}_1\times \widehat{r}_2= 2\times 2$, (ii) underestimated $2\times 1$, and (iii) overestimated rank $2\times 3$ in the second dimension.
For each simulation run, the following steps are carried out:
\begin{enumerate}
    \item A sample of size $T$ is generated from the pseudo-structural model with $p=1, r_1=2, r_2=2$.
    \item The pseudo-structural model is estimated with $p=1$ under each rank setting to obtain the estimate $\widehat{\boldsymbol{\delta}}^{*} = (\widehat{\delta}^*_1, \widehat{\delta}^*_2)^\top$.
    \item The $95$\% confidence interval for $\delta^*_1$ and $\delta^*_2$ is constructed using the observed information matrix, obtained from the Hessian of the log-likelihood at the maximum likelihood estimate in Section \ref{sec:estimation}.
\end{enumerate}

\begin{figure}[t]
    \centering
    \includegraphics[width=\textwidth]{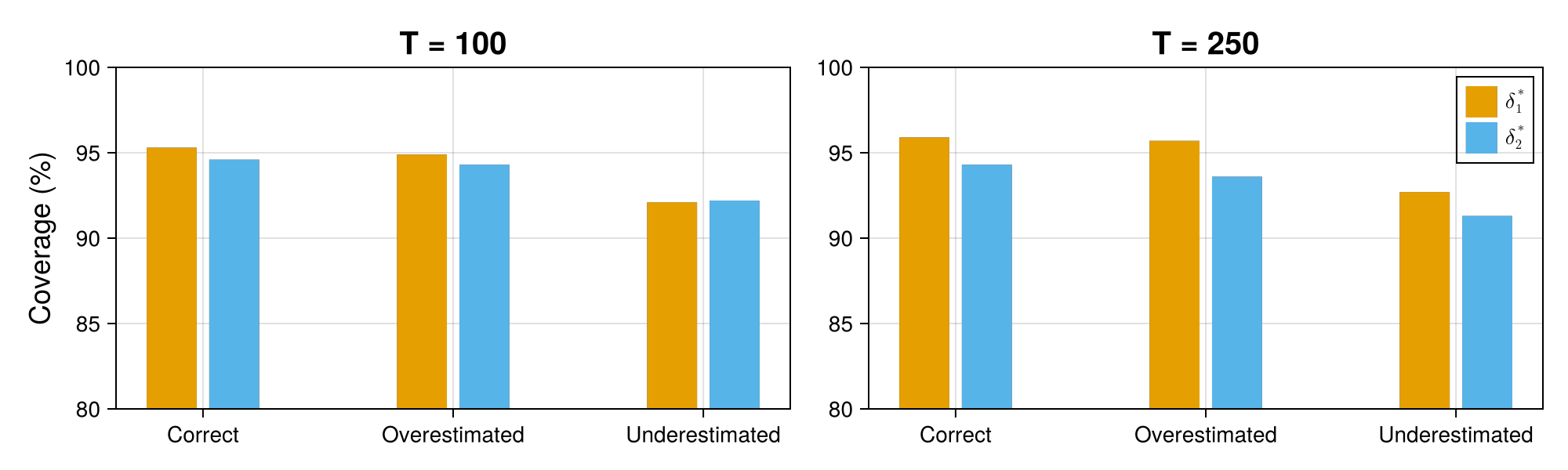}
    \caption{Empirical coverage rates (in \%) of 95\% confidence intervals for $\delta^*_1$ (left bars) and $\delta^*_2$ (right bars)  under correct, over- and under-rank specification in the second dimension, as indicated on the horizontal axis. Left: $T=100$. Right: $T=250$.}
    \label{fig:delta_coverage_plot}
\end{figure}

Figure \ref{fig:delta_densities} displays kernel densities for the two components of $\widehat{\boldsymbol{\delta}}^*$.
When the rank in the second dimension is specified correctly or overestimated, the densities are closely aligned and centered on the true values for both sample sizes.
Under rank underestimation, the densities remain well-behaved but show modestly increased variance.
Figure \ref{fig:delta_coverage_plot} plots the empirical coverage rates of the 95\% confidence intervals for $\delta^*_1$ and $\delta^*_2$.
Both correct and overestimated ranks in the second dimension achieve coverage close to the nominal 95\% level.
When the rank is underestimated, however, coverage deteriorates to around 92\%.

\begin{figure}[t]
  \centering
  \begin{subfigure}[b]{0.32\textwidth}
    \centering
    \includegraphics[width=\textwidth]{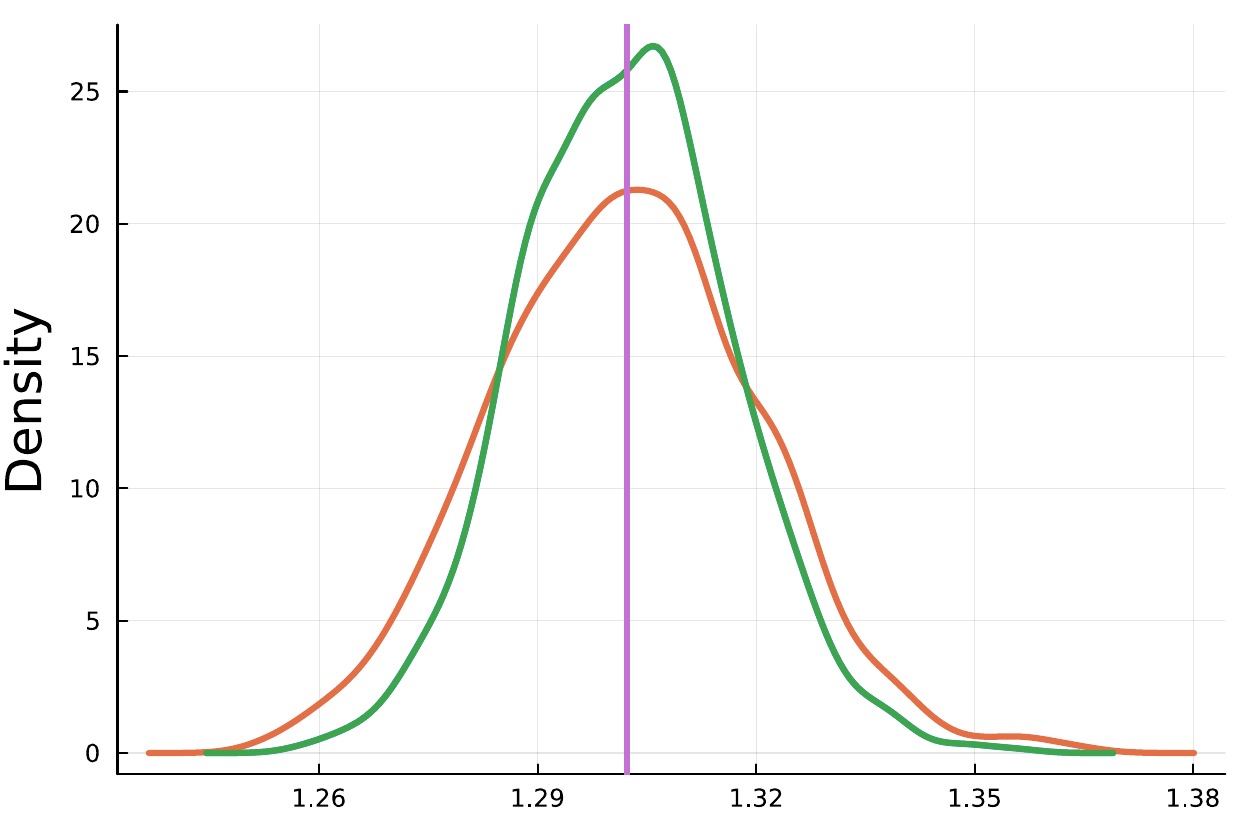}
    \caption{$\widehat{\gamma}_1^*$, $T=100$}
    \label{fig:gamma100_1}
  \end{subfigure}
  \hfill
  \begin{subfigure}[b]{0.32\textwidth}
    \centering
    \includegraphics[width=\textwidth]{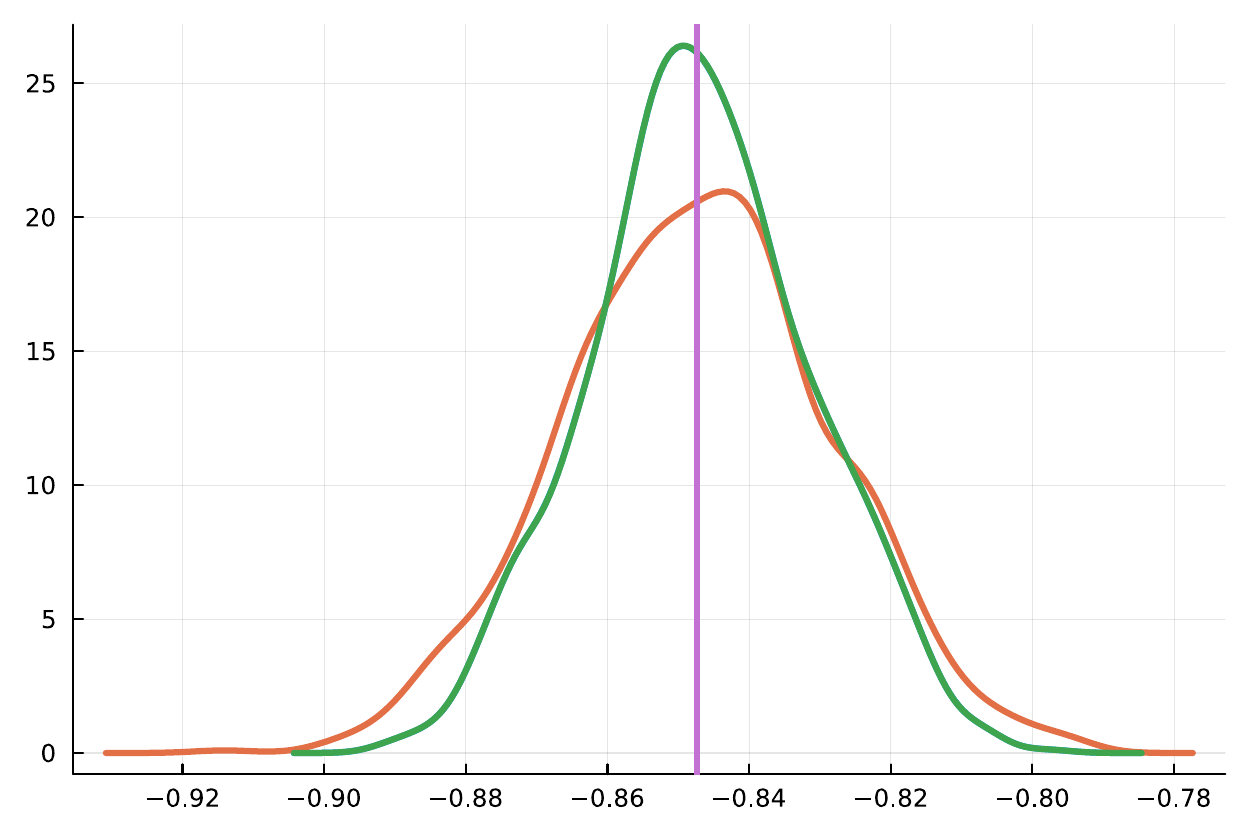}
    \caption{$\widehat{\gamma}_2^*$, $T=100$}
    \label{fig:gamma100_2}
  \end{subfigure}
  \hfill
  \begin{subfigure}[b]{0.32\textwidth}
    \centering
    \includegraphics[width=\textwidth]{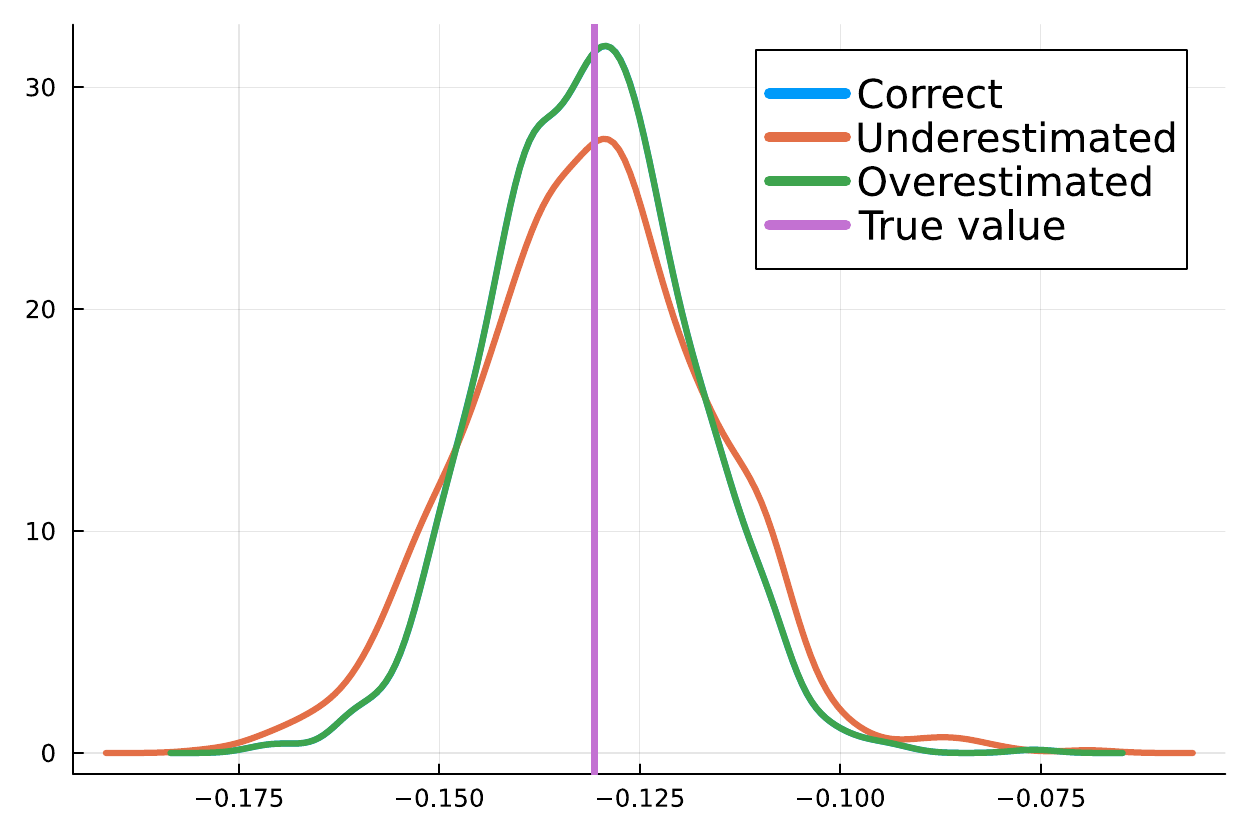}
    \caption{$\widehat{\gamma}_3^*$, $T=100$}
    \label{fig:gamma100_3}
  \end{subfigure}

  \vspace{1em}

  \begin{subfigure}[b]{0.32\textwidth}
    \centering
    \includegraphics[width=\textwidth]{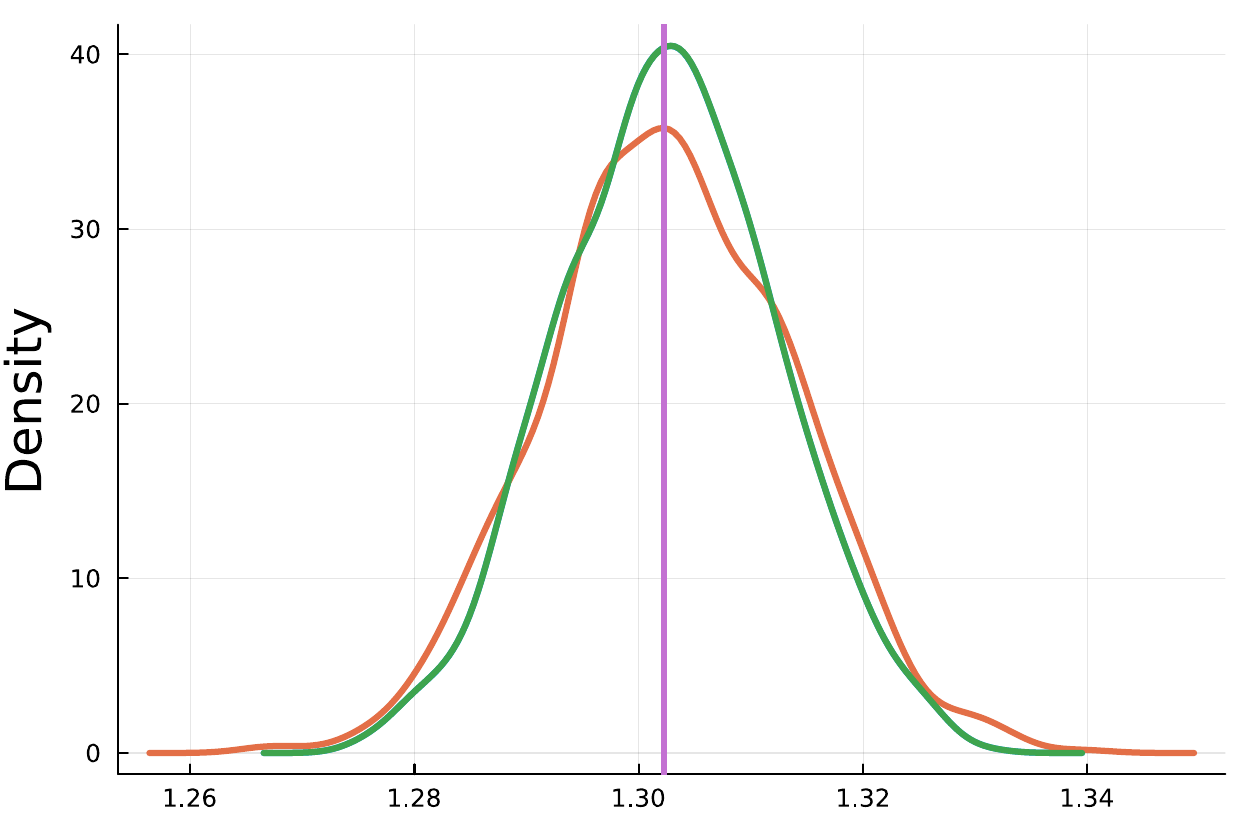}
    \caption{$\widehat{\gamma}_1^*$, $T=250$}
    \label{fig:gamma250_1}
  \end{subfigure}
  \hfill
  \begin{subfigure}[b]{0.32\textwidth}
    \centering
    \includegraphics[width=\textwidth]{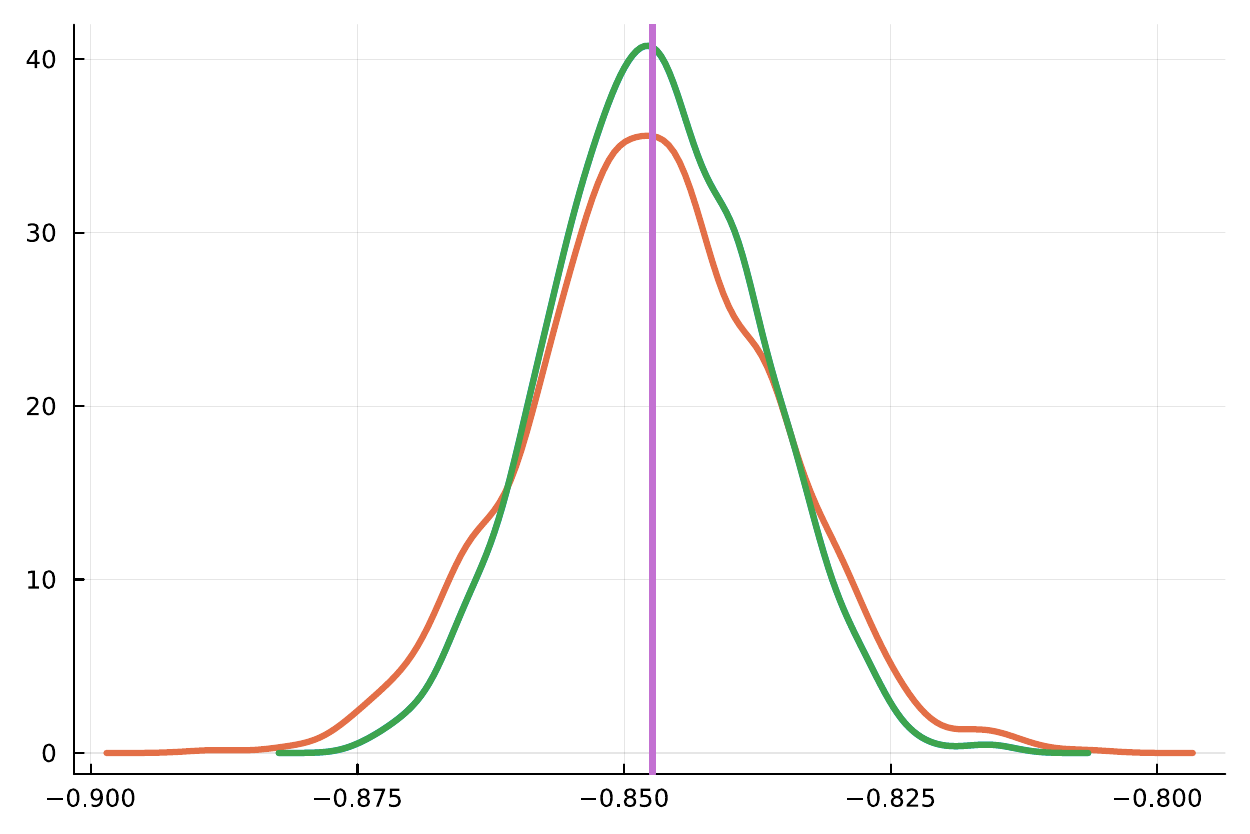}
    \caption{$\widehat{\gamma}_2^*$, $T=250$}
    \label{fig:gamma250_2}
  \end{subfigure}
  \hfill
  \begin{subfigure}[b]{0.32\textwidth}
    \centering
    \includegraphics[width=\textwidth]{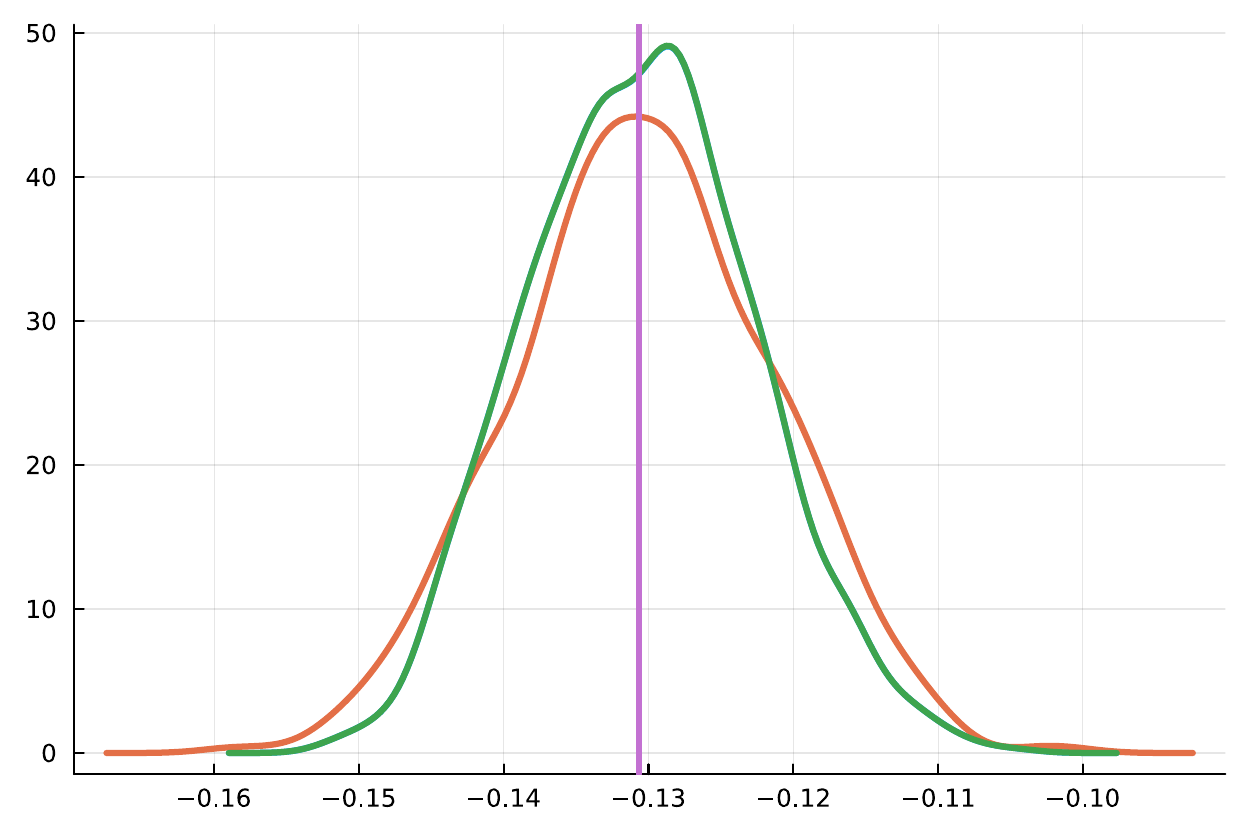}
    \caption{$\widehat{\gamma}_3^*$, $T=250$}
    \label{fig:gamma250_3}
  \end{subfigure}

  \caption{Sampling distributions of the estimated $\gamma$ parameters under correct, under‐ and over‐rank specifications in the first rank for two different sample sizes. Top row: $T=100$. Bottom row: $T=250$. Each panel shows the kernel density of one component of $\widehat{\boldsymbol{\gamma}}$, with the true value marked by a vertical line.}
  \label{fig:gamma_densities_combined}
\end{figure}

\begin{figure}[t]
    \centering
    \includegraphics[width=\textwidth]{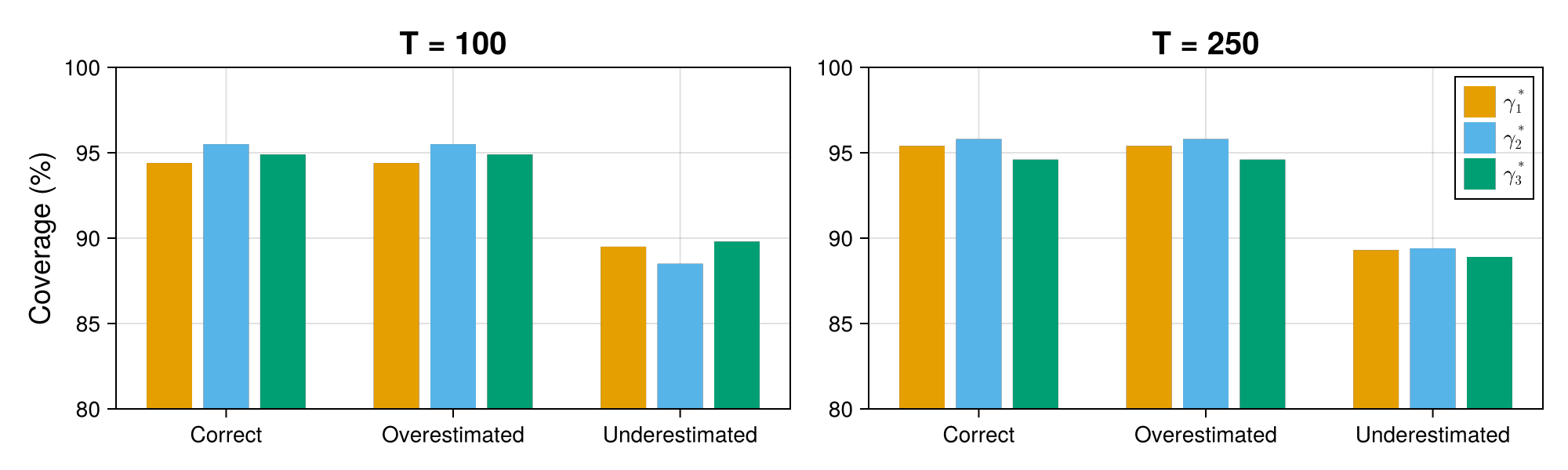}
    \caption{Empirical coverage rates (in \%) of 95\% confidence intervals for $\gamma_1^*$ (left bars), $\gamma_2^*$ (middle bars), and $\gamma_3^*$ (right bars), under correct, over- and under-rank specification in the first dimension, as indicated on the horizontal axis. Left: $T=100$. Right: $T=250$.} 
    \label{fig:gamma_coverage_plot}
\end{figure}

Second, $\boldsymbol{\gamma}^*$ is estimated under rank settings: (i) correctly estimated ranks $2\times 3$, (ii) underestimated $1\times 3$, and (iii) overestimated $3\times 3$ in the first dimension.
Figures \ref{fig:gamma_densities_combined} and \ref{fig:gamma_coverage_plot} present the corresponding density and coverage plots.
As with $\boldsymbol{\delta}$, correct specification or overestimation yields densities centered on the true values.
Underestimating the rank of the first dimension increases the standard deviation of $\widehat{\boldsymbol{\gamma}}$ and reduces coverage to about 89\%.

The estimator is also compared with the reduced-rank matrix autoregressive (RRMAR) estimator of \citet{xiao_reduced_forth}, transforming their coefficient estimates into the implied pseudo-structural parameters.
The resulting densities and a likelihood comparison are reported in Appendix \ref{sec:addsimresults}.
Both estimators are approximately centered at the true values, but the RRMAR initialization can become trapped in local optima, particularly for $\boldsymbol{\delta}^*$, confirming the value of the multi-start procedure.

\subsection{Rank Selection}
\label{sec:rankselectionsim}

To evaluate the rank-selection procedures, the pseudo-structural model is estimated for each simulation design across all possible rank combinations.
In addition to the AIC, BIC, and EBIC criteria from Section \ref{sec:rankselection}, the rank-selection procedure of \citet{xiao_reduced_forth}, labeled ``Bench'', is included as a benchmark.
In this exercise, the autoregressive lag order is fixed at its true value, $p=1$.

For each design, the percentage of correct  selection (``\% Correct'') and the mean absolute deviation  (``MAD'') are reported over 100 simulation runs, a smaller number than in Section \ref{sec:inference_and_coverage} owing to the computational cost of repeated non-convex optimization across all $N_1 \times N_2$ rank combinations.
This cost grows with matrix dimensions $N_1$ and $N_2$ through two channels: the parameter space, which scales as $r_1 (2 N_1 - r_1) + r_2 (2 N_2 - r_2)$, and the number of rank combinations $(r_1, r_2) \in \{1, \dots, N_1\} \times \{1, \dots, N_2\}$ that must each be optimized separately for rank selection.
The results for the different simulation designs and true rank settings are summarized in Table \ref{tab:sim34}.

For $N_1=3$, $N_2=4$ (Table \ref{tab:sim34}), both EBIC and BIC select the true rank at a high rate and outperform AIC.
For example, with true rank $(1,1)$, AIC selects the correct ranks about $89$\% and $81$\% of the time for the first and second dimensions when $T=100$, with little improvement at $T=250$.
BIC, EBIC, and the Bench select the correct ranks at rates of at least $98$\% for $T=100$, rising up to $100$\% for $T=250$.
For partially reduced ranks (reduced over only one dimension), AIC selects the true rank  at least $80$\% of the time for $T=100$, with modest improvements at larger $T$, while BIC, EBIC, and the Bench are accurate across all settings.
When the true rank is full, all information criteria select the correct rank in all reported simulations.

In the setting where the true rank is $(2,1)$, AIC correctly selects the ranks at a rate of $80$\%, with no improvement at larger $T$.
On the other hand, BIC, EBIC and the Bench select the correct rank at a rate of at least $94$\%, going up to at least $96$\% for $T=250$.
In this setting, the performance of the information criteria is also explored when the true rank of the first dimension is underestimated ($\widehat{r}_1 = 1$) while estimating the rank of the second dimension.

In this case, AIC struggles to select the second dimension, selecting the correct rank with $57$\% accuracy, going up to $61$\% accuracy with $T = 250$.
On the other hand, BIC, EBIC, and the Bench maintain $99$\% accuracy, with the EBIC going to $100$\% as the number of observations increases.

Overestimating the true rank of the first dimension ($\widehat{r}_1 = 3$) leads to similar conclusions, the difference being that AIC performs considerably better, selecting the correct rank with $83$\% accuracy and increasing to $84$\% as the number of observations increases.

Furthermore, to assess performance under dimensions matching the empirical application, a design with $N_1 = 9$, $N_2 = 4$, $T=180$, true rank $(2,1)$ and lag one is also considered.
The results are consistent with the findings above: BIC, EBIC, and the Bench all select the correct ranks at rates above $91$\%, while AIC lags behind at around $84$\% and $82$\% for the two dimensions. Overall, BIC, EBIC, and the Bench all perform well, with the EBIC performing marginally better.

Next, the joint performance of rank and lag selection using information criteria is investigated.
To this end, an additional simulation study is conducted, evaluating their ability to recover the true rank-lag specification, with the maximum lag order restricted to two.
The experimental setup is identical to the previous simulations, including the fully reduced case, partially reduced cases and the case with no rank reduction.
Note that the Bench method of \citet{xiao_reduced_forth} is omitted from this analysis, as it does not involve the joint selection of ranks and lag order.

Table \ref{tab:sim34_lag1} reports the results when the true lag is one.
Overall, BIC and EBIC identify the correct rank and lag with high accuracy and consistently outperform AIC.
For example, when the true ranks are $(1,1)$, AIC selects the correct ranks only about $37$\% and $49$\% of the time and the true lag just $8$\% of the time.
By contrast, BIC and EBIC select the correct ranks $100$\% and $99$\% of the time, with the correct lag chosen approximately $90$\% of the time for the BIC, and at least $99$\% of the time for the EBIC.
In the partially reduced cases, AIC struggles to identify the reduced rank dimension, correctly selecting it only about $34$\% of the time in the worst case, although it consistently chooses the full rank dimension ($100$\%).
This misclassification extends to lag selection: AIC tends to favor higher lag orders, while BIC and EBIC select the true lag at rates of at least $90$\%.
A similar pattern emerges in the full rank case, where AIC again favors higher lags, while BIC and EBIC select the true lag with much greater reliability.

The joint rank and lag selection experiment is repeated with true lag $p=2$; all other settings are unchanged.
Table \ref{tab:sim34_lag2} reports the frequency of correct selection and the MAD.
The main findings are, overall, qualitatively similar to the $p=1$ experiment, namely, BIC and EBIC again outperform AIC in jointly recovering both rank and lag, while AIC continues to overselect the reduced rank dimensions.
However, unlike the $p=1$ case, AIC, BIC and EBIC now select the correct lag $100$\% of the time.

For example, when the true ranks are $(1,1)$, AIC correctly identifies the ranks in approximately $75$\% and $80$\% of cases and recovers the true lag ($p=2$) $100$\% of the time.
By contrast, BIC and EBIC correctly recover the ranks about $90$\% and $97$\% of the time and select the correct lag in $100$\% of replications.
In the partially reduced case, AIC continues to struggle with the reduced rank dimension (about $57$\% in the worst case) while always selecting the full rank dimension ($100$\%) and the lag ($100$\%).
BIC and EBIC, however, select the reduced rank dimension and lag correctly at higher rates (approximately $66$\% for BIC and $69$\% for EBIC in the worst case).
For the full rank case, both AIC and BIC select the full rank dimensions and the lag order at near $100$\% of the time.

\begin{landscape}
\setlength{\tabcolsep}{0.8em}
\renewcommand{\arraystretch}{1.2}
\begin{table}[ht]
  \vspace{-2cm}
  \centering
  \caption{Rank selection with matrix-valued time series of dimension 
  $N_1 = 3$, $N_2 = 4$ under different true rank settings. 
  Percentage correct and MAD reported for 
  $T = 100$ and $T = 250$.}
  \label{tab:sim34}
  \begin{tabular}{
    >{\centering\arraybackslash}p{1.8cm}
    >{\centering\arraybackslash}p{2.0cm}
    >{\centering\arraybackslash}p{3.2cm}
    >{\centering\arraybackslash}p{3.2cm}
    >{\centering\arraybackslash}p{3.2cm}
    >{\centering\arraybackslash}p{3.2cm}
  }
    \toprule
    & & \multicolumn{2}{c}{$T = 100$} & \multicolumn{2}{c}{$T = 250$} \\
    \cmidrule(lr){3-4} \cmidrule(lr){5-6}
    \textbf{True Rank} & \textbf{Method} 
      & \textbf{\% Correct} & \textbf{MAD}
      & \textbf{\% Correct} & \textbf{MAD} \\
    \midrule
    \multirow{4}{*}{(1,1)} 
      & AIC       & (89, 81)  & (0.12, 0.21) & (89, 82)   & (0.12, 0.21) \\
      & BIC       & (100, 99) & (0.00, 0.02) & (100, 100) & (0.00, 0.00) \\
      & EBIC      & (100, 99) & (0.00, 0.02) & (100, 100) & (0.00, 0.00) \\
      & Bench     & (98, 98)  & (0.02, 0.03) & (99, 98)   & (0.02, 0.03) \\
    \midrule
    \multirow{4}{*}{(2,1)} 
      & AIC       & (80, 83)  & (0.20, 0.18) & (76, 80)   & (0.24, 0.21) \\
      & BIC       & (94, 100) & (0.06, 0.00) & (96, 100)  & (0.04, 0.00) \\
      & EBIC      & (100, 100) & (0.00, 0.00) & (100, 100)  & (0.00, 0.00) \\
      & Bench     & (98, 99) & (0.02, 0.01) & (97, 100)  & (0.03, 0.00) \\
    \midrule
    \multirow{4}{*}{(3,1)} 
      & AIC       & (100, 85)  & (0.00, 0.15) & (100, 88)  & (0.00, 0.14) \\
      & BIC       & (100, 100) & (0.00, 0.00) & (100, 100) & (0.00, 0.00) \\
      & EBIC      & (100, 100) & (0.00, 0.00) & (100, 100) & (0.00, 0.00) \\
      & Bench     & (100, 100) & (0.00, 0.00) & (100, 100) & (0.00, 0.00) \\
    \midrule
    \multirow{4}{*}{(1,4)} 
      & AIC       & (87, 100)  & (0.14, 0.00) & (91, 100)  & (0.09, 0.00) \\
      & BIC       & (100, 100)  & (0.00, 0.00) & (100, 100) & (0.00, 0.00) \\
      & EBIC      & (100, 100) & (0.00, 0.00) & (100, 100) & (0.00, 0.00) \\
      & Bench     & (100, 100) & (0.00, 0.00) & (100, 100) & (0.00, 0.00) \\
    \midrule
    \multirow{4}{*}{(3,4)} 
      & AIC       & (100, 100) & (0.00, 0.00) & (100, 100) & (0.00, 0.00) \\
      & BIC       & (100, 100) & (0.00, 0.00) & (100, 100) & (0.00, 0.00) \\
      & EBIC      & (100, 100) & (0.00, 0.00) & (100, 100) & (0.00, 0.00) \\
      & Bench     & (100, 100) & (0.00, 0.00) & (100, 100) & (0.00, 0.00) \\
    \bottomrule
  \end{tabular}
\end{table}
\end{landscape}

\begin{landscape}
\setlength{\tabcolsep}{0.6em}
\renewcommand{\arraystretch}{1.2}
\begin{table}[ht]
  \vspace{-2cm}
  \centering
  \caption{Joint rank and lag selection for matrix-valued time series 
  with $N_1 = 3$, $N_2 = 4$ under different true rank and lag settings.
  Percentage correct and MAD reported for ranks $(r_1, r_2)$ and lag 
  $p$ separately. True lag is one.}
  \label{tab:sim34_lag1}
  \begin{tabular}{
    >{\centering\arraybackslash}p{1.8cm}
    >{\centering\arraybackslash}p{2.0cm}
    >{\centering\arraybackslash}p{2.4cm}
    >{\centering\arraybackslash}p{2.4cm}
    >{\centering\arraybackslash}p{1.6cm}
    >{\centering\arraybackslash}p{2.4cm}
    >{\centering\arraybackslash}p{2.4cm}
    >{\centering\arraybackslash}p{1.6cm}
  }
    \toprule
    & & \multicolumn{3}{c}{$T = 100$} 
      & \multicolumn{3}{c}{$T = 250$} \\
    \cmidrule(lr){3-5} \cmidrule(lr){6-8}
    \textbf{True} & \textbf{Method} 
      & \textbf{\% Correct} & \textbf{MAD} & \textbf{Lag \%}
      & \textbf{\% Correct} & \textbf{MAD} & \textbf{Lag \%} \\
    \textbf{Rank/Lag} & & \textbf{$(r_1, r_2)$} & \textbf{$(r_1, r_2)$} & 
      & \textbf{$(r_1, r_2)$} & \textbf{$(r_1, r_2)$} & \\
    \midrule
    \multirow{4}{*}{(1,1)/1}
      & AIC       & (37, 49)   & (0.63, 0.59) & 8
                  & (60, 68)   & (0.40, 0.37) & 16   \\
      & BIC       & (100, 99)  & (0.00, 0.01) & 90  
                  & (100, 100) & (0.00, 0.00) & 97  \\
      & EBIC      & (100, 100) & (0.00, 0.00) & 100 
                  & (100, 100) & (0.00, 0.00) & 100  \\
    \midrule
    \multirow{4}{*}{(3,1)/1}
      & AIC       & (100, 34)  & (0.00, 0.79) & 7  
                  & (100, 43)  & (0.00, 0.64) & 13  \\
      & BIC       & (100, 100) & (0.00, 0.00) & 99 
                  & (100, 100) & (0.00, 0.00) & 100 \\
      & EBIC      & (100, 100) & (0.00, 0.00) & 100 
                  & (100, 100) & (0.00, 0.00) & 100  \\
    \midrule
    \multirow{4}{*}{(1,4)/1}
      & AIC       & (52, 100)  & (0.59, 0.00) & 12  
                  & (35, 100)  & (0.73, 0.00) & 11  \\
      & BIC       & (100, 100) & (0.00, 0.00) & 100 
                  & (100, 100) & (0.00, 0.00) & 100 \\
      & EBIC      & (100, 100) & (0.00, 0.00) & 100 
                  & (100, 100)  & (0.00, 0.00) & 100  \\
    \midrule
    \multirow{4}{*}{(3,4)/1}
      & AIC       & (99, 100)  & (0.02, 0.00) & 60  
                  & (100, 100) & (0.00, 0.00) & 69  \\
      & BIC       & (99, 100)  & (0.02, 0.00) & 100 
                  & (100, 100) & (0.00, 0.00) & 100 \\
      & EBIC      & (100, 100) & (0.00, 0.00) & 100 
                  & (100, 100)  & (0.00, 0.00) & 100  \\
    \bottomrule
  \end{tabular}
\end{table}
\end{landscape}

\newpage

\begin{landscape}
\setlength{\tabcolsep}{0.6em}
\renewcommand{\arraystretch}{1.2}
\begin{table}[ht]
  \vspace{-2cm}
  \centering
  \caption{Joint rank and lag selection for matrix-valued time series 
  with $N_1 = 3$, $N_2 = 4$ under different true rank settings.
  Percentage correct and MAD reported for ranks $(r_1, r_2)$ and lag 
  $p$ separately. True lag is two.}
  \label{tab:sim34_lag2}
  \begin{tabular}{
    >{\centering\arraybackslash}p{1.8cm}
    >{\centering\arraybackslash}p{2.0cm}
    >{\centering\arraybackslash}p{2.4cm}
    >{\centering\arraybackslash}p{2.4cm}
    >{\centering\arraybackslash}p{1.6cm}
    >{\centering\arraybackslash}p{2.4cm}
    >{\centering\arraybackslash}p{2.4cm}
    >{\centering\arraybackslash}p{1.6cm}
  }
    \toprule
    & & \multicolumn{3}{c}{$T = 100$} 
      & \multicolumn{3}{c}{$T = 250$} \\
    \cmidrule(lr){3-5} \cmidrule(lr){6-8}
    \textbf{True} & \textbf{Method} 
      & \textbf{\% Correct} & \textbf{MAD} & \textbf{Lag \%}
      & \textbf{\% Correct} & \textbf{MAD} & \textbf{Lag \%} \\
    \textbf{Rank/Lag} & 
      & \textbf{$(r_1, r_2)$} & \textbf{$(r_1, r_2)$} &
      & \textbf{$(r_1, r_2)$} & \textbf{$(r_1, r_2)$} & \\
    \midrule
    \multirow{4}{*}{(1,1)/2}
      & AIC       & (75, 80) & (0.28, 0.24) & 100
                  & (91, 80) & (0.10, 0.22) & 100 \\
      & BIC       & (90, 97) & (0.12, 0.05) & 100
                  & (95, 97) & (0.05, 0.03) & 100 \\
      & EBIC      & (90, 98) & (0.12, 0.04) & 100
                  & (94, 98) & (0.06, 0.02) & 100 \\
    \midrule
    \multirow{4}{*}{(3,1)/2}
      & AIC       & (100, 57)  & (0.00, 0.60) & 100
                  & (100, 64)  & (0.00, 0.59) & 100 \\
      & BIC       & (100, 66)  & (0.00, 0.49) & 100
                  & (100, 69)  & (0.00, 0.52) & 100 \\
      & EBIC      & (100, 69) & (0.00, 0.40) & 100
                  & (100, 70) & (0.00, 0.49) & 100 \\
    \midrule
    \multirow{4}{*}{(1,4)/2}
      & AIC       & (86, 100)  & (0.14, 0.00) & 100
                  & (91, 100)  & (0.09, 0.00) & 100 \\
      & BIC       & (100, 100) & (0.00, 0.00) & 100
                  & (100, 100) & (0.00, 0.00) & 100 \\
      & EBIC      & (100, 100) & (0.00, 0.00) & 100
                  & (100, 100) & (0.00, 0.00) & 100 \\
    \midrule
    \multirow{4}{*}{(3,4)/2}
      & AIC       & (100, 100) & (0.00, 0.00) & 100
                  & (100, 100) & (0.00, 0.00) & 100 \\
      & BIC       & (100, 98)  & (0.00, 0.02) & 100
                  & (98, 100) & (0.02, 0.00) & 100 \\
      & EBIC      & (100, 83) & (0.00, 0.17) & 100
                  & (98, 100) & (0.02, 0.00) & 100 \\
    \bottomrule
  \end{tabular}
\end{table}
\end{landscape}

\begin{figure}
    \centering
    \includegraphics[width=\linewidth]{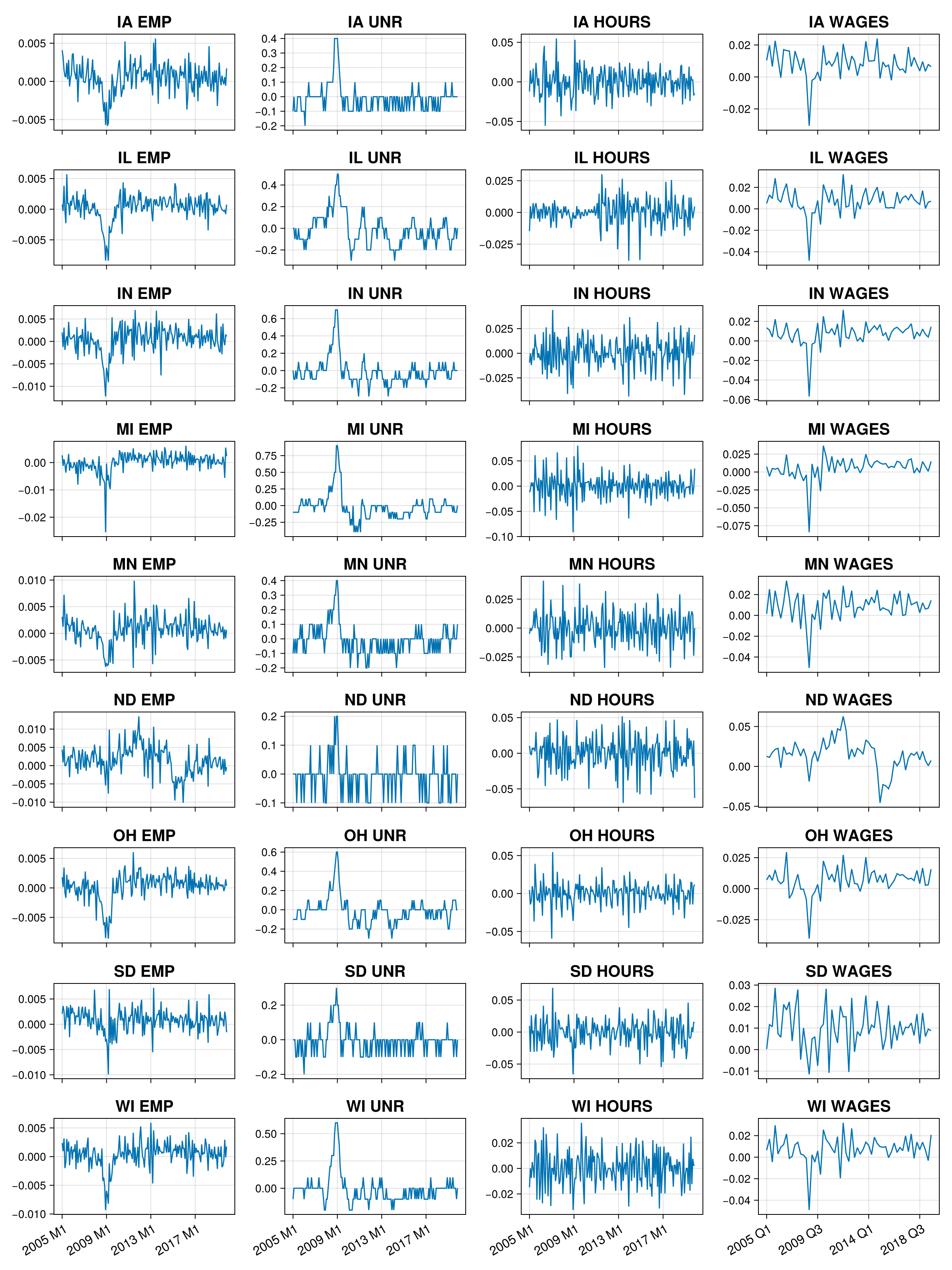}
    \caption{Economic indicators (columns) for nine U.S. states (rows) over the sample period January 2005 to December 2019.
    }
    \label{fig:state_econinds}
\end{figure}

\section{Application to Coincident Indexes of U.S. States}
\label{sec:empirical}

The pseudo-structural framework is applied to distill coincident dynamics from labor market indicators across nine Midwestern U.S. states.
Coincident dynamics move broadly in tandem with the overall state of the economy and are widely used to track regional business cycle conditions \citep{stock1989indexes, crone2005coincident}.
Although U.S. states often respond heterogeneously to shocks such as monetary policy changes and aggregate business cycle fluctuations, states within the same region may share similar economic structures and thus exhibit stronger co-movement \citep{carlino1999differential, owyang2005business}.
The remainder of this section describes the data and presents the empirical results based on this matrix representation.

\subsection{Data and Motivation}
\label{sec:data}

The nine states considered are Iowa (IA),  Illinois (IL), Indiana (IN), Michigan (MI), Minnesota (MN), North Dakota (ND), Ohio (OH), South Dakota (SD), and Wisconsin (WI).
These states span a range of economic structures, from the commodity-driven economies of Iowa, North Dakota, and South Dakota to the manufacturing-intensive labor markets of Indiana, Michigan, and Ohio, and the more diversified economies of Illinois and Minnesota.
This heterogeneity allows meaningful co-movement patterns to be identified along both the indicator and state dimensions of the model.

Following \citet{crone2005coincident}, four indicators are used for each state: nonagricultural employment, the unemployment rate, average hours worked in manufacturing, and real wage and salary disbursements.
These series jointly capture the quantity and price dimensions of the regional labor market, with employment and hours reflecting labor demand intensity, the unemployment rate measuring slack, and wages summarizing labor market tightness.
Let $\mathbf{Y}_t \in \mathbb{R}^{9 \times 4}$ collect the observed indicators in (transformed) levels, where rows index states and columns index indicators.
Logarithms are taken of nonagricultural employment, average hours, and real wages, while the unemployment rate is retained in levels.
As in \citet{stock1989indexes} and \citet{crone2005coincident}, the transformed levels $\mathbf{Y}_t$ are treated as integrated of order one, $I(1)$; the model is therefore specified on the stationary first differences $\Delta \mathbf{Y}_t = \mathbf{Y}_{t} - \mathbf{Y}_{t-1}$, demeaned over the sample prior to estimation:
\begin{align*}
    \Delta y_{ij,t} = y_{ij,t} - y_{ij,t-1}, \qquad \overline{\Delta y_{ij}} = \frac{1}{T-1}\sum_{t=2}^{T} \Delta y_{ij,t},
\end{align*}
so that $\Delta y_{ij,t}$ is a growth rate for the logged indicators and a change in the unemployment rate (in percentage points) for the unemployment column, and the model is estimated on the demeaned series $\Delta y_{ij,t} - \overline{\Delta y_{ij}}$.
Augmented Dickey Fuller tests confirming that the differenced series $\Delta \mathbf{Y}_t$ are stationary, with the unit-root null rejected for 32 of the 36 series (with the few exceptions mostly being borderline cases of employment and unemployment growth in a small number of states), are reported in Appendix \ref{sec:adf}.

All series are obtained from the Bureau of Labor Statistics (BLS, \url{https://www.bls.gov/}) and are available at monthly frequency except real wage and salary disbursements, which are reported quarterly by the Bureau of Economic Analysis (BEA, \url{https://www.bea.gov/data/income-saving/personal-income-by-state}).
This series is converted to monthly frequency by dividing each quarterly observation by three and assigning the resulting value to each month within the quarter.
The sample covers January 2005 to December 2019, yielding $T = 180$ monthly observations and a matrix-valued time series $\mathbf{Y}_t \in \mathbb{R}^{9 \times 4}$, where rows index states and columns index indicators.
Figure \ref{fig:state_econinds} gives a plot of all states with their corresponding transformed economic indicators.

\subsection{Rank Selection}
\label{sec:apprank}

\begin{table}[t]
\centering
\caption{Model comparison across the RRMAR, RRVAR, and unrestricted VAR 
for the nine-state, four-indicator Midwestern dataset ($T = 180$, 
$N_1 = 9$, $N_2 = 4$). The RRMAR is estimated at ranks $(\widehat{r}_1, \widehat{r}_2)$ selected by each criterion; the RRVAR rank is set to $r = \widehat{r}_1 \widehat{r}_2$.}
\label{tab:modelcomparison}
\setlength{\tabcolsep}{8pt}
\renewcommand{\arraystretch}{1.2}
\begin{tabular}{llcccc}
\toprule
\textbf{Model} & \textbf{Rank} & \textbf{Log-lik.} 
    & \textbf{AIC} & \textbf{BIC} & \textbf{EBIC} \\
\midrule
RRMAR  & $(2,1)$  & 20441.26 & -40698.52 & -40404.77 & \textbf{-40075.08} \\
RRMAR  & $(5,4)$  & 20576.80 & -40887.60 & \textbf{-40459.74} & -39979.55 \\
RRMAR  & $(8,4)$  & 20601.63 & -40905.26 & -40429.51 & -39895.57 \\
RRVAR  & $r=2$    & 21521.91 & \textbf{-41359.83} & -38680.77 & -35663.44 \\
RRVAR  & $r=20$   & 22387.47 & -41290.95 & -35748.28 & -29505.79 \\
RRVAR  & $r=32$   & 22459.18 & -40954.36 & -34648.07 & -27545.54 \\
Unres.\ VAR & $r=36$ & 22460.13 & -40924.26 & -34567.05 & -27407.18 \\
\bottomrule
\end{tabular}
\end{table}

Before estimating the pseudo-structural form, the ranks $r_1$ and $r_2$, which determine the number of co-movement equations along the indicator and state dimensions respectively, are selected.
A rank of $r_1 < N_1 = 9$ implies that the nine states share common dynamics across the indicators, while $r_2 < N_2 = 4$ implies that the four indicators share common dynamics across states.
Full rank in either dimension indicates no such reduction is supported by the data.
The information criteria described in Section \ref{sec:rankselection}, namely, AIC, BIC, and EBIC, are applied to jointly select $(r_1, r_2)$ and the lag order $p$, with the maximum lag set to two.

All three criteria select one lag, but differ in their selected ranks.
AIC selects $(\widehat{r}_1, \widehat{r}_2) = (8, 4)$, BIC selects $(\widehat{r}_1, \widehat{r}_2) = (5, 4)$, and EBIC selects $(\widehat{r}_1, \widehat{r}_2) = (2, 1)$.
While AIC and BIC both select full rank along the indicator dimension ($\widehat{r}_2 = 
N_2 = 4$), suggesting that the four labor market indicators contribute independently to regional business cycle dynamics, EBIC selects a single common component along this dimension.
Along the state dimension, all three criteria support some degree of rank reduction.
Given EBIC’s strong performance in the simulation study (Section \ref{sec:rankselectionsim}), together with the fact that its implied rank reduction aligns with the objective of constructing coincident indexes for the states, the ranks selected by EBIC are adopted.

Next, it is assessed whether the matrix autoregressive structure is supported relative to two natural benchmarks: a traditional reduced-rank VAR (RRVAR) applied to the vectorized series $\mathbf{y}_t = \operatorname{vec}(\mathbf{Y}_t)$, and an unrestricted VAR.
The RRVAR imposes a single reduced-rank restriction on the $N_1 N_2 \times N_1 N_2$ coefficient matrix without exploiting the Kronecker structure, while the unrestricted VAR imposes no restrictions at all.
By contrast, the RRMAR additionally restricts the coefficient matrix to have Kronecker product form, thereby reducing the number of free parameters from $r(N_1 N_2 - r)$ in the RRVAR to $2r_1 N_1 - r_1^2 + 2r_2 N_2 - r_2^2$ in the RRMAR.

Table~\ref{tab:modelcomparison} reports the log-likelihood, AIC, BIC, and EBIC for each model.
For a comparison on equal footing, 
the RRVAR rank is set to $r = \hat{r}_1 \hat{r}_2$, the implied Kronecker rank based on the ranks selected for the
 RRMAR under each information criterion.
All information criteria except for AIC favor the RRMAR over the RRVAR and the unrestricted VAR, confirming that the Kronecker product structure is not only parsimonious but also better supported by the data.
The unrestricted VAR achieves the highest log-likelihood by construction, but its information criteria values are substantially worse due to the large number of parameters relative to the sample size of $T = 180$.

\subsection{Pseudo-Structural Estimates}
\label{sec:appestimates}

The pseudo-structural form requires partitioning the nine states and four economic indicators into those populating $\mathbf{y}_{11,t}$, $\mathbf{y}_{12,t}$, $\mathbf{y}_{21,t}$, and $\mathbf{y}_{22,t}$, as given in Section \ref{sec:pseudostructural}.
Here the blocks $\mathbf{y}_{11,t},\mathbf{y}_{12,t},\mathbf{y}_{21,t},\mathbf{y}_{22,t}$ are sub-matrices of the stationary, differenced series $\Delta\mathbf{Y}_t$ of Section~\ref{sec:data} obtained by partitioning its $N_1$ indicator rows and $N_2$ state columns.
The differencing carries through to each block, and we suppress the $\Delta$ on the block symbols to lighten notation.
Since the model is identified up to this partition, any allocation yields observationally equivalent models with identical log-likelihood, and the choice is therefore a matter of convenience rather than a substantive assumption about within-period ordering.
In practice, it suffices to specify which $r_1$ rows and $r_2$ columns populate $\mathbf{y}_{22,t}$, as the remaining blocks are then uniquely determined as the complements.

Iowa (IA) and Ohio (OH) are assigned to the state dimension of $\mathbf{y}_{22,t}$, and real wage and salary disbursements to the indicator dimension.
Iowa and Ohio represent distinct economic structures, agriculture and manufacturing, respectively, and are geographically separated, making them natural candidates for states whose dynamics are not driven contemporaneously by the remaining states.
Wages are included as the indicator dimension of $\mathbf{y}_{22,t}$ because of their well-known stickiness, making them the slowest-moving of the four indicators.

\begin{table}[t]
\centering
\caption{Estimated coefficients in $\widehat{\boldsymbol{\delta}}^*$}
\label{tab:delta_coefs}
\begin{threeparttable}
\begin{tabular}{lcccccccc}
\toprule
 & ND & IN & MI & MN & IL & SD & WI \\
\midrule
 OH & $-0.29$ & $-0.26$ & $-0.58$ & $-0.18$ & $-2.48$ & $0.06$ & $-0.07$ \\
   & {\scriptsize($0.38)$} & {\scriptsize$(0.33)$} & {\scriptsize$(0.10)$} & {\scriptsize$(0.37)$} & {\scriptsize$(0.00)$} & {\scriptsize$(0.77)$} & {\scriptsize$(0.79)$} \\
 IA & $-0.02$ & $-1.39$ & $-1.38$ & $-0.75$ & $2.46$ & $-0.76$ & $-1.47$ \\
   & {\scriptsize$(0.96)$} & {\scriptsize$(0.00)$} & {\scriptsize$(0.01)$} & {\scriptsize$(0.02)$} & {\scriptsize$(0.00)$} & {\scriptsize$(0.03)$} & {\scriptsize$(0.00)$} \\
\bottomrule
\end{tabular}
\begin{tablenotes}
\small \item Note: $p$-values in parentheses.
\end{tablenotes}
\end{threeparttable}
\end{table}

\begin{table}[t]
\centering
\caption{Estimated coefficients in $\widehat{\boldsymbol{\gamma}}^*$}
\label{tab:gamma_coefs}
\begin{threeparttable}
\begin{tabular}{lcccc}
\toprule
 & $\Delta$Employment & $\Delta$Unemployment & $\Delta$Hours \\
\midrule
 $\Delta$Wages & $-0.63$ & $70.58$ & $-0.16$ \\
   & {\scriptsize($0.00)$} & {\scriptsize$(0.00)$} & {\scriptsize$(0.74)$} \\
\bottomrule
\end{tabular}
\begin{tablenotes}
\small \item Note: $p$-values in parentheses.
\end{tablenotes}
\end{threeparttable}
\end{table}

Table \ref{tab:delta_coefs} reports the estimated co-movement coefficients $\widehat{\boldsymbol{\delta}}^*$ for the state dimension, with $p$-values in parentheses.
The two rows correspond to Ohio and Iowa, capturing how the remaining seven states co-move with each of these two anchor states.
The results are asymmetric; Iowa's co-movement coefficients are statistically significant (at the 5\% level) for nearly all remaining states, with the exception of North Dakota, while Ohio's coefficients are significant only for Illinois.
This suggests that Iowa's agricultural cycle transmits broadly across the region, whereas Ohio's manufacturing dynamics are contemporaneously linked only to Illinois, likely reflecting their shared industrial base.
The near zero and insignificant coefficients for Iowa and North Dakota are most likely due to the North Dakota oil boom and subsequent bust from 2012 to 2015, which generated a highly localized economic cycle largely decoupled from the broader agricultural dynamics of the region.

Table \ref{tab:gamma_coefs} reports the estimated co-movement coefficients $\widehat{\boldsymbol{\gamma}}^*$ for the economic indicator dimension, with $p$-values in parentheses.
The single row corresponds to wages, capturing how the remaining three indicators co-move with real wage and salary disbursements.
Employment and unemployment are both significant, while hours worked is not.
The negative coefficient on employment suggests that wage growth and employment growth move in the same direction, while a positive coefficient on unemployment suggests they move in opposite directions.
The insignificance of hours worked suggests that within-period variation in manufacturing hours is not closely tied to wage dynamics, consistent with hours being adjusted more flexibly and at higher frequency than wages.

Table \ref{tab:joint_coefs} reports the estimated joint co-movement coefficients.
The results confirm the pattern suggested by the marginal estimates.
Hours worked is insignificant throughout, directly inheriting the insignificance of $\widehat{\boldsymbol{\gamma}}^*$ on that indicator, while employment and unemployment display significant joint co-movements for the same states that were significant in Table \ref{tab:delta_coefs}.
This illustrates a key feature of the pseudo-structural decomposition.
The joint component does not introduce new co-movement patterns, but rather identifies the relationship between the blocks $\mathbf{y}_{22,t}$ and $\mathbf{y}_{11,t}$, which are the combinations linked through both the row and column dimension.

\begin{table}[t]
\centering
\footnotesize
\caption{Estimated joint co-movement coefficients 
$\widehat{\boldsymbol{\gamma}}^* \otimes \widehat{\boldsymbol{\delta}}^*$. 
These are relations from wages of Ohio and Iowa to labor market indicators of remaining states.}
\label{tab:joint_coefs}
\begin{threeparttable}
\setlength{\tabcolsep}{4pt}
\begin{tabular}{llccccccc}
\toprule
& & \multicolumn{7}{c}{Receiving State} \\
\cmidrule(lr){3-9}
Anchor & Indicator & ND & IN & MI & MN & IL & SD & WI \\
\midrule
\multirow{3}{*}{ $\Delta$OH wages} 
 & $\Delta$Emp.   & $0.18$ & $0.16$ & $0.37$ & $0.11$ & $1.57$ & $-0.04$ & $0.04$ \\
 &        & {\scriptsize$(0.39)$} & {\scriptsize$(0.62)$} & {\scriptsize$(0.04)$} & {\scriptsize$(0.68)$} & {\scriptsize$(0.00)$} & {\scriptsize$(0.91)$} & {\scriptsize$(0.72)$} \\
 & $\Delta$Unemp. & $-20.51$ & $-18.26$ & $-40.67$ & $-12.67$ & $-175.04$ & $4.51$ & $-5.01$ \\
 
 &        & {\scriptsize$(0.38)$} & {\scriptsize$(0.62)$} & {\scriptsize$(0.04)$} & {\scriptsize$(0.68)$} & {\scriptsize$(0.00)$} & {\scriptsize$(0.91)$} & {\scriptsize$(0.72)$} \\
 & $\Delta$Hours  & $0.05$ & $0.04$ & $0.09$ & $0.03$ & $0.40$ & $-0.01$ & $0.01$ \\
 &        & {\scriptsize$(0.76)$} & {\scriptsize$(0.78)$} & {\scriptsize$(0.74)$} & {\scriptsize$(0.79)$} & {\scriptsize$(0.74)$} & {\scriptsize$(0.91)$} & {\scriptsize$(0.80)$} \\
\midrule
\multirow{3}{*}{$\Delta$IA wages} 
 & $\Delta$Emp.   & $0.02$ & $0.88$ & $0.87$ & $0.48$ & $-1.55$ & $0.48$ & $0.93$ \\
 &        & {\scriptsize$(0.94)$} & {\scriptsize$(0.02)$} & {\scriptsize$(0.12)$} & {\scriptsize$(0.00)$} & {\scriptsize$(0.00)$} & {\scriptsize$(0.01)$} & {\scriptsize$(0.00)$} \\
 & $\Delta$Unemp. & $-1.78$ & $-98.44$ & $-97.25$ & $-53.03$ & $173.38$ & $-53.36$ & $-104.05$ \\
 &        & {\scriptsize$(0.94)$} & {\scriptsize$(0.02)$} & {\scriptsize$(0.12)$} & {\scriptsize$(0.00)$} & {\scriptsize$(0.00)$} & {\scriptsize$(0.01)$} & {\scriptsize$(0.00)$} \\
 & $\Delta$Hours  & $0.00$ & $0.23$ & $0.22$ & $0.12$ & $-0.39$ & $0.12$ & $0.23$ \\
 &        & {\scriptsize$(0.94)$} & {\scriptsize$(0.74)$} & {\scriptsize$(0.75)$} & {\scriptsize$(0.74)$} & {\scriptsize$(0.74)$} & {\scriptsize$(0.74)$} & {\scriptsize$(0.74)$} \\
\bottomrule
\end{tabular}
\begin{tablenotes}
\small \item Note: $p$-values in parentheses. Each cell reports the 
co-movement from the wages of the anchor state (Ohio or Iowa) to the labor market indicator of the receiving state, computed as $\widehat{\gamma}_j^* \cdot \widehat{\delta}_k^*$.
\end{tablenotes}
\end{threeparttable}
\end{table} 

\subsection{Comparison with \citet{crone2005coincident}}
\label{sec:ccm}

The coincident signal extracted by \citet{crone2005coincident} is next compared with that of our pseudo-structural method.
The two approaches differ fundamentally in construction, with \citet{crone2005coincident} using a state-space model to extract a single latent factor from the same four labor market indicators, while the pseudo-structural method recovers coincident dynamics through the reduced-rank decomposition, which additionally identifies the co-movement structure across both the state and indicator dimensions.
Both methods are estimated on stationary first-differenced data, and their estimated common components are themselves stationary growth-rate signals.
The two are therefore compared at the stationary stage.
In particular, each series is standardized to zero mean and unit variance, and their agreement is assessed through the pairwise correlation, which is well defined precisely because both series are stationary.
High correlation would indicate that the pseudo-structural approach recovers the same underlying business cycle signal despite the methodological differences.

\begin{figure}[t]
    \centering
    \includegraphics[width=\linewidth]{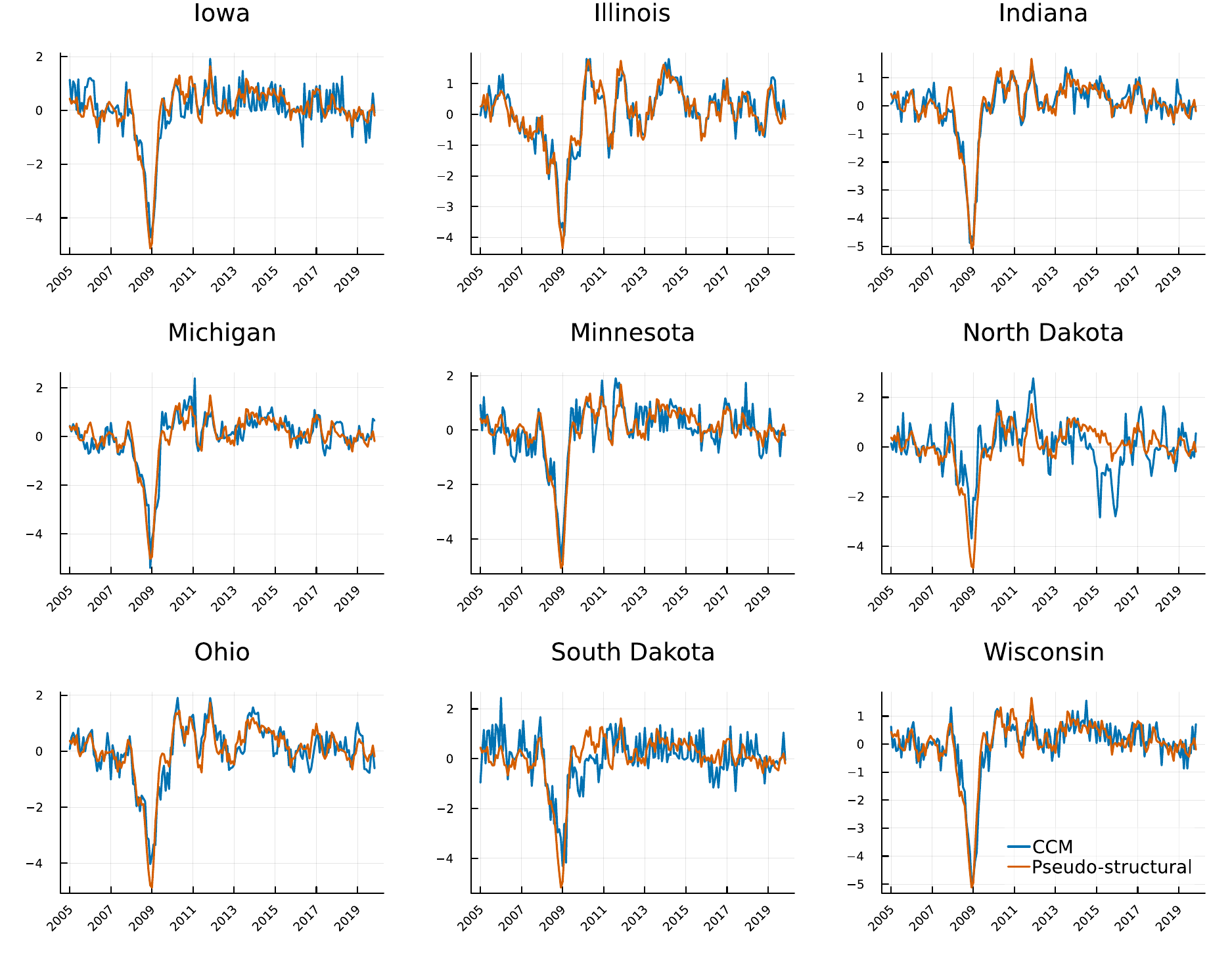}
    \caption{Standardized coincident growth rates for nine Midwestern U.S. states over the
sample period January 2005 to December 2019. Each panel displays the
Crone and Clayton-Matthews common component (blue, CCM) alongside the pseudo-structural
coincident component (orange). Both series are in stationary first differences and standardized to zero mean and unit variance.}
    \label{fig:coincident_indexes}
\end{figure}

Figure \ref{fig:coincident_indexes} plots the standardized pseudo-structural coincident component alongside that of \citet{crone2005coincident} for each of the nine states over the sample period.
The two series track each other closely for most states, suggesting that the pseudo-structural model captures the broad business cycle dynamics identified by the benchmark.
Table \ref{tab:correlations} reports the pairwise correlation between the two standardized series for each state.
Correlations are high for most states, ranging from $0.85$ to $0.94$, with the notable exceptions of North Dakota ($0.59$) and South Dakota ($0.68$), where the localized oil boom dynamics discussed in Section \ref{sec:appestimates} likely drive the divergence.
The strong co-movement between the two indexes for the remaining seven states provides informal validation that the pseudo-structural model recovers economically meaningful coincident dynamics from the underlying labor market indicators.

\begin{table}[t]
\centering
\caption{Correlations between pseudo-structural and 
\citet{crone2005coincident} coincident indexes.}
\label{tab:correlations}
\begin{tabular}{lc}
\toprule
\textbf{State} & \textbf{Correlation} \\
\midrule
Iowa (IA)         & $0.855$ \\
Illinois (IL)     & $0.933$ \\
Indiana (IN)      & $0.935$ \\
Michigan (MI)     & $0.885$ \\
Minnesota (MN)    & $0.857$ \\
North Dakota (ND) & $0.587$ \\
Ohio (OH)         & $0.895$ \\
South Dakota (SD) & $0.679$ \\
Wisconsin (WI)    & $0.890$ \\
\bottomrule
\end{tabular}
\end{table}

\section{Conclusion}
\label{sec:conclusion}

A pseudo-structural framework for reduced-rank matrix-valued time series models is introduced, enabling the decomposition of contemporaneous co-movements into row-specific, column-specific, and joint components.
By leveraging the Kronecker structure of the reduced-rank matrix autoregressive model, interpretable linear combinations that annihilate serial correlation are derived, akin to common feature analysis in vector autoregressions.
The approach provides explicit inference on these co-movement structures, supported by an estimation procedure to navigate the non-convex landscape and a rank-lag selection criterion validated in simulations.
An empirical application to coincident indicators across nine Midwestern U.S. states reveals distinct co-movement patterns along both the state and indicator dimensions, with Iowa's agricultural cycle transmitting broadly across the region while Ohio's manufacturing dynamics link primarily to Illinois.
All code and replication material are available in the Julia repository \texttt{PseudoStructuralComovements} on the second author's GitHub page \url{https://github.com/ivanuricardo/pseudostructuralcomovements}.

The proposed methodology can be extended in several directions.
One could consider a tensor-valued time series instead of a matrix-valued time series.
The left null space would then require additional terms to further disaggregate the co-movement structure into three co-movement dimensions instead of simply row- and column-wise co-movement relations.
Another interesting future research direction is the investigation of non-contemporaneous co-movements in matrix-valued time series (e.g., \citealp{cubadda2001noncontemporaneous}).
This would allow for adjustment delays in the co-movement relations across the different rows and columns of the matrix-valued time series.

\bigskip
\noindent
{\bf Acknowledgements.}
We thank the editor, associate editor and referees for their constructive and detailed comments which substantially improved the quality of the manuscript.
The last author was financially supported by the Dutch Research Council (NWO) under grant number VI.Vidi.211.032.

\appendix

\section{Appendix}

\subsection{Proof of Proposition 1}
\label{sec:proofprop1}

\begin{proof}
    Consider the left null space of $\mathbf{U}_{2} \otimes \mathbf{U}_{1}$.
    There exist left null space matrices $\boldsymbol{\delta}$ and $\boldsymbol{\gamma}$ that  annihilate $\mathbf{U}_{1}$ and $\mathbf{U}_{2}$ respectively.
    By definition of the annihilator, there exists an orthogonal splitting
    \begin{align*}
        \mathbf{W}_{1} &= \operatorname{ColSpan}(\mathbf{U}_1) \oplus \operatorname{RowSpan}(\boldsymbol{\delta}), \\
        \mathbf{W}_{2} &= \operatorname{ColSpan}(\mathbf{U}_2) \oplus \operatorname{RowSpan}(\boldsymbol{\gamma}),
    \end{align*}
    where $\mathbf{W}_{1}$ makes up the entire space associated with $\mathbf{U}_{1}$ and 
    likewise $\mathbf{W}_{2}$ makes up the entire space associated with $\mathbf{U}_{2}$.
    Taking the Kronecker product of these spaces and distributing over the direct sum gives
    \begin{align*}
        \mathbf{W}_{2} \otimes \mathbf{W}_{1} &= \left(\operatorname{ColSpan}(\mathbf{U}_1) 
        \oplus \operatorname{RowSpan}(\boldsymbol{\delta})\right) \otimes 
        \left(\operatorname{ColSpan}(\mathbf{U}_2) \oplus \operatorname{RowSpan}(\boldsymbol{\gamma})\right) \\
        &= \left(\operatorname{ColSpan}(\mathbf{U}_{1}) \otimes \operatorname{ColSpan}(\mathbf{U}_{2})\right) 
        \oplus \left(\operatorname{ColSpan}(\mathbf{U}_{2}) \otimes \operatorname{RowSpan}(\boldsymbol{\delta})\right) \\
        &\quad \oplus \left(\operatorname{RowSpan}(\boldsymbol{\gamma}) \otimes 
        \operatorname{ColSpan}(\mathbf{U}_{1})\right) \oplus \left(\operatorname{RowSpan}(\boldsymbol{\gamma}) 
        \otimes \operatorname{RowSpan}(\boldsymbol{\delta})\right).
    \end{align*}
    By \citet[][formula 1.11]{greub1978multilinear}, the image of the Kronecker product satisfies 
    $\operatorname{Im}(\mathbf{U}_2 \otimes \mathbf{U}_1) = \operatorname{ColSpan}(\mathbf{U}_1) \otimes 
    \operatorname{ColSpan}(\mathbf{U}_2)$, so the first term corresponds to the column space of the 
    reduced-rank matrix autoregressive model.
    The remaining three terms therefore constitute the orthogonal complement, i.e., the null space 
    of $(\mathbf{U}_2 \otimes \mathbf{U}_1)^\top$.
    This refines the result of \citet[][formula 1.12]{greub1978multilinear}, who establishes the 
    kernel of a tensor product of two linear maps as a non-orthogonal sum of two subspaces; here, 
    the orthogonal splitting of each factor space yields a three-way \emph{direct} sum, 
    giving the decomposition stated in the proposition.
\end{proof}
\subsection{Proof of Proposition 2}

\begin{proof}
The proof is written in terms of the BIC penalty $c_T = \ln T$, but applies equally to the EBIC with $c_T = \ln(T) + 2\gamma\ln(N_1 N_2)$, since the additional term $2\gamma\ln(N_1 N_2)$ is fixed as $T \to \infty$ and thus preserves both $c_T \to \infty$ and $c_T/T \to 0$.
BIC is shown to yield weakly consistent estimators for ranks $(r_1,r_2)$.
With $N=N_1N_2$ and $p$ fixed and $T\to\infty$, the penalty $c_T=\ln T$ satisfies $c_T\to\infty$ and $c_T/T\to0$ (e.g. \citealp{lutkepohl2005new}).
Standard results for reduced-rank MLE in the pseudo-structural model \eqref{eq:lagp_pseudostructural} imply that the full-information MLE in \eqref{eq:fiml} attains the usual $\sqrt T$ rate (cf. \citealp{xiao_reduced_forth}).

Define
\[  
\text{BIC}(r_1,r_2)=-2\mathcal L(\widehat\theta_{r_1,r_2}) + \ln(T)\,\phi(r_1,r_2),  
\]
where $\phi(r_1,r_2)=2r_1N_1+2r_2N_2-r_1^2-r_2^2$.
Let $(r_1,r_2)$ be the true ranks and $(r_1',r_2')$ an arbitrary candidate.
Define
\[
\Delta\ell_T=\mathcal L(\widehat\theta_{r_1',r_2'})-\mathcal L(\widehat\theta_{r_1,r_2}),
\quad
\Delta\phi=\phi(r_1',r_2')-\phi(r_1,r_2),
\]
and
\[
\Delta\text{BIC}=-2\Delta\ell_T+\ln(T)\,\Delta\phi.
\]

\textbf{1. Underestimation:}  If $r_i'<r_i$ for some $i$, model misspecification yields a likelihood loss of order $T$, i.e.\ $\Delta\ell_T=-O_p(T)$ \citep{white1982misspecified}.
Since reducing a rank lowers $\phi$, it follows that $\Delta\phi<0$.
Hence
\[
\Delta\text{BIC}=O_p(T)-O_p(\ln T)>0
\]
as $T\to\infty$ w.p.\ $\to1$.

\textbf{2. Overestimation:}  If $r_i'>r_i$ for some $i$, the extra parameters do not improve the limit likelihood, so $\Delta\ell_T=o_p(1)$, while $\Delta\phi>0$.  Thus
\[
\Delta\text{BIC}=o_p(1)+O_p(\ln T)>0
\]
with probability tending to one.

\textbf{3. Mixed case:}  If one rank is under and the other overestimated, misspecification in one dimension still incurs $\Delta\ell_T=-O_p(T)$.  Whether $\Delta\phi$ is positive, negative, or zero, the $O_p(T)$ term dominates $O(\ln T)$, so $\Delta\text{BIC}>0$ w.p.\ $\to1$.

In all cases any misspecified $(r_1',r_2')\neq(r_1,r_2)$ yields larger BIC than at the true ranks, proving weak consistency.
\end{proof}

\subsection{Companion Form for the Pseudo-Structural Model}
\label{sec:companionform}

This section details how to construct the companion form for the pseudo-structural model.
Given the pseudo-structural model with one lag in equation \eqref{eq:pseudo-structural}, the pseudo-structural model with $p$ lags is
\begin{align*}
    \boldsymbol{\Omega}^* \mathbf{y}_t = \boldsymbol{\Pi}_1 \mathbf{y}_{t-1} + \boldsymbol{\Pi}_2 \mathbf{y}_{t-2} + \dots + \boldsymbol{\Pi}_p \mathbf{y}_{t-p} + \mathbf{e}_t
\end{align*}
where $\boldsymbol{\Omega}^* = \boldsymbol{\Omega} \mathbf{P}$ and $\mathbf{P}$ is defined as follows. Define the row- and column-selection matrices
\begin{align*}
            \mathbf{S}_{1} =
	\begin{bmatrix}
	    \mathbf{I}_{N_{1}-r_{1}} & \mathbf{0} 
	\end{bmatrix} \in \mathbb{R}^{N_{1} - r_{1} \times N_{1}}, \qquad 
	\mathbf{S}_{2} = 
	\begin{bmatrix}
	    \mathbf{0} & \mathbf{I}_{r_{1}}
	\end{bmatrix} \in \mathbb{R}^{r_{1} \times N_{1}}, \\
	\mathbf{T}_{1} = 
	\begin{bmatrix}
	    \mathbf{I}_{N_{2} - r_{2}} & \mathbf{0}
	\end{bmatrix} \in \mathbb{R}^{N_{2} - r_{2} \times N_{2}}, \qquad 
	\mathbf{T}_{2} = 
	\begin{bmatrix}
	    \mathbf{0} & \mathbf{I}_{r_{2}}
	\end{bmatrix} \in \mathbb{R}^{r_{2} \times N_{2}},
\end{align*}
    so that each block satisfies $\mathbf{Y}_{ij,t} = \mathbf{S}_{i} \mathbf{Y}_{t} \mathbf{T}_{j}^{\top}$.
    Applying the standard vec identity $\operatorname{vec}(\mathbf{A} \mathbf{B} \mathbf{C}) = (\mathbf{C}^{\top} \otimes \mathbf{A}) \operatorname{vec}(\mathbf{B})$ gives $\operatorname{vec}(\mathbf{Y}_{ij,t}) = (\mathbf{T}_{j} \otimes \mathbf{S}_{i}) \operatorname{vec}(\mathbf{Y}_{t})$.
    Stacking the four blocks then yields $\mathbf{y}_{t}^{*} = \mathbf{P} \operatorname{vec}(\mathbf{Y}_{t})$, with
    \begin{align*}
        \mathbf{P} = 
	\begin{bmatrix}
	    \mathbf{T}_{1} \otimes \mathbf{S}_{1} \\
	    \mathbf{T}_{1} \otimes \mathbf{S}_{2} \\
	    \mathbf{T}_{2} \otimes \mathbf{S}_{1} \\
	    \mathbf{T}_{2} \otimes \mathbf{S}_{2}
	\end{bmatrix}.
    \end{align*}
    This is a permutation matrix because $\begin{bmatrix} \mathbf{S}_{1}^{\top}, \mathbf{S}_{2}^{\top} \end{bmatrix}^{\top} = \mathbf{I}_{N_{1}}$ and $\begin{bmatrix} \mathbf{T}_{1}^{\top}, \mathbf{T}_{2}^{\top} \end{bmatrix}^{\top} = \mathbf{I}_{N_{2}}$, thus each row and column of $\mathbf{P}$ contain exactly one unit entry.

Additionally, the matrices $\boldsymbol{\Pi}_{i}$ for $i = 1, \dots, p$ are lagged autoregressive coefficient matrices with a Kronecker product restriction 
\begin{align*}
    \boldsymbol{\Pi}_{i} = 
    \begin{bmatrix}
	\mathbf{0} \\
	\vdots \\
	(\mathbf{U}_{4,i} \otimes \mathbf{U}_{3,i})^\top
    \end{bmatrix}.
\end{align*}

The companion form is then given by
\begin{align*}
    \begin{bmatrix}
	\boldsymbol{\Omega} \mathbf{P} & \dots & \mathbf{0} \\
	\vdots & \ddots & \vdots \\
	\mathbf{0} & \dots & \mathbf{I}
    \end{bmatrix}
    \begin{bmatrix}
        \mathbf{y}_{t} \\
	\vdots \\
        \mathbf{y}_{t-p+1}
    \end{bmatrix}
    = 
    \begin{bmatrix}
	\boldsymbol{\Pi}_{1} & \dots & \boldsymbol{\Pi}_{p-1} & \boldsymbol{\Pi}_{p} \\
	\mathbf{I} &  & \mathbf{0} & \mathbf{0} \\
	 & \ddots &  & \vdots \\
	\mathbf{0} &  & \mathbf{I} & \mathbf{0} \\
    \end{bmatrix}
    \begin{bmatrix}
        \mathbf{y}_{t-1} \\
	\vdots \\
	\mathbf{y}_{t-p}
    \end{bmatrix}
    + 
    \begin{bmatrix}
	\boldsymbol{\Omega} \mathbf{P} & \dots & \mathbf{0} \\
	\vdots & \ddots & \vdots \\
	\mathbf{0} & \dots & \mathbf{I}
    \end{bmatrix}
    \begin{bmatrix}
        \mathbf{e}_{t} \\
	\vdots \\
        \mathbf{0}
    \end{bmatrix}.
\end{align*}

\subsection{Rotational Invariance}
\label{sec:norm_invariance}
    For any two invertible matrices $\mathbf{Q}_{1}$ and $\mathbf{Q}_{2}$, the representations
    \begin{align*}
        \mathbf{Y}_{t} = \mathbf{U}_{1} \mathbf{U}_{3}^{\top} \mathbf{Y}_{t-1} \mathbf{U}_{4} \mathbf{U}_{2}^{\top} + \mathbf{E}_{t},
    \end{align*}
    and
    \begin{align*}
    \mathbf{Y}_{t} = \underbrace{\mathbf{U}_{1} \mathbf{Q}_{1}}_{\mathbf{U}_1^*} \underbrace{\mathbf{Q}_{1}^{-1} \mathbf{U}_{3}^{\top}}_{\mathbf{U}_3^{*\top}} \mathbf{Y}_{t-1} \underbrace{\mathbf{U}_{4} \mathbf{Q}_{2}^{-\top}}_{\mathbf{U}_4^{*}} \underbrace{\mathbf{Q}_2^{\top} \mathbf{U}_{2}^\top}_{\mathbf{U}_2^{*\top}} + \mathbf{E}_{t},
    \end{align*}
    lead to the same result.
    The matrices $\mathbf{Q}_{1}$ and $\mathbf{Q}_{2}$ are defined as the inverse of the bottom $r_{1} \times r_{1}$ or $r_{2} \times r_{2}$ submatrix of $\mathbf{U}_{1}$ and $\mathbf{U}_{2}$ respectively, assuming this inverse exists (this inverse may not always exist, but one may always rearrange the matrix time series such that this inverse does exist).
    Thus, this yields
    \begin{align*}
	\mathbf{U}_{1}^* = 
    \begin{bmatrix}
	- \boldsymbol{\delta}^{*\top} \\
	\mathbf{I}
    \end{bmatrix} \qquad \text{and} \qquad 
    \mathbf{U}_{2}^* = 
    \begin{bmatrix}
	- \boldsymbol{\gamma}^{*\top} \\
	\mathbf{I}
    \end{bmatrix}.
    \end{align*}
    Taking the left nullspace of $\mathbf{U}_{1}^*$ and $\mathbf{U}_{2}^*$ then yields 
    \begin{align*}
	\mathcal{N}(\mathbf{U}_{1}^{*\top}) = 
    \begin{bmatrix}
	\mathbf{I} &
	\boldsymbol{\delta}^{*\top}
    \end{bmatrix} \qquad \text{and} \qquad 
	\mathcal{N}(\mathbf{U}_{2}^{*\top}) = 
    \begin{bmatrix}
	\mathbf{I} & 
	\boldsymbol{\gamma}^{*\top}
    \end{bmatrix}.
    \end{align*}
Because left multiplying by the left nullspace yields the zero vector, this is a valid nullspace for each of $\mathbf{U}_{1}^*$ and $\mathbf{U}_{2}^*$.
    
    To illustrate, consider $N_1 = N_2 = 2$ with true rank $(r_1, r_2) = (1,1)$, 
so that $\boldsymbol{\delta}^*$ and $\boldsymbol{\gamma}^*$ are scalars. 
Suppose the true reduced-form dynamics are given by
\begin{align*}
    \mathbf{A} = \mathbf{U}_1\mathbf{U}_3^\top = 
    \begin{bmatrix} 2 & 2 \\ 1 & 1 \end{bmatrix}, \qquad
    \mathbf{B} = \mathbf{U}_4\mathbf{U}_2^\top = 
    \begin{bmatrix} 6 & 6 \\ 4 & 4 \end{bmatrix}.
\end{align*}
\textbf{Normalization 1} (bottom row set to identity, as in the paper):
\begin{align*}
    \mathbf{U}_1^{(1)} = \begin{bmatrix} 2 \\ 1 \end{bmatrix}, \quad
    \mathbf{U}_3^{(1)} = \begin{bmatrix} 1 \\ 1 \end{bmatrix}, \quad
    \mathbf{U}_2^{(1)} = \begin{bmatrix} 3/2 \\ 1 \end{bmatrix}, \quad
    \mathbf{U}_4^{(1)} = \begin{bmatrix} 4 \\ 4 \end{bmatrix},
\end{align*}
yielding $\boldsymbol{\delta}^{*(1)} = -2$ and $\boldsymbol{\gamma}^{*(1)} = -3/2$, and thus
\begin{align*}
    \boldsymbol{\Omega}_1^{(1)} = \begin{bmatrix} 1 & -2 \\ 0 & 1 \end{bmatrix}, \qquad
    \boldsymbol{\Omega}_2^{(1)} = \begin{bmatrix} 1 & -3/2 \\ 0 & 1 \end{bmatrix}.
\end{align*}
\textbf{Normalization 2} (top row set to identity):
\begin{align*}
    \mathbf{U}_1^{(2)} = \begin{bmatrix} 1 \\ 1/2 \end{bmatrix}, \quad
    \mathbf{U}_3^{(2)} = \begin{bmatrix} 2 \\ 2 \end{bmatrix}, \quad
    \mathbf{U}_2^{(2)} = \begin{bmatrix} 1 \\ 2/3 \end{bmatrix}, \quad
    \mathbf{U}_4^{(2)} = \begin{bmatrix} 6 \\ 4 \end{bmatrix},
\end{align*}
yielding $\boldsymbol{\delta}^{*(2)} = -1/2$ and $\boldsymbol{\gamma}^{*(2)} = -2/3$, and thus
\begin{align*}
    \boldsymbol{\Omega}_1^{(2)} = \begin{bmatrix} -1/2 & 1 \\ 1 & 0 \end{bmatrix}, \qquad
    \boldsymbol{\Omega}_2^{(2)} = \begin{bmatrix} -2/3 & 1 \\ 1 & 0 \end{bmatrix}.
\end{align*}
Despite $\boldsymbol{\delta}^{*(1)} \neq \boldsymbol{\delta}^{*(2)}$ and 
$\boldsymbol{\gamma}^{*(1)} \neq \boldsymbol{\gamma}^{*(2)}$, both normalizations 
recover identical reduced-form dynamics:
\begin{align*}
    \mathbf{U}_1^{(1)}\mathbf{U}_3^{(1)\top} = \mathbf{U}_1^{(2)}\mathbf{U}_3^{(2)\top} = 
    \begin{bmatrix} 2 & 2 \\ 1 & 1 \end{bmatrix}, \qquad
    \mathbf{U}_4^{(1)}\mathbf{U}_2^{(1)\top} = \mathbf{U}_4^{(2)}\mathbf{U}_2^{(2)\top} = 
    \begin{bmatrix} 6 & 6 \\ 4 & 4 \end{bmatrix}.
\end{align*}
This confirms that while the pseudo-structural parameters $\boldsymbol{\delta}^*$ and 
$\boldsymbol{\gamma}^*$, and hence $\boldsymbol{\Omega}_1$ and $\boldsymbol{\Omega}_2$, depend on the normalization choice, the reduced-form dynamics are invariant to it.

\subsection{Algorithm}
\label{sec:algorithm}

\RestyleAlgo{ruled}
\begin{algorithm}[ht]
\caption{Pseudocode that maximizes the log likelihood objective in equation \eqref{eq:pseudo-structural}.}
\label{alg:maxloglike}
\KwIn{Ranks $(r_{1}, r_{2})$, lag order $p$, convergence tolerance $\epsilon = 10^{-10}$, number of initializations $K = 100$, continued initializations $L = 10$, maximum iterations $t_{max}= 1000$.}
\KwOut{$\boldsymbol{\theta}^{(t)}$ that maximizes $\mathcal{L}(\theta)$ in equation \eqref{eq:pseudo-structural}.}

\For{$i \in \{1, \dots, K\}$}{
  Initialize $\boldsymbol{\theta}_{i}^{(0)}$ at random start \\
  Run BFGS for the optimization problem in equation \eqref{eq:lagp_pseudostructural} for $5$ iterations \\
  Save $\boldsymbol{\theta}_{i}^{(5)}$ and the maximized value $\mathcal{L}(\boldsymbol{\theta}_{i}^{(5)})$
}

Filter the best $L = 10$ initializations as measured by the maximized value $\mathcal{L}(\boldsymbol{\theta}_{i}^{(5)})$ for $i = 1, \dots, K$

\For{$i \in \{1, \dots, L\}$}{
  Run BFGS for the optimization problem in equation \eqref{eq:pseudo-structural} under $\boldsymbol{\theta}_{i}^{(5)}$ for $i = 1, \dots, L$, convergence criteria $\| \mathcal{L}(\boldsymbol{\theta}_{i}^{(t)}) - \mathcal{L}(\boldsymbol{\theta}_{i}^{(t-1)}) \| < \epsilon$, and maximum iterations $t_{max}$ \\
  Save $\boldsymbol{\theta}_{i}^{(t)}$ and the maximized value $\mathcal{L}(\boldsymbol{\theta}_{i}^{(t)})$
}

\Return $\boldsymbol{\theta}_{i}^{(t)}$ which maximizes the value of $\mathcal{L}(\boldsymbol{\theta}_{i}^{(t)})$

\end{algorithm}

Algorithm \ref{alg:maxloglike} details how the log likelihood criterion in equation \eqref{eq:lagp_pseudostructural} is maximized.
The optimization is difficult, due to the nonconvexity of the objective function; a strategy is therefore used that leverages multiple initializations along with continuing with the best initializations.
The initializations may include an initialization based on the RRMAR given in \citet{xiao_reduced_forth}.
Because the parameters for the RRMAR also provide consistent estimates for the pseudo-structural parameters, this can be leveraged by starting at the parameter values for an estimated RRMAR.
The rotated parameter values can then serve as an initialization for Algorithm \ref{alg:maxloglike}.
This can be further improved by adding a permutation on the rotated parameter values to explore the surrounding region.

Throughout the algorithm, basic checks are also considered to ensure that the estimate is not at a saddle point.
This includes checking whether the Hessian of the objective function at the converged estimate $\boldsymbol{\theta}_{i}^{(t)}$ has negative eigenvalues.
If the Hessian does have negative eigenvalues, this parameter estimate is discarded.
However, due to numerical instability, extremely small negative eigenvalues may arise (for instance, when the rank is overestimated).
In this case, the Hessian is projected to the nearest positive semidefinite matrix.
An additional check may be to monitor the Frobenius norm of the gradient of the objective function as a surrogate to the normal convergence criteria.
This indicates whether the optimization is stuck in a shallow but steep region where some progress is made but at a very slow rate.
Both evaluating the Hessian of the objective function and monitoring the gradient norm are therefore recommended to reduce the chance of becoming stuck at saddle points.

\subsection{Stationarity Test Results}
\label{sec:adf}

\begin{table}[ht]
\centering

\caption{Augmented Dickey Fuller tests for the differenced series $\Delta \mathbf{Y}_t$.
Each cell reports the ADF statistic with the $p$-value below in parentheses.
The test is run with an intercept and a lag order selected by BIC.
An asterisk denotes rejection of the unit-root null at the 5\% level.}
\label{tab:adf}
\setlength{\tabcolsep}{6pt}
\renewcommand{\arraystretch}{1.1}
\begin{tabular}{lcccc}
\toprule
\textbf{State} & \textbf{Employment} & \textbf{Unemployment} & \textbf{Hours} & \textbf{Wages} \\
\midrule
Iowa (IA) & $-10.88^{*}$ & $-3.95^{*}$ & $-17.45^{*}$ & $-4.33^{*}$ \\
 & {\scriptsize(0.00)} & {\scriptsize(0.00)} & {\scriptsize(0.00)} & {\scriptsize(0.00)} \\
Illinois (IL) & $-3.38^{*}$ & $-3.78^{*}$ & $-13.44^{*}$ & $-4.11^{*}$ \\
 & {\scriptsize(0.01)} & {\scriptsize(0.00)} & {\scriptsize(0.00)} & {\scriptsize(0.00)} \\
Indiana (IN) & $-4.00^{*}$ & $-2.52$ & $-18.27^{*}$ & $-4.08^{*}$ \\
 & {\scriptsize(0.00)} & {\scriptsize(0.11)} & {\scriptsize(0.00)} & {\scriptsize(0.00)} \\
Michigan (MI) & $-2.09$ & $-2.86$ & $-6.03^{*}$ & $-4.20^{*}$ \\
 & {\scriptsize(0.25)} & {\scriptsize(0.05)} & {\scriptsize(0.00)} & {\scriptsize(0.00)} \\
Minnesota (MN) & $-11.48^{*}$ & $-3.54^{*}$ & $-13.72^{*}$ & $-7.25^{*}$ \\
 & {\scriptsize(0.00)} & {\scriptsize(0.01)} & {\scriptsize(0.00)} & {\scriptsize(0.00)} \\
North Dakota (ND) & $-3.66^{*}$ & $-6.00^{*}$ & $-19.03^{*}$ & $-3.32^{*}$ \\
 & {\scriptsize(0.01)} & {\scriptsize(0.00)} & {\scriptsize(0.00)} & {\scriptsize(0.02)} \\
Ohio (OH) & $-2.77$ & $-3.73^{*}$ & $-6.19^{*}$ & $-4.07^{*}$ \\
 & {\scriptsize(0.06)} & {\scriptsize(0.00)} & {\scriptsize(0.00)} & {\scriptsize(0.00)} \\
South Dakota (SD) & $-14.95^{*}$ & $-3.61^{*}$ & $-16.89^{*}$ & $-2.99^{*}$ \\
 & {\scriptsize(0.00)} & {\scriptsize(0.01)} & {\scriptsize(0.00)} & {\scriptsize(0.04)} \\
Wisconsin (WI) & $-3.98^{*}$ & $-2.91^{*}$ & $-13.95^{*}$ & $-4.62^{*}$ \\
 & {\scriptsize(0.00)} & {\scriptsize(0.05)} & {\scriptsize(0.00)} & {\scriptsize(0.00)} \\
\bottomrule
\end{tabular}

\end{table}

The application is estimated on the first-differenced series $\Delta \mathbf{Y}_t$ described in Section~\ref{sec:data}, rather than on the levels $\mathbf{Y}_t$.
As in the coincident-index literature \citep{stock1989indexes, crone2005coincident}, the logged quantity indicators and the unemployment rate are treated as integrated of order one, so that the differenced series are the stationary objects on which the model is built.
This appendix documents the unit-root behavior of those differenced series and confirms that the stationarity assumption underlying the model is supported by the data.

Table~\ref{tab:adf} reports augmented Dickey Fuller (ADF) tests for each of the $9 \times 4 = 36$ differenced series.
Each test includes an intercept, with the lag order selected by BIC.
The unit-root null is rejected at the 5\% level for 32 of the 36 series.
The four exceptions are employment growth in Michigan and Ohio and the change in the unemployment rate in Indiana and Michigan.
With the exception of Michigan employment and Indiana Unemployment, the test statistic lies close to the 5\% critical value.

\color{black}

\subsection{Additional Simulation Results}
\label{sec:addsimresults}

\begin{figure}[t]
  \centering

  \begin{subfigure}[b]{0.45\textwidth}
    \centering
    \includegraphics[width=\textwidth]{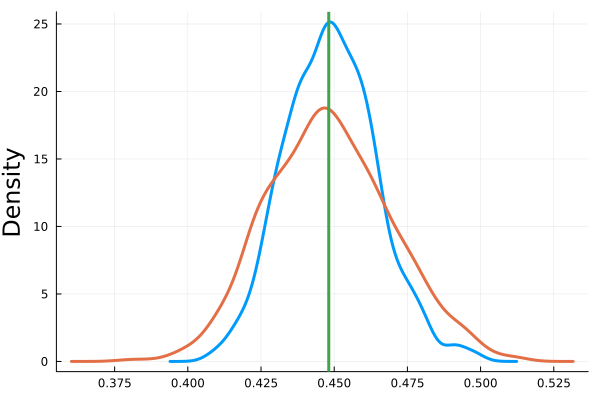}
    \caption{First $\widehat{\delta}$, $T=100$}
  \end{subfigure}
  \hfill
  \begin{subfigure}[b]{0.45\textwidth}
    \centering
    \includegraphics[width=\textwidth]{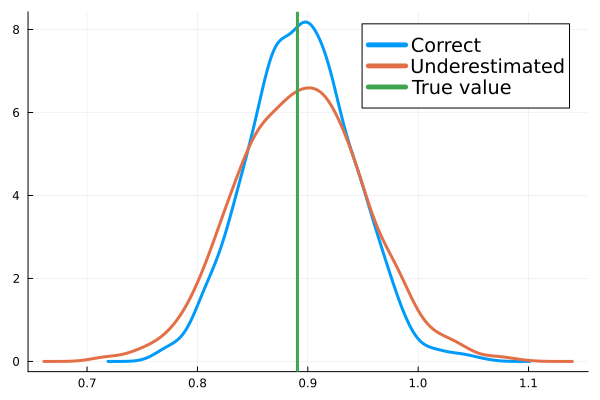}
    \caption{Second $\widehat{\delta}$, $T=100$}
  \end{subfigure}

  \vspace{1em}

  \begin{subfigure}[b]{0.45\textwidth}
    \centering
    \includegraphics[width=\textwidth]{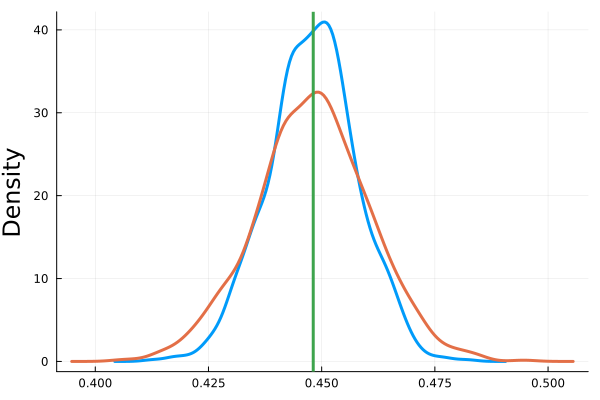}
    \caption{First $\widehat{\delta}$, $T=250$}
  \end{subfigure}
  \hfill
  \begin{subfigure}[b]{0.45\textwidth}
    \centering
    \includegraphics[width=\textwidth]{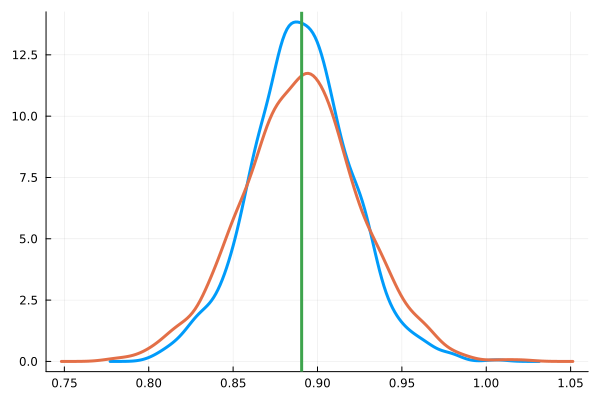}
    \caption{Second $\widehat{\delta}$, $T=250$}
  \end{subfigure}

  \caption{Sampling distributions of the estimated $\delta$ parameters under correct, under‐ and over‐rank specifications in the first rank and for two different sample sizes. Top row: $T=100$. Bottom row: $T=250$. Each panel shows the kernel density of one component of $\widehat{\boldsymbol{\delta}}$, with the true value marked by a vertical line.}
  \label{fig:additional_delta_results}
\end{figure}

\begin{figure}[t]
    \centering
    \includegraphics[width=\textwidth]{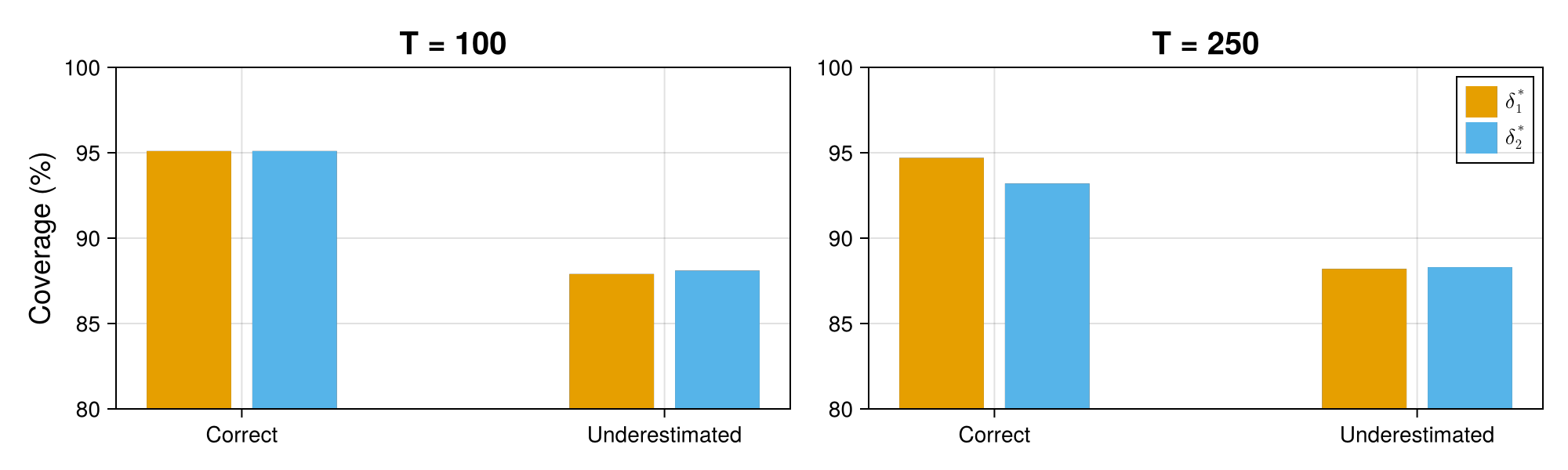}
    \caption{Empirical coverage rates (in \%) of 95\% confidence intervals for $\delta^*_1$ (left bars) and $\delta^*_2$ (right bars) under correct, under‐ and over‐rank specification in the second dimension, as indicated on the horizontal axis. Top row: $T=100$. Bottom row: $T=250$.
    }
    \label{fig:add_delta_cov_results}
\end{figure}

Additional estimation and coverage results are detailed for the case with matrix-valued time series of dimension $3 \times 6$.
The $\boldsymbol{\delta}^{*}$ parameter is examined with the same simulation setup as detailed in Section \ref{sec:inference_and_coverage}.
With dimensions $3 \times 6$, this allows the examination of a more severe rank reduction in the second dimension.
The data are generated according to the pseudo-structural model in equation \eqref{eq:pseudo-structural} with ranks $r_{1} \times r_{2} = 2 \times 5$ and one lag.
The kernel density of $\boldsymbol{\delta}^{*}$ is thus plotted in two cases: (i) correctly estimated ranks $\widehat{r}_{1} \times \widehat{r}_{2} = 2 \times 5$ and (ii) underestimated rank $\widehat{r}_{1} \times \widehat{r}_{2} =  2 \times 1$ in the second dimension.
The overestimated case is omitted, as it still aligns closely with the correctly estimated rank.
The results are given in Figure \ref{fig:additional_delta_results} and \ref{fig:add_delta_cov_results}.

As can be seen, a large efficiency gain is obtained when correctly estimating the rank.
However, underestimating the second dimension still results in a density that is centered and symmetric around the true value.
This effect becomes more pronounced as the number of observations increases.
Moreover, as expected, correctly estimating the rank leads to coverage close to 95\%.
Underestimating the rank, on the other hand, drops the coverage to approximately 87\%.
Thus, little coverage is lost even if the rank is underestimated by a substantial amount.

\begin{figure}[t]
  \centering
  \begin{subfigure}[b]{0.48\textwidth}
    \centering
    \includegraphics[width=\textwidth]{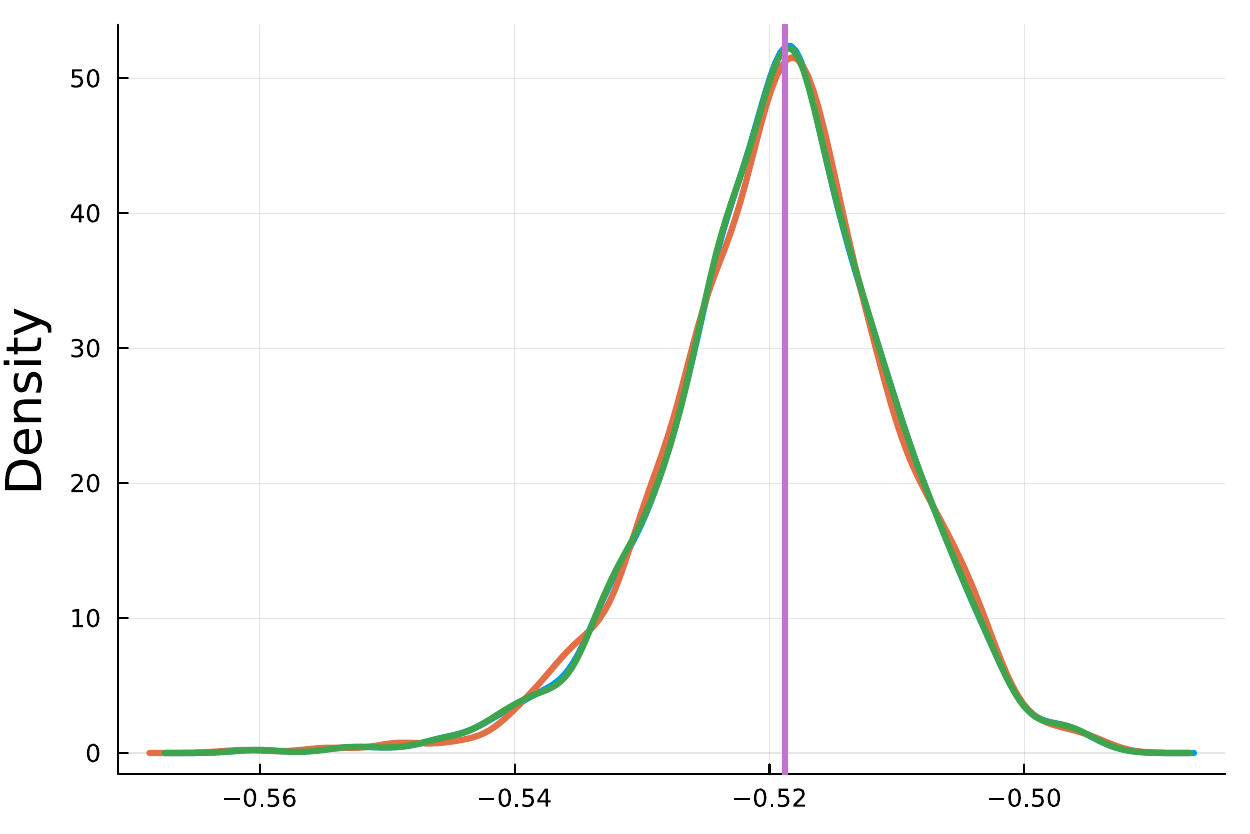}
    \caption{$\widehat{\delta}_1^*$, $T=100$}
  \end{subfigure}
  \hfill
  \begin{subfigure}[b]{0.48\textwidth}
    \centering
    \includegraphics[width=\textwidth]{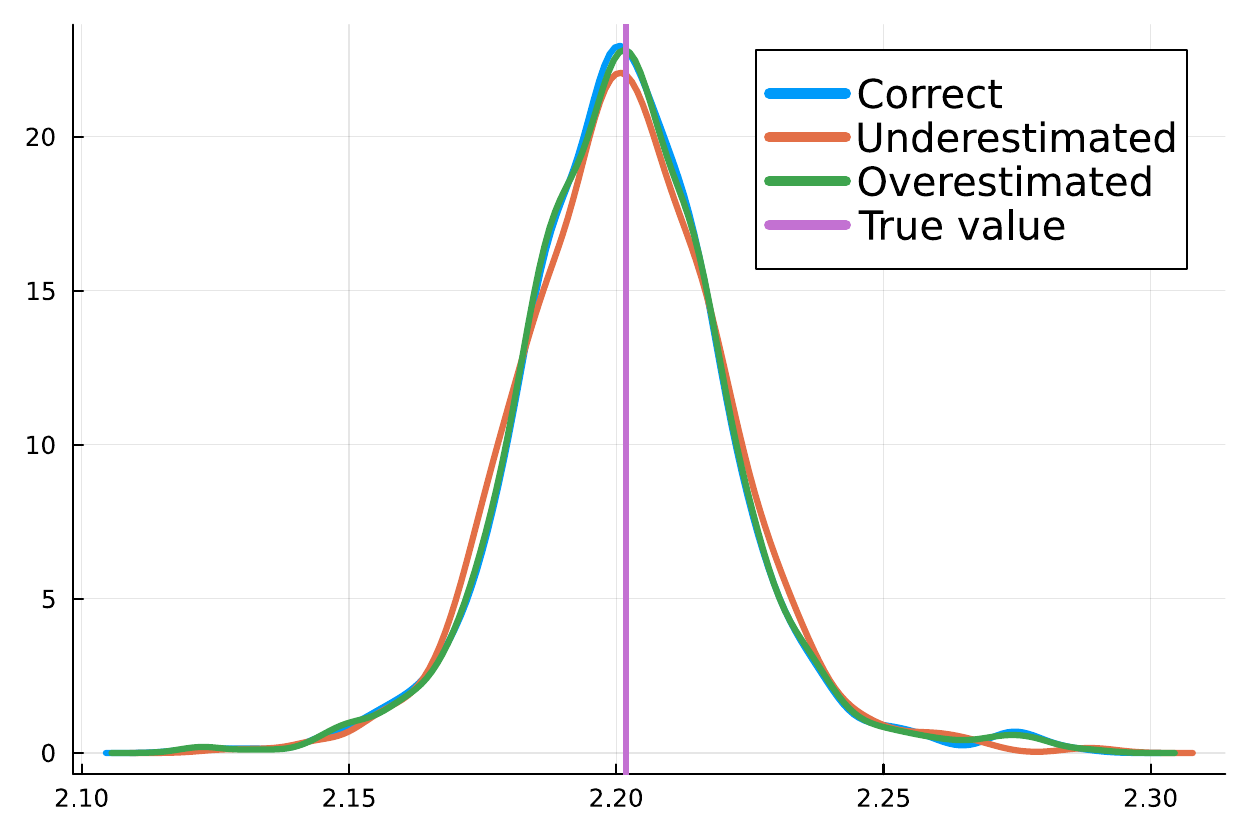}
    \caption{$\widehat{\delta}_2^*$, $T=100$}
  \end{subfigure}

  \vspace{0.5em}

  \begin{subfigure}[b]{0.48\textwidth}
    \centering
    \includegraphics[width=\textwidth]{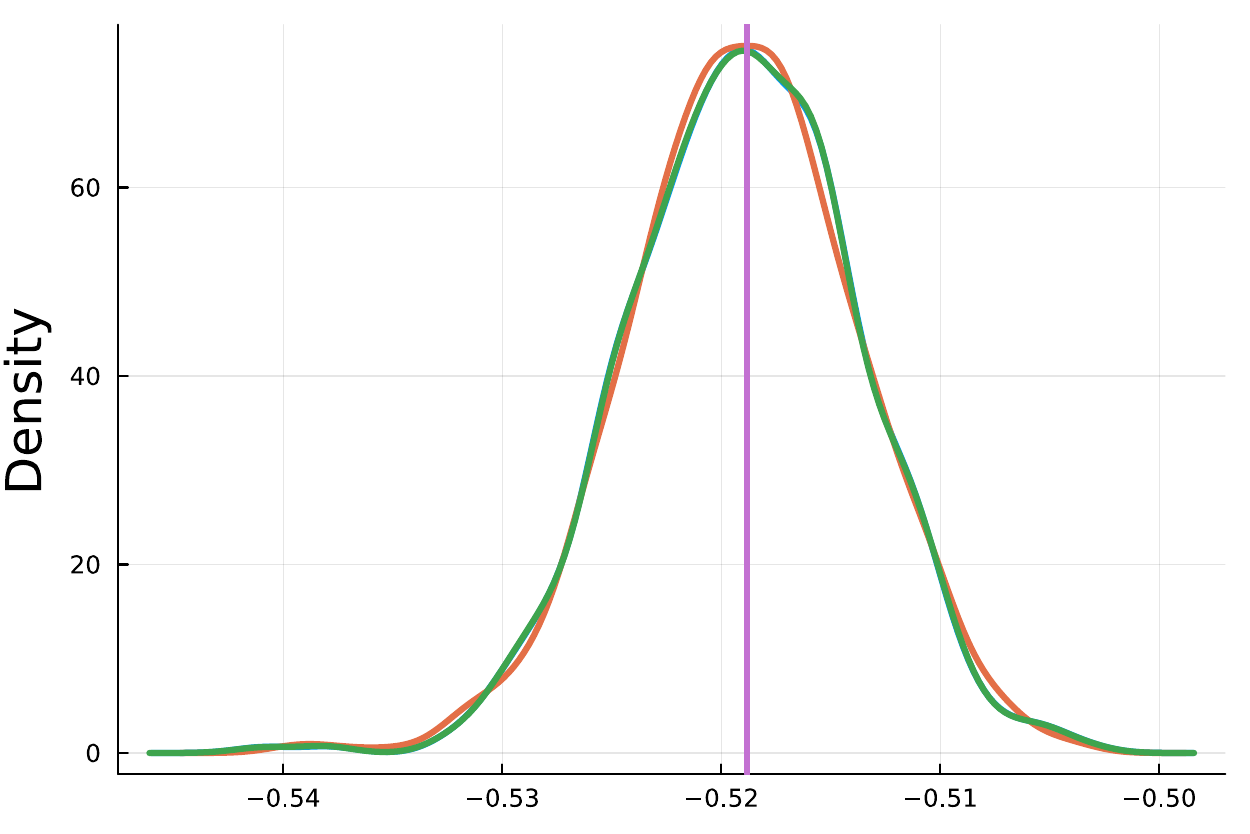}
    \caption{$\widehat{\delta}_1^*$, $T=250$}
  \end{subfigure}
  \hfill
  \begin{subfigure}[b]{0.48\textwidth}
    \centering
    \includegraphics[width=\textwidth]{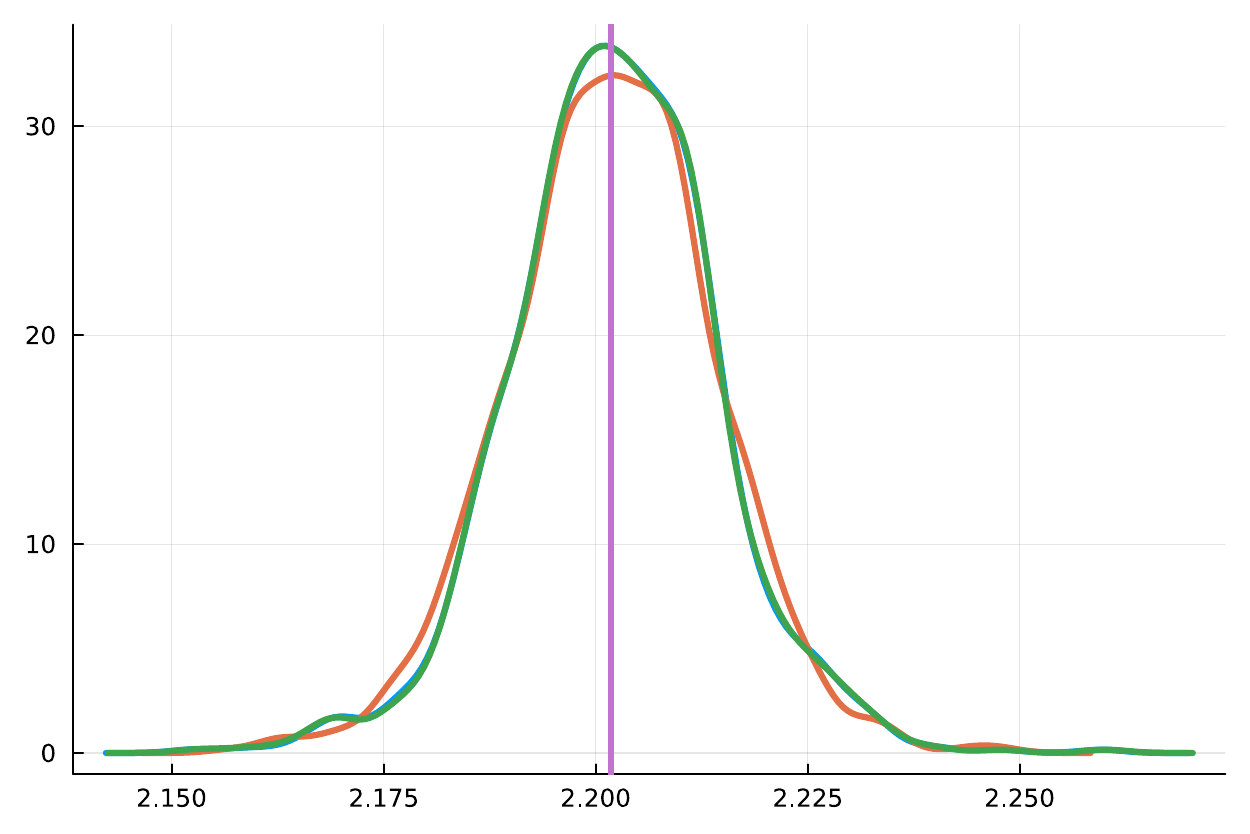}
    \caption{$\widehat{\delta}_2^*$, $T=250$}
  \end{subfigure}
  \caption{Sampling distributions of $\widehat{\delta}_1^*$ (left) and $\widehat{\delta}_2^*$ (right) under correct, under- and over-rank specifications in the second rank with the RRMAR initialization. Each panel shows the kernel density with the true value marked by a vertical line. Top row: $T=100$; bottom row: $T=250$.}
  \label{fig:rrmar_delta_densities}
\end{figure}

\begin{figure}[t]
  \centering
  \begin{subfigure}[b]{0.32\textwidth}
    \centering
    \includegraphics[width=\textwidth]{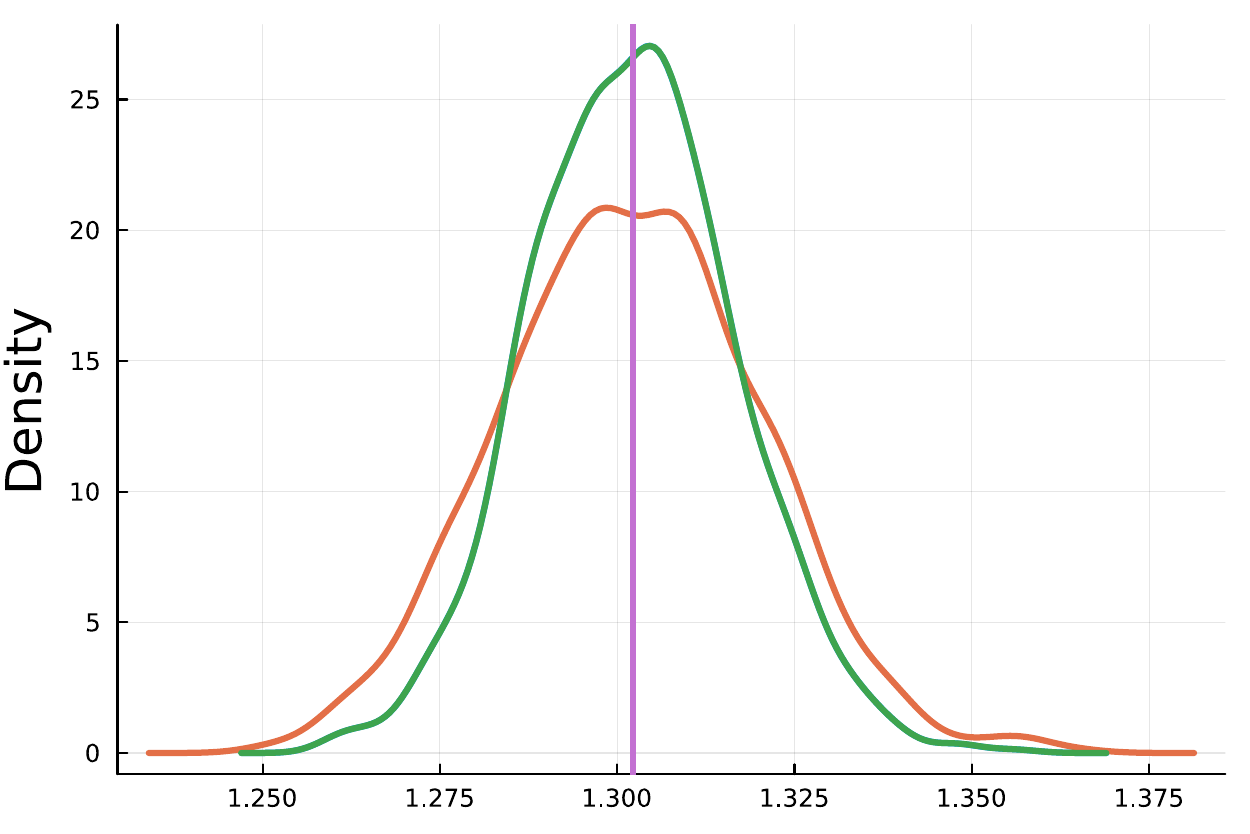}
    \caption{$\widehat{\gamma}_1^*$, $T=100$}
  \end{subfigure}
  \hfill
  \begin{subfigure}[b]{0.32\textwidth}
    \centering
    \includegraphics[width=\textwidth]{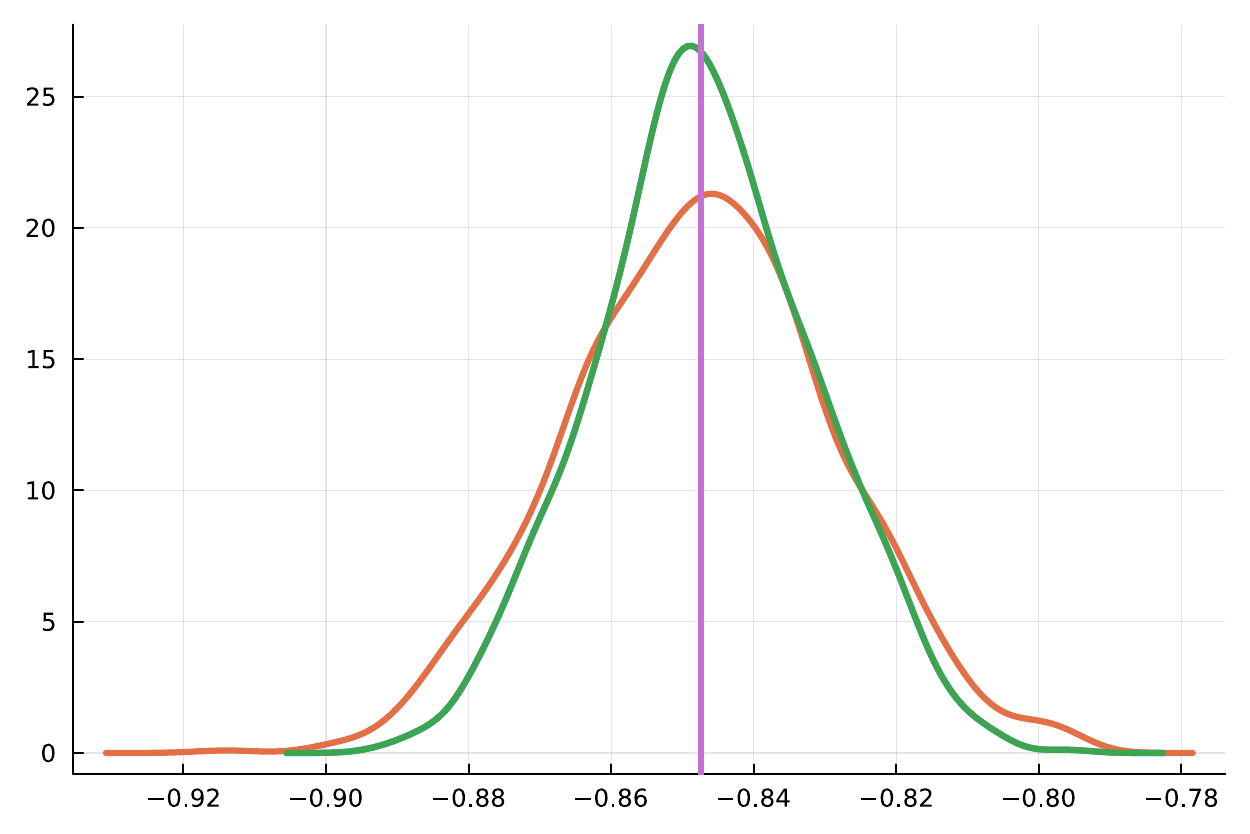}
    \caption{$\widehat{\gamma}_2^*$, $T=100$}
  \end{subfigure}
  \hfill
  \begin{subfigure}[b]{0.32\textwidth}
    \centering
    \includegraphics[width=\textwidth]{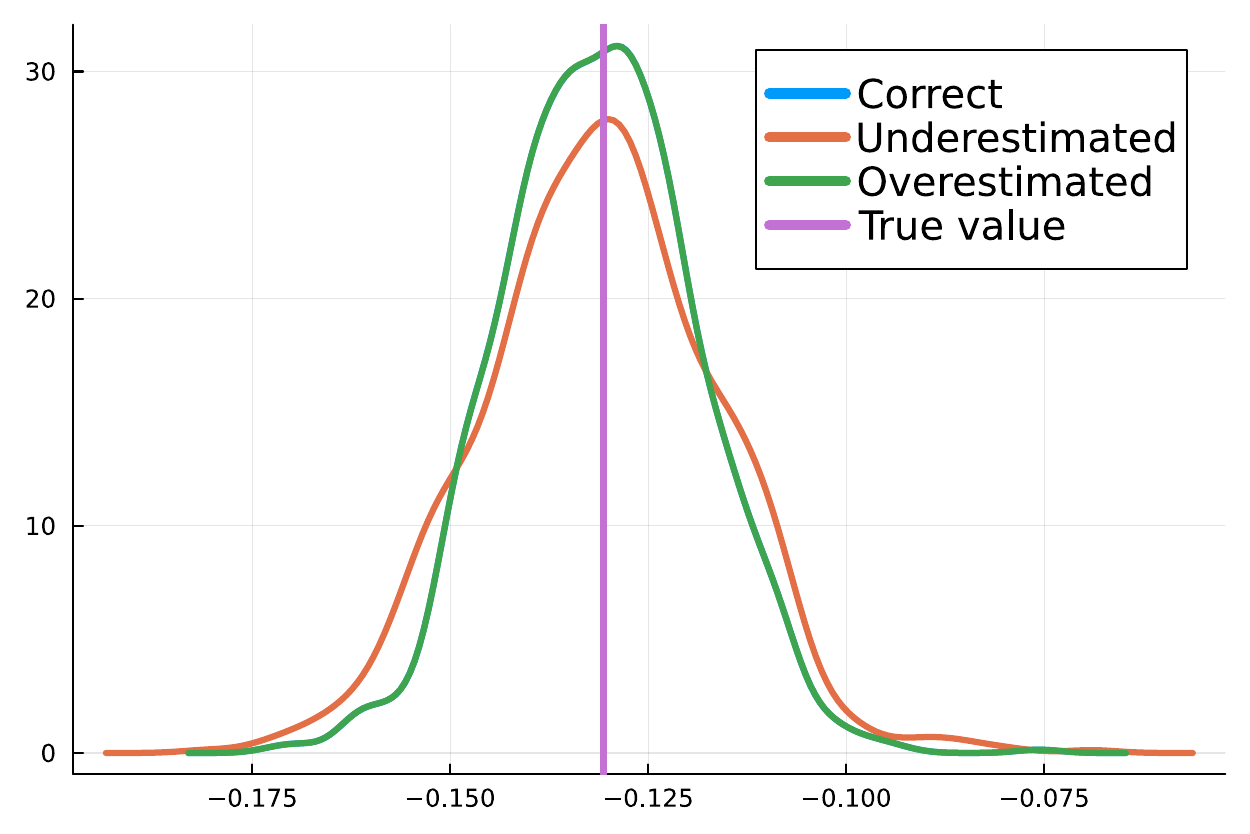}
    \caption{$\widehat{\gamma}_3^*$, $T=100$}
  \end{subfigure}

  \vspace{1em}

  \begin{subfigure}[b]{0.32\textwidth}
    \centering
    \includegraphics[width=\textwidth]{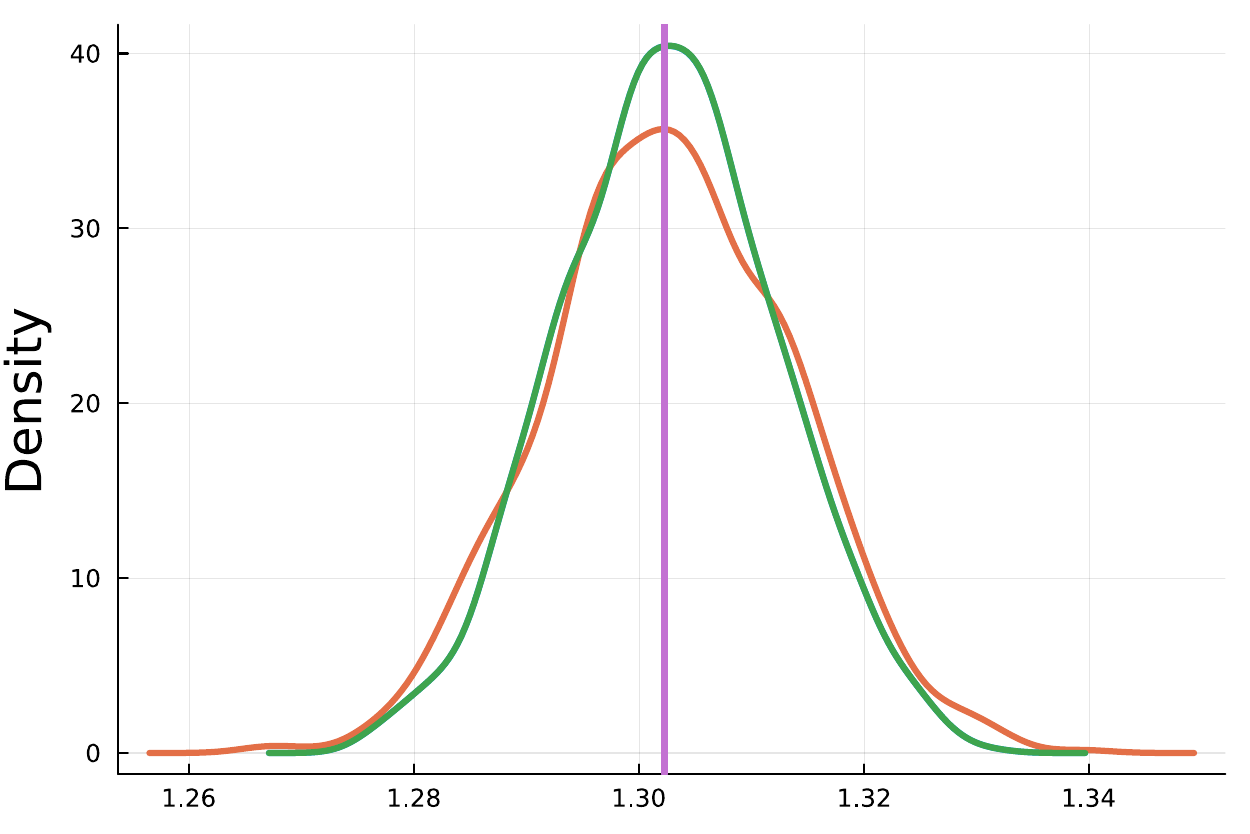}
    \caption{$\widehat{\gamma}_1^*$, $T=250$}
  \end{subfigure}
  \hfill
  \begin{subfigure}[b]{0.32\textwidth}
    \centering
    \includegraphics[width=\textwidth]{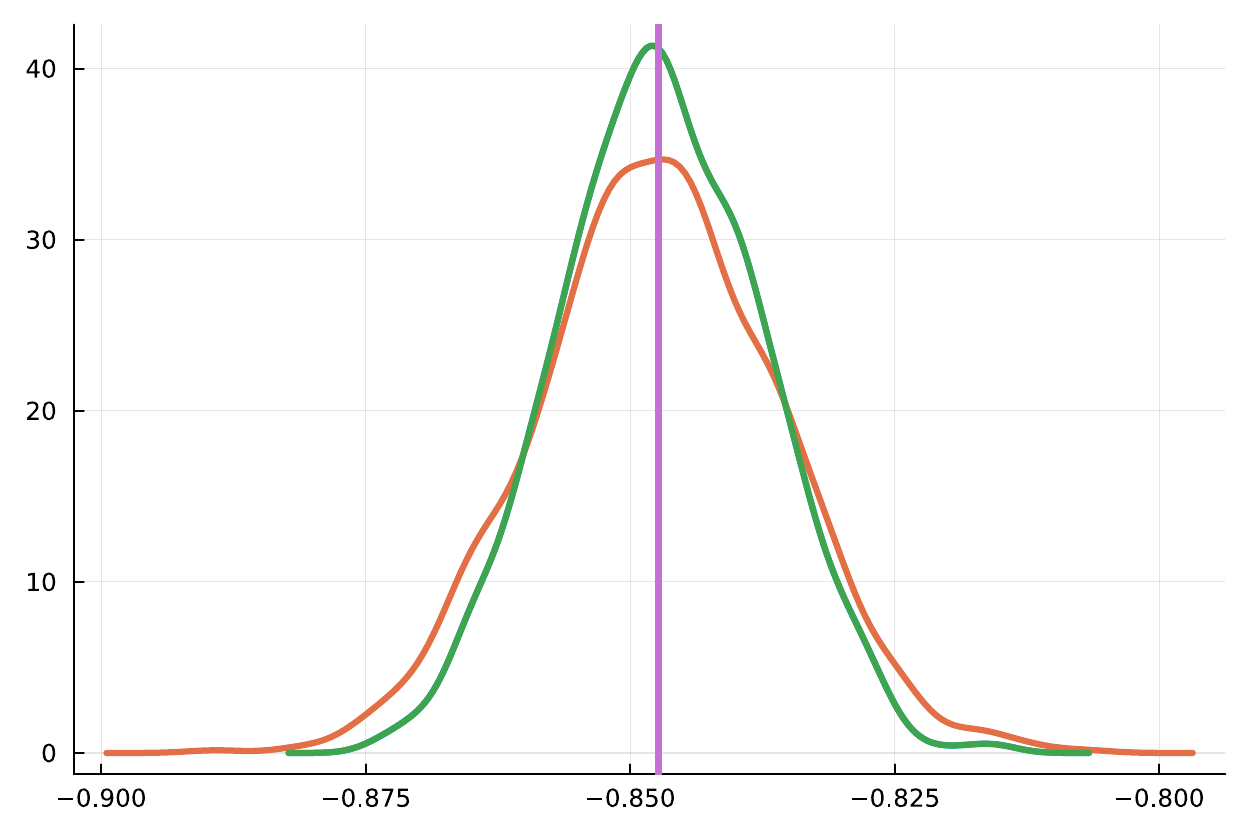}
    \caption{$\widehat{\gamma}_2^*$, $T=250$}
  \end{subfigure}
  \hfill
  \begin{subfigure}[b]{0.32\textwidth}
    \centering
    \includegraphics[width=\textwidth]{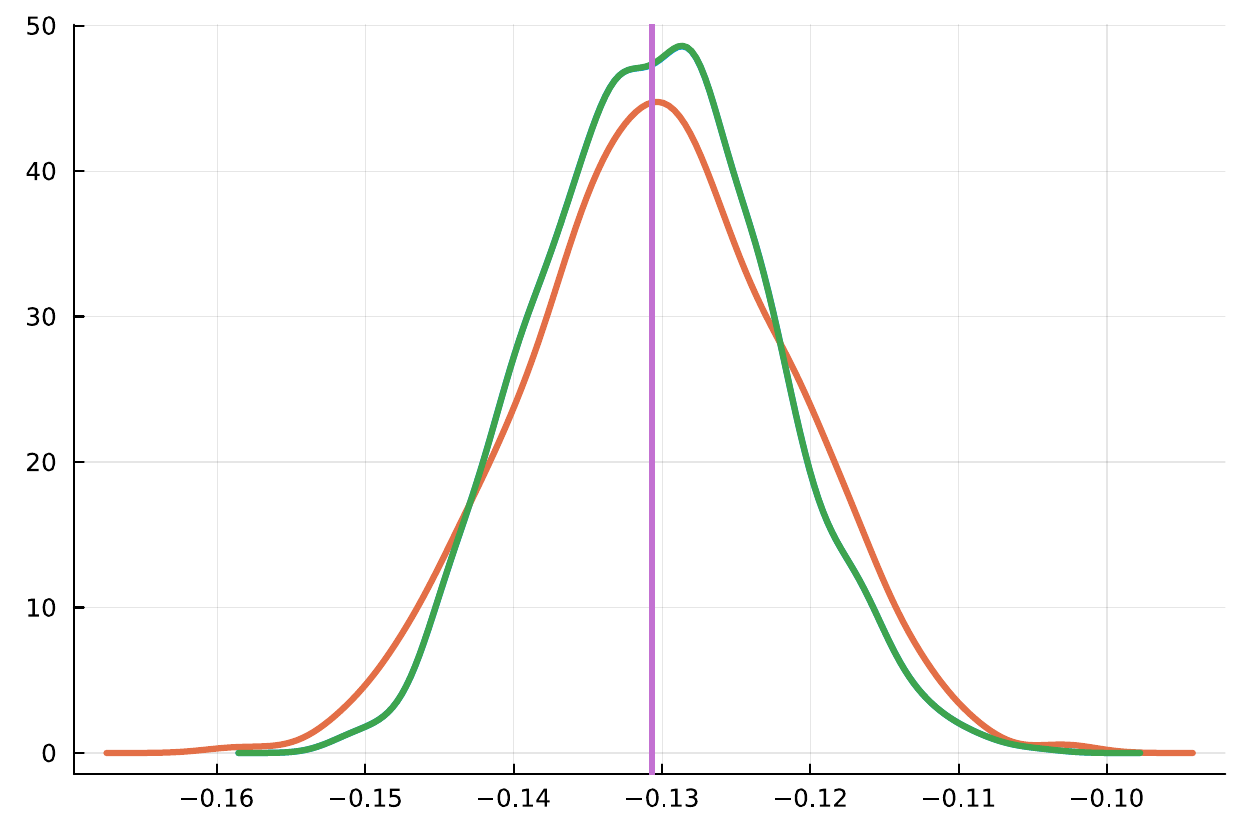}
    \caption{$\widehat{\gamma}_3^*$, $T=250$}
  \end{subfigure}

  \caption{Sampling distributions of the estimated $\gamma$ parameters under correct, under‐ and over‐rank specifications in the first rank for two different sample sizes with the RRMAR initialization. Top row: $T=100$. Bottom row: $T=250$. Each panel shows the kernel density of one component of $\widehat{\boldsymbol{\gamma}}$, with the true value marked by a vertical line.}
  \label{fig:rrmar_gamma_densities_combined}
\end{figure}

\begin{figure}[t]
    \centering
    \begin{subfigure}[b]{0.48\textwidth}
        \centering
        \includegraphics[width=\linewidth]{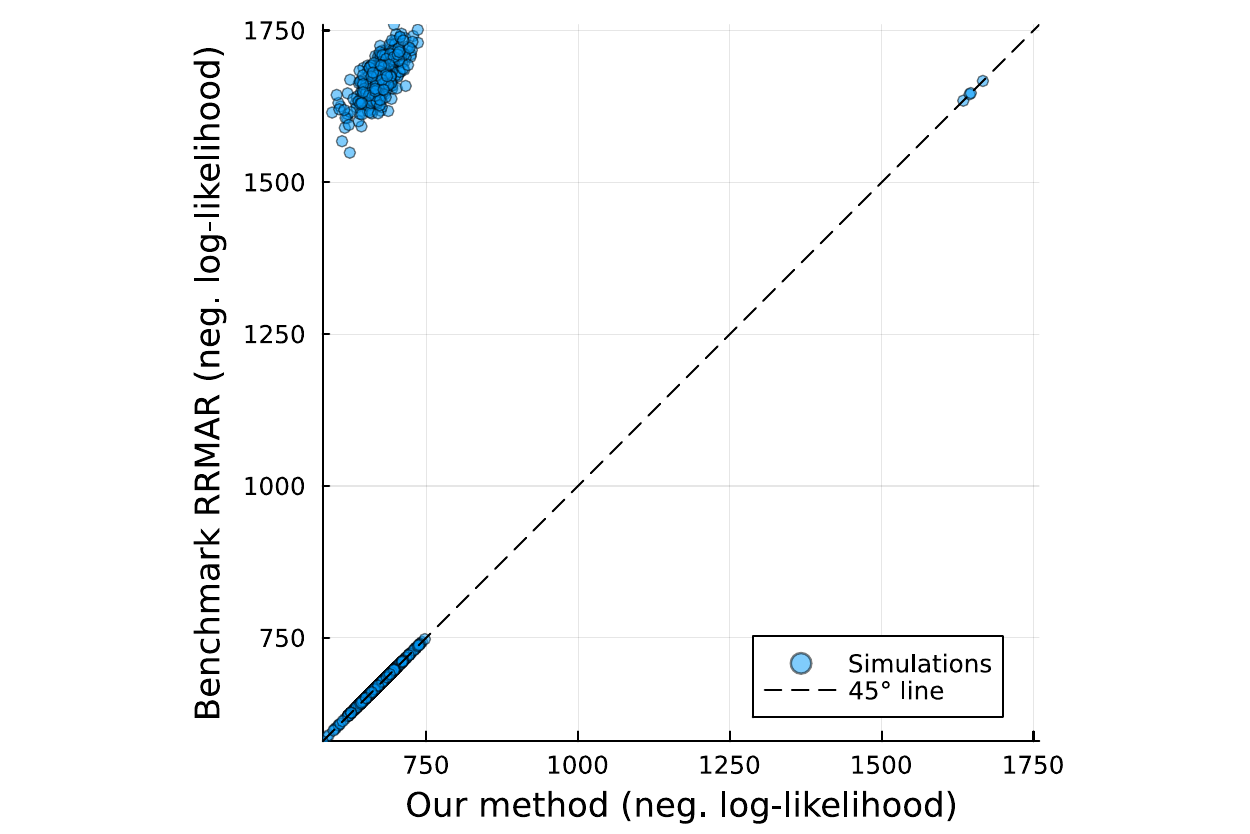}
        \caption{$\boldsymbol{\delta}^*$, $T=100$}
        \label{fig:sub1}
    \end{subfigure}
    \hfill
    \begin{subfigure}[b]{0.48\textwidth}
        \centering
        \includegraphics[width=\linewidth]{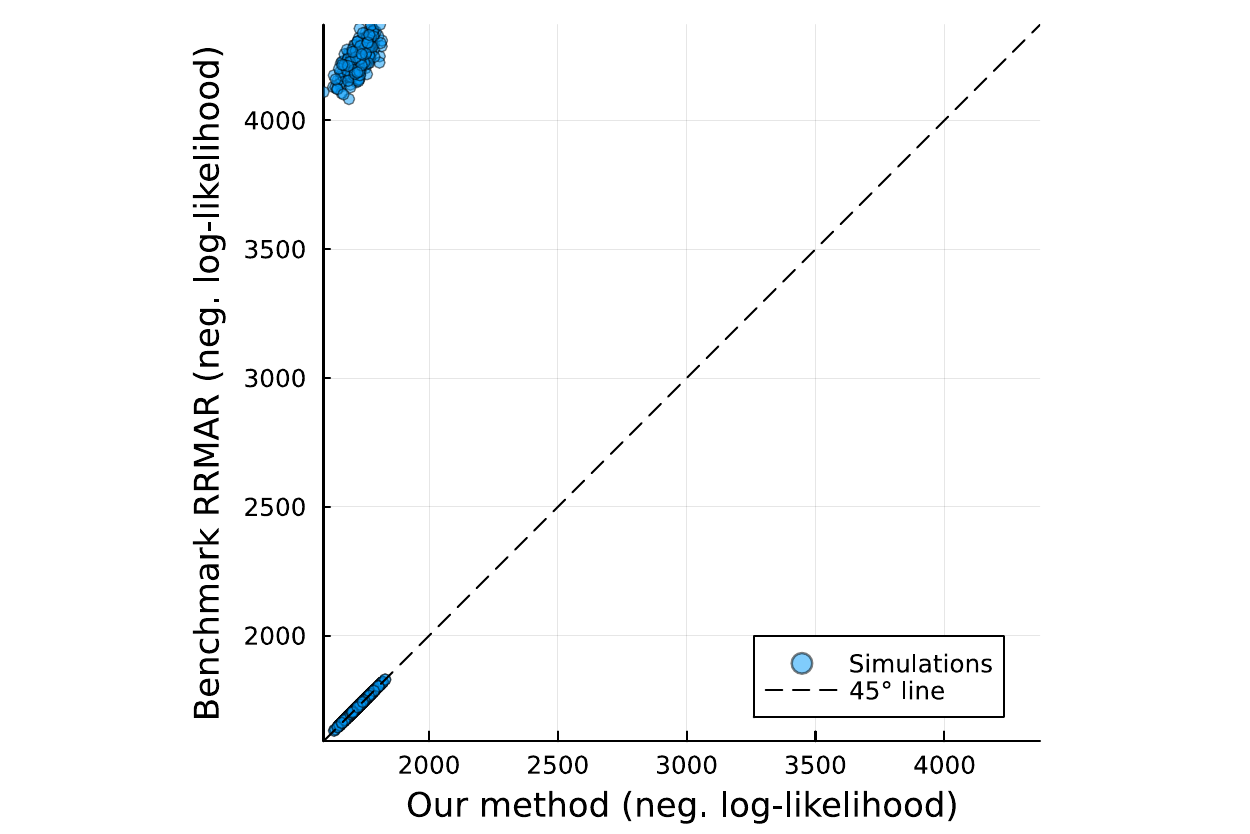}
        \caption{$\boldsymbol{\delta}^*$, $T=250$}
        \label{fig:sub2}
    \end{subfigure}

    \vspace{0.5em}

    \begin{subfigure}[b]{0.48\textwidth}
        \centering
        \includegraphics[width=\linewidth]{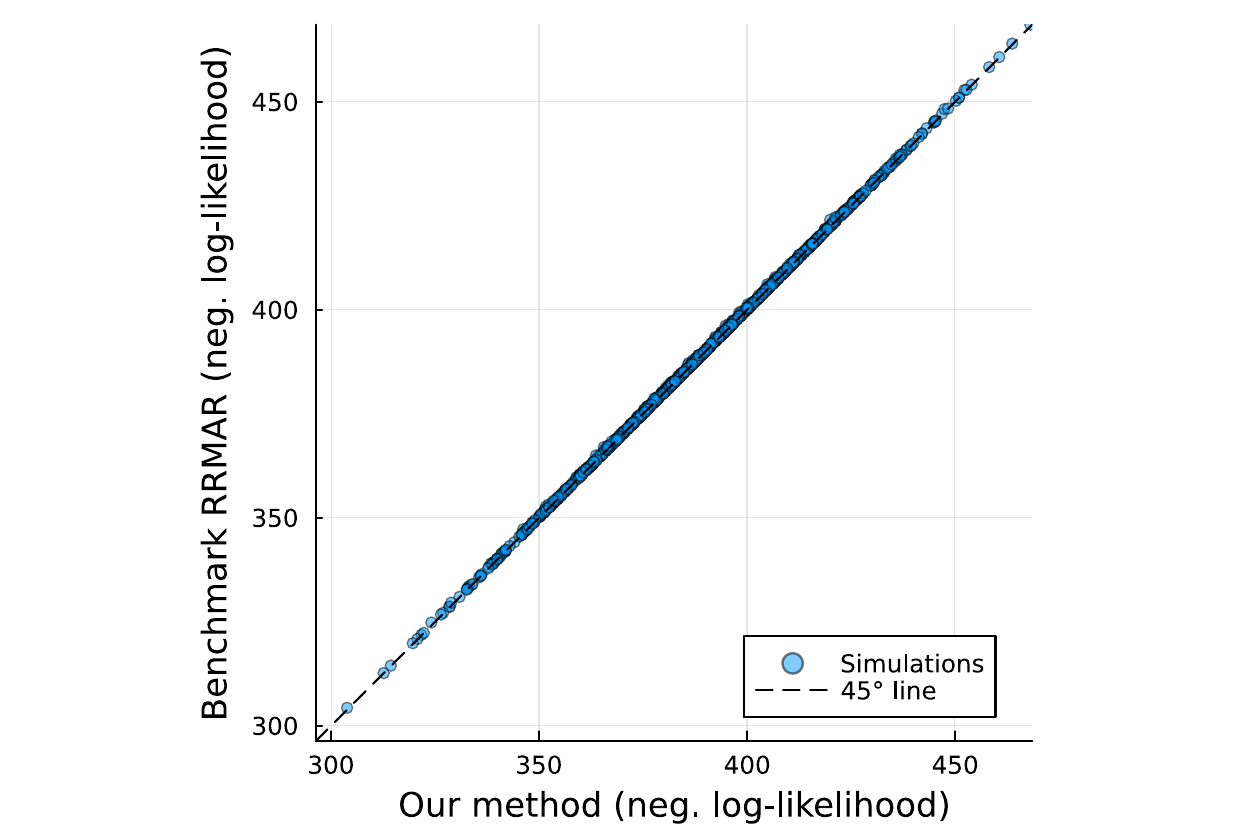}
        \caption{$\boldsymbol{\gamma}^*$, $T=100$}
        \label{fig:sub3}
    \end{subfigure}
    \hfill
    \begin{subfigure}[b]{0.48\textwidth}
        \centering
        \includegraphics[width=\linewidth]{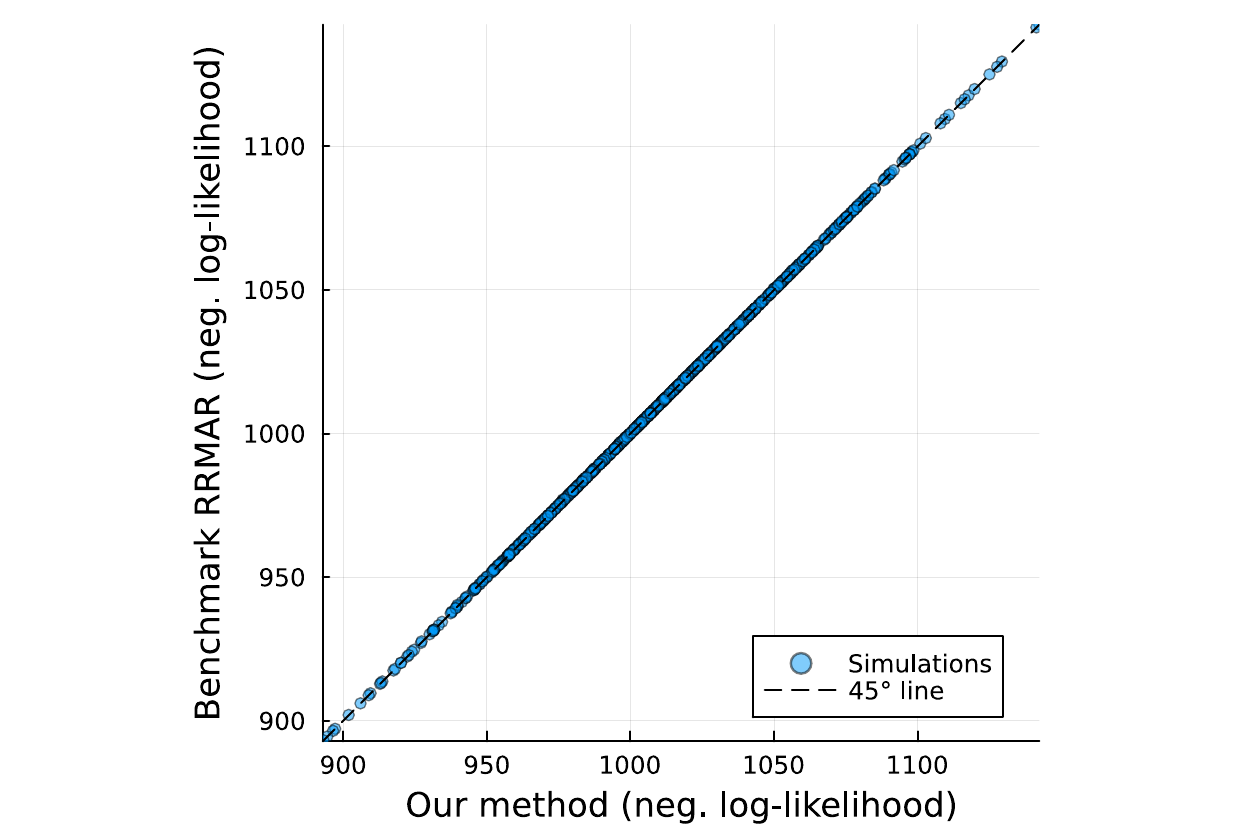}
        \caption{$\boldsymbol{\gamma}^*$, $T=250$}
        \label{fig:sub4}
    \end{subfigure}
    \caption{Negative log-likelihood achieved by RRMAR initialization versus the pseudo-structural estimator across 1000 simulations. Points on the 45$^\circ$ line indicate identical objective values; points above indicate simulations in which the pseudo-structural estimator achieves a strictly lower negative log-likelihood.}
    \label{fig:likelihood_comparison}
\end{figure}

Simulation results for the RRMAR initialization given by \citet{xiao_reduced_forth} are additionally provided.
The co-movement parameters $\delta_1^*$, $\delta_2^*$, $\gamma_1^*$, $\gamma_2^*$, and $\gamma_3^*$ are obtained by rotating the matrices $\mathbf{U}_1$ and $\mathbf{U}_2$ such that the bottom of each matrix is the identity, as in Section \ref{sec:pseudostructural}.
These are densities prior to the multi-start procedure, and thus can get stuck at local optima.
For the $\boldsymbol{\delta}$ simulations in Figure \ref{fig:rrmar_delta_densities}, \citet{xiao_reduced_forth} yield similar densities for overspecifying, underspecifying, and correctly specifying the second dimension, masking the efficiency gains from correctly or overspecifying the second dimension.
The $\boldsymbol{\gamma}$ simulations in Figure \ref{fig:rrmar_gamma_densities_combined} do not show this, and the multi-start procedure yields similar results.
Figure \ref{fig:likelihood_comparison} compares the negative log-likelihood values attained by the two approaches.
While the RRMAR initialization typically reaches the same optimum as the multi-start procedure, it becomes trapped in local optima in roughly a third of the $\boldsymbol{\delta}^*$ simulations.

\bibliographystyle{elsarticle-harv} 
\bibliography{references}

\end{document}